\global\def\draftcontrol{0}

   \def\versionno{ Holographic Kondo DpDq -- draft   }
\catcode`\@=11

\expandafter\ifx\csname draftcontrol\endcsname\relax\global\def\draftcontrol{0}
\fi

{\count255=\time\divide\count255 by 60
\xdef\hourmin{\number\count255}
\multiply\count255 by-60\advance\count255 by\time
\xdef\hourmin{\hourmin:\ifnum\count255<10 0\fi\the\count255}}
\def\draftdate{\number\month/\number\day/\number\year\ \ \ \hourmin }

\newcommand\makepapertitle{\par
  \begingroup
    \renewcommand\thefootnote{\@fnsymbol\c@footnote}%
    \def\@makefnmark{\rlap{\@textsuperscript{\normalfont\@thefnmark}}}%
    \long\def\@makefntext##1{\parindent 1em\noindent
            \hb@xt@1.8em{%
                \hss\@textsuperscript{\normalfont\@thefnmark}}##1}%
     \newpage
     \global\@topnum\z@   % Prevents figures from going at top of page.
     \@makepapertitle
     \thispagestyle{empty}\@thanks
  \endgroup
  \setcounter{footnote}{0}%
  \global\let\thanks\relax
  \global\let\makepapertitle\relax
  \global\let\@makepapertitle\relax
  \global\let\@thanks\@empty
  \global\let\@author\@empty
  \global\let\@date\@empty
  \global\let\@title\@empty
  \global\let\title\relax
  \global\let\author\relax
  \global\let\date\relax
  \global\let\and\relax
  \def\version{\let\version\@version\@gobble}
}
\def\@makepapertitle{%
  \newpage
   \ifnum\draftcontrol=1 {}
   \version\versionno
   \vskip 3em%
   \else
   \hfill\hbox to 3cm {\parbox{4cm}{\@pubnum}\hss}%
   \vskip 3em%
   \fi
   \begin{center}%
   \let \footnote \thanks
     {\LARGE {\@title}}%
     \vskip 1.5em%
     {\normalsize%\large
       \lineskip .5em%
       \begin{tabular}[t]{c}%
         \@author
       \end{tabular}\par}%
     \vskip 1.5em%
     {\@bstract}%
     \end{center}%
     \vskip 1.5em
     \@date%
   \par
}

\gdef\@pubnum{}
\def\pubnum#1{%
  \gdef\@pubnum{#1}}

\gdef\@bstract{}
\def\Abstract#1{%
  \gdef\@bstract{%
   \parbox{\textwidth-0pc}{%
   \centerline{\bf Abstract}\penalty1000%
\kern.2cm%
\noindent%\abstractfont \baselineskip=12pt
\renewcommand\baselinestretch{1.0}%
{#1}}}
}

\def\ps@paper{\let\@mkboth\@gobbletwo%
     \ifnum\draftcontrol=1
    \def\@oddfoot{\hbox to \textwidth{\tiny \versionno \hfil\tiny\draftdate}%
    \hskip -\textwidth \hbox to \textwidth{\hfil\rm\thepage\hfil}}%
     \else\def\@oddfoot{\hbox to \textwidth{\hfil\rm\thepage\hfil}}
     \fi
     \let\@evenfoot\@oddfoot
}
\def\body{\clearpage
          \pagestyle{paper}
    }
\def\@version#1{\ifnum\draftcontrol=1
\typeout{}\typeout{#1}\typeout{}
\vskip3mm\centerline{\hbox{\fbox{\normalsize{\tt DRAFT -- #1 -- }
                   {\draftdate}}}}\vskip3mm
\fi}
\let\version\@version
\long\def\eqlabel#1{\ifnum\draftcontrol=1
                    \tag@false  % there are some problems with multline without this
                    \tag*{(\theequation) \hbox to -0.2cm{\hspace{0cm}\small{#1}\hss}}
                    \refstepcounter{equation}
                    \edef\@currentlabel{\theequation}
                    \ltx@label{#1}          % use old LaTeX \label instead of new definition
                    \else
                    \label{#1}
                    \fi
                    }
\let\st@bibitem\@bibitem
\let\st@lbibitem\@lbibitem
\ifnum\draftcontrol=1
  \def\@bibitem#1{%
    \st@bibitem{#1}\a@@label{#1}\ignorespaces}
  \def\@lbibitem[#1]#2{%
    \st@lbibitem[#1]{#2}\a@@label{#2}\ignorespaces}
  \def\a@@label#1{%
    \gdef\a@lab{\smash{\normalfont\small#1}}
    \ifvmode
      \if@inlabel
        \global\setbox\@labels\hbox{%
          \llap{\a@lab\let\a@lab\relax
                \kern\@totalleftmargin\kern\marginparsep}%
          \box\@labels}%
      \fi
    \fi}
\fi
\documentclass[11pt,letterpaper]{article}
\usepackage{amsmath,amssymb,array,calc,epsfig}
\usepackage{cite}
\usepackage[
      colorlinks=true,
      linkcolor=red,  
      urlcolor=blue,    
      filecolor=red,     
      linkcolor=blue,
      citecolor = red,
      pdfstartview=FitV,
      pdftitle={},
        pdfauthor={Paolo Benincasa},
        pdfsubject={},
        pdfkeywords={},
        pdfpagemode=None,
        bookmarksopen=true    
      ]{hyperref}
      
\usepackage{graphicx}
\usepackage{subfigure}
\usepackage{epstopdf}
\usepackage{ccaption}

\ifnum\draftcontrol=1
\fi
\renewcommand\baselinestretch{1.25}
\renewcommand\section{\@startsection {section}{1}{\z@}%
                                   {-3.5ex \@plus -1ex \@minus -.2ex}%
                                   {2.3ex \@plus.2ex}%
                                   {\normalfont\large\bfseries}}
\renewcommand\subsection{\@startsection{subsection}{2}{\z@}%
                                   {-3.25ex\@plus -1ex \@minus -.2ex}%
                                   {1.5ex \@plus .2ex}%
                                   {\normalfont\normalsize\bfseries}}
\renewcommand\subsubsection{\@startsection{subsubsection}{3}{\z@}%
                                   {-3.25ex\@plus -1ex \@minus -.2ex}%
                                   {1.5ex \@plus .2ex}%
                                   {\normalfont\normalsize\it}}
\renewcommand\paragraph{\@startsection{paragraph}{4}{\z@}%
                                   {-3.25ex\@plus -1ex \@minus -.2ex}%
                                   {1.5ex \@plus .2ex}%
                                   {\normalfont\normalsize\bf}}

\numberwithin{equation}{section}

\def\ie{{\it i.e.}}

\def\revise#1       {\raisebox{-0em}{\rule{3pt}{1em}}%
                     \marginpar{\raisebox{.5em}{\vrule width3pt\
                     \vrule width0pt height 0pt depth0.5em
                     \hbox to 0cm{\hspace{0cm}{%
                     \parbox[t]{4em}{\raggedright\footnotesize{#1}}}\hss}}}}

\def\sqr#1#2{{\vcenter{\vbox{\hrule height.#2pt
 \hbox{\vrule width.#2pt height#1pt \kern#1pt
 \vrule width.#2pt}\hrule height.#2pt}}}}

\newcommand{\beq}{\begin{equation}}
\newcommand{\eeq}{\end{equation}}
\newcommand{\bea}{\begin{eqnarray}}
\newcommand{\eea}{\end{eqnarray}}
\newcommand{\bear}{\begin{eqnarray}}
\newcommand{\eear}{\end{eqnarray}}
\newcommand{\rc}{\nonumber\\}

\def\aa1{\phi}
\def\cc1{\psi}

\catcode`\@=12

\begin{document}

%%%
%%%%%% text starts here
%%%%%%%%%

\title{\bf Holographic Kondo Model in Various Dimensions}

%\pubnum{%
%arXiv:11xx.xxxx}
\date{April 2012}

\author{
\scshape Paolo Benincasa${}^{\dagger}$, Alfonso V. Ramallo${}^{\ddagger}$\\[0.4cm]
\ttfamily Departamento de F{\'i}sica de Part{\'i}culas,\\
\ttfamily Universidade de Santiago de Compostela\\
\ttfamily E-15782 Santiago de Compostela, Spain\\[0.2cm]
\small \ttfamily ${}^{\dagger}$paolo.benincasa@usc.es, ${}^{\ddagger}$alfonso@fpaxp1.usc.es
}

\Abstract{We study the addition of localised impurities to $U(N)$ Supersymmetric Yang-Mills theories in $(p+1)$-dimensions by using the
 gauge/gravity correspondence. From the gravity side, the impurities are introduced by considering probe D$(8-p)$-branes extending
 along the time and radial directions and wrapping an $(7-p)$-dimensional submanifold of the internal $(8-p)$-sphere, so that the
 degrees of freedom are point-like from the gauge theory perspective. We analyse both the configuration in which the branes
 generate straight flux tubes -- corresponding to actual single impurities -- and the one in which connected flux tubes are created
 -- corresponding to dimers. We discuss the thermodynamics of both the configurations and the related phase transition. 
 In particular, the specific heat of the straight flux-tube configuration is negative for $p\,<\,3$,
 while it is never the case for the connected one. We study the  stability of the system by looking at the impurity fluctuations. Finally, we characterise the theory 
 by computing one- and two-point  correlators of the gauge theory operators dual to the impurity fluctuations. Because of the underlying generalised conformal 
 structure, such correlators can be expressed in terms of an effective coupling constant (which runs because of its dimensionality) and a
 generalised conformal dimension. 
}

\makepapertitle

\body

\version\versionno

\section{Introduction}\label{sec:Intro}

The gauge/gravity correspondence \cite{Maldacena:1997re, Gubser:1998bc, Witten:1998qj, Aharony:1999ti} provides a powerful handle
on the strongly coupled regime of gauge theories. As originally formulated \cite{Maldacena:1997re}, it conjectures a 
relation between the $AdS_5\times S^5$ ``near-horizon'' geometry generated by a stack of $N$ coincident D$3$-branes and
the four-dimensional $\mathcal{N}\,=\,4$ $SU(N)$ Supersymmetric Yang-Mills theory living on the boundary of $AdS_5$.
Such an equivalence between supergravity on $AdS_5\times S^5$ on one side and $\mathcal{N}\,=\,4$ SYM theory on the other, 
can be made precise by identifying the supergravity partition function with the generating function for the gauge theory 
correlators, with the boundary value of the bulk modes acting as a source of the corresponding gauge theory operator
\cite{Witten:1998qj}.

The correspondence can be extended to the case of generic D$p$-branes ($p\,\neq\,3$), whose ``near-horizon'' geometry 
describes the adjoint degrees of freedom of $U(N)$ Supersymmetric Yang-Mills theory in $(p+1)$-dimensions 
\cite{Itzhaki:1998dd}. One main difference with the $p\,=\,3$ case is that, while the latter is characterised by a 
dimensionless coupling constant, the coupling constant turns out to be dimensionful for $p\not= 3$ and the effective coupling constant 
runs with the energy scale, making the theory non-conformal. However, there exists a frame \cite{Duff:1994fg} -- named 
{\it dual frame} -- in which the background induced by the stack of $N$ D$p$-branes is conformally $AdS_{p+1}\times S^{8-p}$
with the conformal factor depending on a non-trivial dilaton \cite{Boonstra:1997dy, Boonstra:1998yu, Boonstra:1998mp}. 
This frame is further characterised by a manifest generalised conformal symmetry 
\cite{Jevicki:1998yr, Jevicki:1998qs, Jevicki:1998ub}. Specifically, the generalised conformal structure emerges by allowing
the string/Yang-Mills coupling to transform as a background field under conformal transformation, providing, on the other hand, 
diffeomorphism and trace Ward identities. Moreover, in this frame the radial direction plays the role of the energy scale of
the dual gauge-theory \cite{Boonstra:1998mp, Skenderis:1999dq}. The holographic RG flow is, in any case, 
trivial since the theory flows just because of the dimensionality of the coupling constant. The {\it dual frame} has been
crucial to make holography for D$p$-branes precise and to extend the holographic renormalisation procedure\footnote{For more details
about holographic renormalisation, see \cite{Skenderis:2002wp}.}
\cite{Wiseman:2008qa, Kanitscheider:2008kd, Kanitscheider:2009as}. In particular, it has been observed 
\cite{Kanitscheider:2009as} that the $(p+2)$-effective supergravity action, obtained by the Kaluza-Klein reduction on
the $(8-p)$-dimensional sphere, can be recovered by dimensional reduction of the theory on pure asymptotically 
$AdS_{2\sigma+1}$ on a torus, where $\sigma$ is a parameter related to the power of the radial direction in the dilaton
and, generically, takes fractional values for generic values of $p$. Such a parameter $\sigma$ can be considered as
an integer and, after the reduction on the torus, can be analytically continued to take its actual $p$-dependent value.
One can therefore map the problem to a pure $AdS_{2\sigma+1}$ theory and then obtain the $(p+2)$-dimensional answer
through Kaluza-Klein reduction. This way to rephrase the problem allows to drastically simplify the computation of
the counterterms needed for holographic renormalisation and of quantities such as, for example, the correlators of the 
stress-energy tensor.

As we mentioned earlier, the backgrounds generated by a stack of D$p$-branes describe the adjoint degrees of freedom
of the dual gauge theory. The correspondence can however be generalised by inserting extra degrees of freedom in the theory.
In particular, one can add a certain number of branes as probes, so that they do not backreact on the geometry, introducing
a fundamental hyper-multiplet and partially or completely breaking the initial supersymmetries \cite{Karch:2002sh}.

In this paper we are concerned with the introduction of D$(8-p)$-branes in the conformally $AdS_{p+1}\times S^{8-p}$
background generated by a stack of D$p$-branes, in such a way that they wrap an $S^{7-p}\,\subset\,S^{8-p}$ and the
induced metric on their world-volume turns out to be conformally $AdS_{2}\times S^{7-p}$. Such configurations were
analysed in \cite{Pawelczyk:2000hy,Camino:2001at} and they  describe bound states of fundamental strings stretching along the 
radial direction. From the point of view of the dual SYM theory, this is equivalent to introducing point like objects 
acting as impurities in an interacting (trivially) non-conformal field theory. The conformal case $p\,=\,3$ has been 
recently studied to holographically realise Kondo models \cite{Kachru:2009xf, Kachru:2010dk, Mueck:2010ja, Harrison:2011fs, Faraggi:2011ge}, and such idea
has been extended to the case of ABJM theory for which the impurities have been introduced via D$6$-branes extended along 
$AdS_2\,\subset\,AdS_4$ and wrapping a squashed $T^{1,1}$ space inside the internal $\mathbb{CP}^3$ \cite{Benincasa:2011zu}.

We mainly study two classes of configurations: one in which a straight flux tube is formed and the other in which two flux
tubes ending on two different points at the boundary are connected in the bulk. In order to establish   the stability of the
brane configurations, we analyse the fluctuations of the probe branes at zero temperature. After decoupling the modes,
we can characterise the operators dual to such fluctuations through the effective coupling constant and the
generalised scaling dimension. In particular, we can read them off by computing one-point and two-point correlation 
functions in the coordinate space. The calculation is made straightforward by observing that such fluctuations 
satisfy the equation of motion of a free massive scalar propagating in a higher dimensional $AdS_{q+1}$ space, where
the dimension $q$ needs to be analytically continued to a (generically) fractional value\footnote{This has already been observed
in the case of probe flavour branes \cite{Benincasa:2009ze} in these non-conformal backgrounds, for which mainly the embedding functions have been considered. 
For a discussion about the holographic renormalisation of probe flavour branes in the conformal case, see \cite{Karch:2005ms}.}. 
The dual operators are irrelevant, so we can make sense of the correlation functions by using the observation made in \cite{vanRees:2011fr}
stating that it is possible to holographically renormalise the bulk fields in a perturbative way up to some fixed order
$n$. 
Given that we are interested into one-point and two-point correlators, it is enough to holographically renormalise
the fluctuation action up to quadratic order, where it can always be written as the action of a free massive
scalar in $AdS_{q+1}$. The results we find for the generalised scaling dimensions coincide with the known conformal
dimensions for such operators in the case $p\,=\,3$.

We further analyse the basic thermodynamic properties of the impurities, such as their  free energy and entropy. 
Interestingly, we find that, in the class of systems of interest, the impurity entropy is generically non-analytic in
the filling fraction, except for the D$4$/D$4$-system for which it is possible to obtain a simple closed expression.
This is actually the only system in which the impurity specific heat turns out to be positive, while it is zero for $p\,=\,3$
and it is negative for $p\,<\,3$.
%However, in such a system, the impurity specific heat turns out to be negative, while it is not the case for $p<4$.

The paper is organised as follows. 
In Section \ref{sec:GenKondo} we review the basic features of the Kondo models. In particular, we discuss the effect of a single impurity-spin in a gas of free
fermions in $(3+1)$-dimensions with one or $k$ identical channels and with an $SU(2)$ or $SU(N)$ spin-symmetry group. We also review the case in which the 
impurity spin is inserted in $(1+1)$-dimensional systems, in which the interactions cannot be generally neglected. 
In Section \ref{sec:HolSetUp} we discuss the holographic setup for the ambient non-conformal theory, emphasising the existence of a frame in which the bulk is 
conformally $AdS$, and we also introduce the impurities as probe D$(8-p)$-branes wrapping a conformally $AdS_2\,\times\,S^{7-p}$ submanifold of the bulk geometry.
We also comment on the underlying generalised conformal structure in the systems we are considering. 
In Sections \ref{sec:Straight} and \ref{sec:Hanging} we discuss the two classes of configurations of interests, respectively
the one in which the probe branes generate straight flux tubes and the one in which two flux
tubes ending on two different points at the boundary are connected in the bulk. In particular, Section \ref{sec:Straight} is dedicated to the study of
the thermodynamics of the straight-flux tube/impurity by computing explicitly the free energy, the entropy, the internal energy, the specific heat and
the susceptibility as functions of the temperature and of the filling fraction. Section \ref{sec:Hanging} is instead dedicated to a preliminary study of 
the hanging flux tube configurations, for which the free energy and the one-point function of the operator dual to the embedding function are computed.
In Section \ref{sec:dimer_thermo} we study the thermodynamics of such connected configurations. We again compute the free energy, entropy, internal energy, and
specific heat and we study the competition between the dimer configurations and the disconnected impurities. We identify a first order phase transition and
the related temperature at which it occurs.
Section \ref{sec:FluctRev} is devoted to the detailed analysis of the fluctuations for the probe branes. We identify two decoupled channels and we compute
the one-point and two-point correlators of the operators dual to such fluctuations, emphasising their generalised conformal structure. 
Finally, Section \ref{sec:Concl} contains our conclusion.

\section{Generalities of the Kondo Model}\label{sec:GenKondo}

The Kondo model \cite{Kondo:1964aa, Affleck:1995aa, Hewson:1997aa, Affleck:2008aa} is one of the examples of quantum impurity problems 
\cite{Affleck:2008aa}, where by impurity one refers to point-like degrees of freedom
inserted in and interacting with a (generally non-interacting) gas. This class of systems is characterised by two main features: 
the existence of gapless excitations far away from the impurity and the localisation at one point in position space of the interaction
with the impurity. Furthermore, even if the impurities are introduced in (generally) ($3+1$)-dimensional theories, such systems can be mapped
to ($1+1$)-dimensional one -- except for the case in which the bulk theory is already ($1+1$)-dimensional --, and the impurities appear located
at the origin of the one-dimensional position space.

Typically, one talks about Kondo problem when a free fermion gas is coupled to an impurity with a spin degree of freedom. The interesting feature of such a 
system is that the resistivity shows a minimum as a function of the temperature: when the temperature approaches the so-called Kondo temperature, 
the interaction of the bulk electrons with the impurity starts to compensate the effect of the interaction with the lattice phonons
(which leads to a decrease of the resistivity as the temperature is lowered) so that, as the temperature becomes lower than the Kondo temperature, 
the effect of the impurity dominates.

\subsection{Single-Channel Kondo Model}\label{subsec:Kondo1C}

The Hamiltonian density describing the Kondo model with just one impurity spin is 
\begin{equation}\eqlabel{KondoHamDens}
 \mathcal{H}_{\mbox{\tiny $(3+1)$}}\:=\:\psi^{\dagger}_{\alpha}\left[-\frac{\nabla^2}{2m}-\epsilon_{\mbox{\tiny F}}\right]\psi_{\alpha} + 
   J\delta^{\mbox{\tiny $(3)$}}(\overrightarrow{x})\psi^{\dagger}_{\alpha}\frac{\overrightarrow{\sigma}}{2}\psi_{\alpha}\cdot\overrightarrow{S},
\end{equation}
where $\psi_{\alpha}$ is the bulk electron field with spin-index $\alpha$, $\epsilon_{\mbox{\tiny F}}$ is the Fermi energy, $\overrightarrow{\sigma}$ are the Pauli 
matrices, $\overrightarrow{S}$ is the spin-$1/2$ impurity operator and $J$ is the coupling for a Heisenberg-type exchange interaction between the impurity spin and
the electron spin density. Moreover, a sum over the spin-index $\alpha$ is understood.

The coupling constant $J$ is dimensionful. Its dimensionless version can be defined as $\lambda\,\overset{\mbox{\tiny def}}{=}\,J\rho$, $\rho$ being the density of 
states per unit of energy, per unity volume and per spin. In the case of a free fermion gas $\rho\,=\,m\,k_{\mbox{\tiny F}}/\pi^2$, with $k_{\mbox{\tiny F}}$ being the 
Fermi momentum.

The delta function in the coupling selects just the s-wave to interact with impurity. This allows the $(3+1)$-dimensional model described by the
Hamiltonian density \eqref{KondoHamDens} to be reduced to an effective $(1+1)$-dimensional model defined on half-line with the impurity located at the origin:
\begin{equation}\eqlabel{KondoHamDens1}
 \mathcal{H}_{\mbox{\tiny $(1+1)$}}\:=\:\psi^{\dagger}_{\alpha}\,\,i\frac{d}{dx}\psi_{\alpha}+
  \lambda\,\delta(x)\,\psi^{\dagger}_{\alpha}\frac{\overrightarrow{\sigma}}{2}\psi_{\alpha}\cdot\overrightarrow{S}\,\,.
\end{equation}
This is the Hamiltonian density for a massless Dirac fermion in
$(1+1)$-dimensions and therefore the theory is conformal, with a boundary. A characteristic of this system is that the renormalised coupling constant
$\lambda(E)$ increases as the energy scale $E$ is lowered, so that at low enough energies perturbation theory breaks down. The $\beta$-function turns out to be
\begin{equation}\eqlabel{KondoBeta}
 \frac{d\lambda}{d\ln{\Lambda}}\:=\:-\lambda^2+\ldots,\qquad\Longrightarrow\qquad
 \lambda(\Lambda)\:\sim\:\frac{\lambda_{\mbox{\tiny $0$}}}{1-\lambda_{\mbox{\tiny $0$}}\ln{\left(\Lambda_{\mbox{\tiny $0$}}/\Lambda\right)}}\,\,,
\end{equation}
$\Lambda$ being a scale, and $\Lambda_{\mbox{\tiny $0$}}$ is the scale at which the Kondo coupling constant acquires the bare value $\lambda_{\mbox{\tiny $0$}}$. 
In the case in which the bare coupling $\lambda_{\mbox{\tiny $0$}}$ is negative (ferromagnetic behaviour), the coupling $\lambda$ decreases as the energy scale decreases, 
being therefore well-behaved. If instead $\lambda_{\mbox{\tiny $0$}}$ is positive, there is a value for the energy scale at which perturbation theory breaks down
and it is represented by the so-called Kondo temperature $T_{\mbox{\tiny K}}$
\begin{equation}\eqlabel{KondoTemp}
 T_{\mbox{\tiny K}}\:\sim\:\Lambda_{\mbox{\tiny $0$}}\,e^{-1/\lambda_{\mbox{\tiny $0$}}}.
\end{equation}
In this case, the Kondo coupling constant can be thought to renormalise to infinity \cite{Anderson:1970aa, Wilson:1975aa, Nozieres:1975aa}. This can be understood
by looking at the lattice version of the $(1+1)$-dimensional Hamiltonian density \eqref{KondoHamDens1}
\begin{equation}\eqlabel{KondoHamDens1Lat}
 \mathcal{H}_{\mbox{\tiny $(1+1)$}}^{\mbox{\tiny Lat}}\:=\:
  -t\sum_{i}\left[\psi^{\dagger}_{i}\psi_{i+1}+\mbox{ h.c. }\right]+J\,\psi_{0}^{\dagger}\,\frac{\overrightarrow{\sigma}}{2}\psi_{0}\cdot\overrightarrow{S},
\end{equation}
where the strong coupling regime is at $J\,\gg\,t$.
For $t\,=\,0$, the electron configuration in a general site is arbitrary, except at the origin where the electron form a singlet with the impurity. For
relatively small $t$ compared with the Kondo coupling $\lambda$, the electrons in a general site are in a Bloch state whose single-particle wave-function
vanishes at the origin in order to preserve the singlet-condition.

The interesting feature of this system is that, in the even-parity sector, the wave-functions at zero and infinity Kondo couplings differ from each other,
while in the odd-parity sector they do not. The strong coupling fixed point is the same as the weak coupling one, with the difference that the impurity is 
screened and substituted by the boundary condition $\psi(0)\,=\,0$. For finite/small Kondo coupling this is still true but only at low energies and long distances.
The boundary condition is a fixed point and this is a feature of all quantum impurity systems.

The flow to the low energy fixed point is controlled by the leading irrelevant operator, which is constructed out of the fermion fields and it is $SU(2)$-invariant.
Such a symmetry allows for two dimension-$2$ operators, $\left(\psi^{\dagger\alpha}(0)\psi_{\alpha}(0)\right)^2$ and 
$\left((\psi^{\dagger\alpha}(0)\overrightarrow{\sigma}_{\alpha}^{\phantom{\alpha}\beta}\psi_{\alpha}(0))/2\right)^2$. However,
the first one is suppressed given that its coefficient turns out to be of order $1/\Lambda_{\mbox{\tiny $0$}}$, while the coefficient of the latter operator of order
$1/T_{\mbox{\tiny K}}$, with $\Lambda_{\mbox{\tiny $0$}}\,\gg\,T_{\mbox{\tiny K}}$.

Using perturbation theory in $1/T_{\mbox{\tiny K}}$, one finds that, at low temperature, the impurity susceptibility -- defined as the difference between 
susceptibility with and without impurity -- turns out to behave as a constant. Furthermore, at high temperature and in the scaling limit of small bare Kondo coupling, 
it is proportional to the inverse of the temperature. Actually,  the impurity susceptibility can be written as
\begin{equation}\eqlabel{KondoMagnSusc}
 \chi^{\mbox{\tiny imp}}(T)\:=\:\frac{1}{4T_{\mbox{\tiny K}}}f\left(\frac{T}{T_{\mbox{\tiny K}}}\right),\qquad
 \left\{
  \begin{array}{l}
   \chi^{\mbox{\tiny imp}}(T)\:\longrightarrow\: (4T_{\mbox{\tiny K}})^{-1}, \mbox{ as } T\,\rightarrow\,0\,\,,\\
   \phantom{\ldots}\\
   \chi^{\mbox{\tiny imp}}(T)\:\longrightarrow\: ( 4T)^{-1}, \mbox{ \,\,for } T\,\gg\,T_{\mbox{\tiny K}}\,\,.
  \end{array}
 \right.
\end{equation}

\subsection{Multi-Channel Kondo Model}\label{subsec:KondoMC}

The single-impurity model discussed in the previous section can be generalised by considering $k$ conduction electron channels interacting with the impurity
spin operator $\overrightarrow{S}$ \cite{Nozieres:1980aa, Cox:1997aa}
\begin{equation}\eqlabel{KondoMHamDens}
 \mathcal{H}_{\mbox{\tiny $(1+1)$}}^{\mbox{\tiny $(k)$}}\:=\:\psi^{\dagger j\alpha}\,\,i\frac{d}{dx}\psi_{j\alpha}+
  \lambda\,\psi^{\dagger j\alpha}(0)\frac{\overrightarrow{\sigma}}{2}\psi_{j\alpha}(0)\cdot\overrightarrow{S},  
\end{equation}
where $\alpha$ is a spin-index (for the moment considered to run on two possible spin-states), the sum over the index $j$ is understood and $j\,=\,1,\ldots,k$. 
In this Hamiltonian density, the channels are assumed to be identical, preserving a $SU(k)$-symmetry. 

The $\beta$-function turns out to be
\begin{equation}\eqlabel{KondoMBeta}
 \frac{d\lambda}{d\ln{\Lambda}}\:=\:-\lambda^2+\frac{k}{2}\lambda^3+\ldots
\end{equation}
As in the single-channel case, the Kondo coupling $\lambda$ renormalises to infinity for an anti-ferromagnetic bare coupling $\lambda_{\mbox{\tiny $0$}}$,
as long as $S\,\ge\,k/2$. More precisely, one electron per channel is expected to go into a symmetric state near the origin forming a total spin $k/2$ and the ground 
state has size $|S-k/2|$. 
The low temperature features in the under-screened case $S\,>\,k/2$ are similar to the ones of the single-channel Kondo model, where the strong
coupling fixed point is stable. The same occurs in the exactly screened case $S\,=k/2$. In the over-screened case $S\,<\,k/2$, the RG-flow leads to a non-trivial
finite coupling fixed point, for which the free energy is non-analytic as function of temperature and magnetic field. Let us look more in detail to the 
thermodynamic properties. In particular, the impurity specific heat and the impurity magnetic susceptibility in the over-screened case turn out to be
\begin{equation}\eqlabel{KondoMTherm}
 C^{\mbox{\tiny imp}}(T)\:\propto\:
 \left\{
  \begin{array}{l}
   T^{4/(k+2)},\qquad k\,\neq\,2\\
   \phantom{\ldots}\\
   \frac{T}{T_{\mbox{\tiny K}}}\ln{\frac{T}{T_{\mbox{\tiny K}}}},\qquad k\,=\,2
  \end{array}
 \right. ,
 \qquad
 \chi^{\mbox{\tiny imp}}(T)\:\propto\:
 \left\{
  \begin{array}{l}
   T^{-\frac{k-2}{k+2}},\qquad k\,\neq\,2\\
   \phantom{\ldots}\\
   \ln{\frac{T_{\mbox{\tiny K}}}{T}},\qquad k\,=\,2
  \end{array}
 \right. .
\end{equation}
In the under-screened case, the impurity magnetic susceptibility diverges as the temperature $T$ is lowered, leading to a non-Fermi-liquid behaviour. In the large-$k$
case, instead, one has
\begin{equation}\eqlabel{KondoMThermLargeK}
 C^{\mbox{\tiny imp}}(T)\:\propto\:1+\frac{4}{k}\ln{T}+\mathcal{O}\left(k^{-2}\right),
 \qquad
 \chi^{\mbox{\tiny imp}}(T)\:\propto\:\frac{1}{T}\left[1+\frac{4}{k}\ln{T}+\mathcal{O}\left(k^{-2}\right)\right].
\end{equation}
The magnetic susceptibility diverges as $T^{-1}$ as the temperature is lowered, while a low-temperature logarithmic divergence in the impurity specific heat appears
as a $1/k$-effect.

Let us now consider the case in which the spin-symmetry group is $SU(N)$, so that the spin-index $\alpha$ in the Hamiltonian density \eqref{KondoMHamDens} run
from $1$ to $N$ \cite{Parcollet:1997aa}. The bulk fermions transform under the fundamental representation of $SU(N)$, while the impurity-spin transforms under
an anti-symmetric representation with $Q$ indices. 
As in the $SU(2)$ multichannel model discussed above, also its $SU(N)$ generalisation presents a fixed point at intermediate Kondo
coupling. In this fixed point, the local impurity spin two-point function has the following zero-temperature behaviour
\begin{equation}\eqlabel{ImpSpin2ptFun}
 \langle S(t)S(0)\rangle\:\sim\:\frac{1}{t^{2\Delta_{\mbox{\tiny imp}}}},\qquad
 \Delta_{\mbox{\tiny imp}}\:=\:\frac{N}{N+k},
\end{equation}
and the impurity susceptibility behaves as the local susceptibility (which is obtained integrating the above correlation function)
\begin{equation}\eqlabel{ImpSuscN}
 \chi^{\mbox{\tiny imp}}(T)\:\sim\:\chi^{\mbox{\tiny loc}}(T)\:\sim\:
 \left\{
  \begin{array}{l}
   T^{-\frac{k-N}{k+N}},\qquad k\,>\,N,\\
   \phantom{\ldots}\\
   \ln{T^{-1}},\qquad k\,=\,N,\\
   \phantom{\ldots}\\
   \mbox{const.},\qquad k\,<\,N
  \end{array}
 \right.
\end{equation}
Notice that the expression \eqref{ImpSuscN} is the natural generalisation to the $SU(N)$-spin symmetry of Eq. \eqref{KondoMTherm}, where the spin symmetry is 
$SU(2)$ -- the impurity susceptibility acquires a logarithmic behaviour when the number of channels is equal to the rank of the spin symmetry group.

It is interesting to consider the large-$N$ limit of this class of systems. There are two ways to take such a limit: one can take just the rank of the 
spin-symmetry group to be large (with $k\,\ll\,N$) or taking it to be large and keeping the ratio $\gamma\,\overset{\mbox{\tiny def}}{=}\,k/N$ between the number
of channels and the rank of the spin-symmetry group to be fixed. In the latter case, the expression for the scaling dimension of the local impurity spin and the
behaviour of the impurity susceptibility with the temperature do not change -- one can just conveniently re-write them in terms of the parameter $\gamma$ as
$\Delta_{\mbox{\tiny imp}}\,=\,(1+\gamma)^{-1}$ and $\chi^{\mbox{\tiny imp}}\,\sim\,T^{-(\gamma-1)/(\gamma+1)}$ for $\gamma\,>\,1$. In case $k$ is instead not
taken to be large (and of the same order of $N$), the large-$N$ limit of the local impurity spin scaling dimension and of the impurity susceptibility become
\begin{equation}\eqlabel{ImpSuscLargeN}
 \Delta_{\mbox{\tiny imp}}\:=\:1-\frac{k}{N}\,+\,{\cal O}\left(\left(k/ N\right)^2\right)\,,
  \qquad \chi^{\mbox{\tiny imp}}(T)\,\sim\,\mbox{const.},\qquad k\,\ll\,N.
\end{equation}

Finally, it is interesting to write down the explicit expression for the impurity entropy
\begin{equation}\eqlabel{ImpEntr}
 \mathcal{S}_{\mbox{\tiny imp}}\:=\:\ln{\prod_{v=1}^{Q}\frac{\sin{\left[\pi(1-v+N)/(k+N)\right]}}{\sin{\left[\pi\,v/(k+N)\right]}}},
\end{equation}
which vanishes in the single-channel case.

%In the next section we will explore a holographic realisation of the impurity systems. In such an approach, the role of the bulk electrons will be played by
%the adjoint degrees of freedom provided by the background branes, which enjoy a $SU(N)$-symmetry. The rank $N$ of the gauge group provides the number of channels

\subsection{Impurities in Luttinger Liquids}\label{subsec:ImpLut}

The impurity systems reviewed in the previous two sections are characterised by the fact that the bulk fermions are non-interacting and the localisation of
the impurity at a point allows to reduce the original $(3+1)$-dimensional problem to a $(1+1)$-dimensional one. However, one can think to introduce an impurity spin
directly in a $(1+1)$-dimensional system. In this case, the bulk degrees of freedom cannot always be considered as non-interacting and, moreover, they cannot behave
neither as a Bose liquid nor as a Fermi liquid. Rather, they are thought to be described by Tomonaga-Luttinger liquids 
\cite{Tomonaga:1950aa, Luttinger:1963aa, Mattis:1965aa, Haldane:1981aa, Schulz:1998aa, Giamarchi:2009aa}.

This class of systems is characterised by the absence of quasi-particle excitations and the presence of plasmons and spin density waves, which are independent of each
other. Furthermore, while for Bose and Fermi liquids the specific heat at low temperature scales as $T^{d}$, $d\,>\,1$ being the number of spacial dimensions, and
$T$ respectively, in $(1+1)$-dimensions it can scale either as $\sim T^{\alpha(\lambda^{\mbox{\tiny (L)}})}$ or linearly with the temperature, depending on whether the
system is interacting or not -- $\alpha(\lambda^{\mbox{\tiny (L)}})$ is a function of the Luttinger parameter $\lambda^{\mbox{\tiny (L)}}$ and its form depends on the
attractive or repulsive nature of the interaction. 

The Hamiltonian density for a single impurity in a Luttinger liquid can be written as
\begin{equation}\eqlabel{LutHamDens}
 \mathcal{H}_{\mbox{\tiny Lut}}\:=\:-v_{\mbox{\tiny F}}\sum_{j}
  \left[
   \left(
    \psi_{j}^{\dagger}\psi_{j+1}+\mbox{ h.c. }
   \right)
   + U\hat{n}^2_{j}
  \right]
  +\mathcal{H}_{\mbox{\tiny int}}^{\mbox{\tiny bulk}}+\mathcal{H}^{\mbox{\tiny imp}},  
\end{equation}
where the first term is the Hubbard model describing the bulk degrees of freedom, with $\hat{n}_j$ being the total-number operator at site $j$, 
$v_{\mbox{\tiny F}}\,=\,2at\sin{\left(a\,k_{\mbox{\tiny F}}\right)}$, $U$ and $t$ are the
usual Hubbard parameters, $\mathcal{H}_{\mbox{\tiny int}}^{\mbox{\tiny bulk}}$ is the bulk interaction term, and $\mathcal{H}^{\mbox{\tiny imp}}$ is the bulk-impurity
interaction. Several bulk interactions are possible. Examples are the operators 
$\hat{\mathcal{O}}_{1}\:\overset{\mbox{\tiny def}}{=}\:\lambda^{\mbox{\tiny (L)}}J_{\mbox{\tiny L}}J_{\mbox{\tiny R}}$, 
$\hat{\mathcal{O}}_{2}\:\overset{\mbox{\tiny def}}{=}\:\lambda^{\mbox{\tiny (L)}}\psi_{\mbox{\tiny R,$\alpha$}}^{\dagger}\psi_{\mbox{\tiny L,$\alpha$}}
  \psi_{\mbox{\tiny R,$-\alpha$}}^{\dagger}\psi_{\mbox{\tiny L,$-\alpha$}}$, 
$\hat{\mathcal{O}}_{3}\:\overset{\mbox{\tiny def}}{=}\:-(\lambda^{\mbox{\tiny (L)}}/2\pi)\overrightarrow{J}_{\mbox{\tiny L}}\cdot\overrightarrow{J}_{\mbox{\tiny R}}$, as well as
spin anisotropic interactions of zero conformal spin.
In the  case  $\mathcal{H}_{\mbox{\tiny int}}^{\mbox{\tiny bulk}}\sim\hat{\mathcal{O}}_{1}$, through a field redefinition, the Luttinger-liquid Hamiltonian density acquires the non-interacting form, at the price
that the scaling dimensions of various operators change. If
$\mathcal{H}_{\mbox{\tiny int}}^{\mbox{\tiny bulk}}\sim\hat{\mathcal{O}}_{2}$
 is considered, its effect is not-negligible just in the half-filling case and
produces a charge gap for values of the Luttinger parameter for which the interaction is repulsive.

Let us now turn to the bulk-impurity term of $\mathcal{H}_{\mbox{\tiny Lut}}$, which can be taken as
\begin{equation}\eqlabel{LutHamImp}
 \mathcal{H}^{\mbox{\tiny imp}}\:=\:\lambda_{kl}\,\psi_{\mbox{\tiny $k,\alpha$}}^{\dagger}
  \frac{\overrightarrow{\sigma}^{\mbox{\tiny $\alpha$}}_{\mbox{\tiny $\beta$}}}{2}\,\psi_{\mbox{\tiny $l$}}^{\mbox{\tiny $\beta$}},
\end{equation}
where the indices $k,l$ run over L,R, which indicates left- and right-movers, while the indices $\alpha,\,\beta$ run over the spin values -- in the case 
of $SU(2)$ spin-symmetry, they can take just two values. For $\lambda_{\mbox{\tiny LL}}\:=\:\lambda_{\mbox{\tiny RR}}$ and 
$\lambda_{\mbox{\tiny LR}}\:=\:\lambda_{\mbox{\tiny RL}}$ one gets respectively the amplitudes for forward and backward scattering of the bulk degrees of freedom
with the impurity. The Kondo interaction is obtained when all these couplings are equal.

In a model with such a bulk-impurity interaction with the bulk-interaction switched off, the Hamiltonian density can be mapped to the one of a 
two-channel model with the impurity coupled to the bulk fermions just in one of the channels. Like in the (3+1)-dimensional Kondo problem, the model renormalises
to a local Fermi liquid and the response function scales analytically with the temperature \cite{Nozieres:1980aa}. 

When a bulk interaction such as $\hat{\mathcal{O}}_{2}$ is switched on, any impurity interaction is expected to be substituted by a renormalised boundary condition on the critical
bulk theory. The theory turns out to be characterised by the impurity specific heat and the impurity susceptibility at low temperature given by
\cite{Frojdh:1995aa}
\begin{equation}\eqlabel{LutSpHeatSusc}
 C^{\mbox{\tiny imp}}(T)\:=\:
  \mathtt{c}_{1}\left[\frac{1}{\alpha\left(\lambda^{\mbox{\tiny (L)}}\right)}-1\right]^2 T^{\frac{1}{\alpha\left(\lambda^{\mbox{\tiny (L)}}\right)}-1}+\mathtt{c}_2 T,
 \qquad
 \chi^{\mbox{\tiny imp}}(T)\:=\:\mathtt{c}_3 T^0,
\end{equation}
with $\alpha\left(\lambda^{\mbox{\tiny (L)}}\right)\:=\:\left(1+2\lambda^{\mbox{\tiny (L)}}/v_{\mbox{\tiny F}}\right)^{-1/2}$.

\section{The Holographic Set-up}\label{sec:HolSetUp}

Let us start with a brief review of the holographic realisation of the ``ambient'' theory where the impurities will be
later introduced. Consider the background generated by a stack of $N$ D$p$-branes in the string frame:

\begin{equation}\eqlabel{DpbraneBkg}
 ds_{\mbox{\tiny $10$}}^2\:=\:\left(1+\frac{r_{p}^{7-p}}{r^{7-p}}\right)^{-1/2}\eta_{\mu\nu}dx^{\mu}dx^{\nu}+
  \left(1+\frac{r_{p}^{7-p}}{r^{7-p}}\right)^{1/2}\,ds_{\mbox{\tiny T}}^2,
\end{equation}
where $\mu,\nu\,=\,0,\ldots,p$, $\eta_{\mu\,\nu}$ is the flat Minkowski metric in $p+1$ dimensions,  $r$ is the radial coordinate of the transverse space (with 
$ds_{\mbox{\tiny T}}^2$ being its line element) and the constant  $r_p$ is defined through
\begin{equation}\eqlabel{rp}
 r_p^{7-p}\:\overset{\mbox{\tiny def}}{=}\:\left(2\sqrt{\pi}\right)^{5-p}\Gamma\left(\frac{7-p}{2}\right)\,g_{\mbox{\tiny s}}\,N
  \left(\alpha'\right)^{(7-p)/2}\:\equiv\:d_{p}\,g_{\mbox{\tiny s}}N\left(\alpha'\right)^{(7-p)/2}\,\,,
\end{equation}
where, in the last step, we have introduced  the numerical constant $d_p$, defined as:
\beq
d_p\,=\,\left(2\sqrt{\pi}\right)^{5-p}\Gamma\left(\frac{7-p}{2}\right)\,\,.
\label{dp}
\eeq
The decoupling limit 
\begin{equation}\eqlabel{declim}
 g_{\mbox{\tiny s}}\,\rightarrow\,0,\qquad \alpha'\,\rightarrow\,0,\qquad 
 U\:\overset{\mbox{\tiny def}}{=}\frac{r}{\alpha'}\:\equiv\:\mbox{ fixed},\qquad
 g_{\mbox{\tiny YM}}^2 N\:\equiv\:\mbox{ fixed},
\end{equation}
where the coupling constant $g_{\mbox{\tiny YM}}$ is dimensionful and defined by
\begin{equation}\eqlabel{gYM}
  g_{\mbox{\tiny YM}}^2\:\overset{\mbox{\tiny def}}{=}\:
   2\left(2\pi\right)^{p-2}g_{\mbox{\tiny s}}\left(\alpha'\right)^{(p-3)/2},
\end{equation}
corresponds to the ``near-horizon'' geometry for D$p$-branes. In this limit the metric becomes
\begin{equation}\eqlabel{DpbraneNH}
 \begin{split}
  ds_{\mbox{\tiny $10$}}^2\:&=\:g_{MN} dx^M dx^N\:=\\
  &=\:\alpha'
   \left\{
    \left(\frac{U}{U_p}\right)^{(7-p)/2}\eta_{\mu\nu}dx^{\mu}dx^{\nu}+\left(\frac{U_p}{U}\right)^{(7-p)/2}
    \left[dU^2 + U^2 d\Omega^2_{8-p}\right]
   \right\}.
 \end{split}
\end{equation}
In (\ref{DpbraneNH}) $d\Omega^2_{8-p}$ is the line element of a $S^{8-p}$ sphere and we have introduced the constant $U_p$, which is defined as:
\begin{equation}\eqlabel{Up}
 U_p^{7-p}\:\overset{\mbox{\tiny def}}{=}\:\frac{d_p}{2(2\pi)^{p-2}}g_{\mbox{\tiny YM}}^2 N\,\,.
\end{equation}
The D$p$-brane background is also endowed with a non-trivial dilaton $\phi$ and a RR ($p+1$)-form potential $C^{(p+1)}$, which are given by:
\begin{equation}\eqlabel{DilCp1}
 e^{\phi}\:=\:\frac{g_{\mbox{\tiny YM}}^2 N}{2(2\pi)^{p-2} N}\left(\frac{U}{U_p}\right)^{(7-p)(p-3)/4},\qquad
 C^{(p+1)}_{0\ldots p}\:=\:\frac{2(2\pi)^{p-2}(\alpha')^{(p+1)/2}N}{g_{\mbox{\tiny YM}}^2 N}\left(\frac{U}{U_p}\right)^{7-p}.
\end{equation}
As it has just been mentioned, for $p\,\neq\,3$ the coupling constant is dimensionful and, therefore, the effective
coupling  runs with energy scale as:
\begin{equation}\eqlabel{effcc}
 g_{\mbox{\tiny eff}}^2\:=\:g_{\mbox{\tiny YM}}^2 N U^{p-3}.
\end{equation}
It turns out that the background metric \eqref{DpbraneNH} is conformal to an $AdS_{p+2}\,\times\,S^{8-p}$ space for 
$p\,\neq\,5$ \cite{Duff:1994fg, Boonstra:1998mp, Skenderis:1999dq}. This feature can be seen explicitly by redefining the 
radial coordinate as follows
\begin{equation}\eqlabel{u}
 \frac{u^2}{u_p^2}\:\overset{\mbox{\tiny def}}{=}\:\left(\frac{d_p}{2(2\pi)^{p-2}}g_{\mbox{\tiny YM}}^2 N\right)^{-1}
  U^{5-p}, \qquad
 u_p\,=\,\frac{5-p}{2},
\end{equation}
and rewriting the line element \eqref{DpbraneNH} as
\begin{equation}\eqlabel{DualMetric}
 ds_{\mbox{\tiny $10$}}^2\:=\:\left(N\,e^{\phi}\right)^{2/(7-p)}B_p\,d\tilde{s}_{\mbox{\tiny $10$}}^2,
 \qquad
 B_p\:\overset{\mbox{\tiny def}}{=}\:\alpha' \frac{d_p^{2/(7-p)}}{u_p^2}\,\,,
\end{equation}
so that the line element $d\tilde{s}_{\mbox{\tiny $10$}}^2$ describes an $AdS_{p+2}\,\times\,S^{8-p}$ geometry in the
Poincar{\'e} patch
\begin{equation}\eqlabel{DualAdSxS}
 d\tilde{s}_{\mbox{\tiny $10$}}^2\:=\:\tilde{g}_{\mbox{\tiny MN}}dx^{\mbox{\tiny M}}dx^{\mbox{\tiny N}}\:=\:
   u^2\eta_{\mu\nu}dx^{\mu}dx^{\nu}+\frac{du^2}{u^2}+u_p^2 d\Omega_{8-p}^2.
\end{equation}
In such a frame, the radial coordinate $u$ plays the role of the energy scale of the boundary theory and its rescalings
are just the dilatations in the boundary theory \cite{Skenderis:1999dq}.
One can also use the Fefferman-Graham coordinates by introducing a new radial coordinate $\rho$ as the inverse of the 
current one, namely
\beq
\rho\,=\,u^{-1}\,\,.
\eeq
In terms of $\rho$  the metric $d\tilde{s}_{\mbox{\tiny $10$}}^2$  of (\ref{DualAdSxS}) can be written as
\begin{equation}\eqlabel{FGcoords}
 d\tilde{s}_{\mbox{\tiny $10$}}^2\:=\
  \frac{\eta_{\mu\nu}dx^{\mu}dx^{\nu}+d\rho^2}{\rho^2}+u_p^2 d\Omega_{8-p}^2\,\,,
\end{equation}
\begin{figure}
 \centering%
 {\scalebox{.85}{\includegraphics{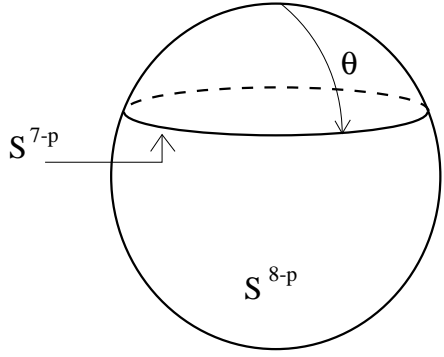}}}
 \caption{The points of the $S^{(8-p)}$ sphere with the same latitude $\theta$, measured from one of its poles, define a $S^{(7-p)}$ sphere. }
 \label{spheres}
\end{figure}
with dilaton and $(8-p)$-form field strength $F^{(8-p)}={}^* d C^{(p+1)}$
which become
\begin{equation}\eqlabel{FGcoords2}
 e^{\phi}\:=\:\frac{g_{\mbox{\tiny YM}}^2 N}{2(2\pi)^{p-2}N}\left(\frac{u_p}{U_p}\rho\right)^{\frac{(3-p)(7-p)}{2(5-p)}},
 \qquad
 F^{(8-p)}\:=\:(7-p)d_p N (\alpha')^{\frac{7-p}{2}}\,\,{\rm Vol}\,(S^{8-p})\,=\:dC^{(7-p)},
\end{equation}
where ${\rm Vol}\,(S^{q})$ denotes the volume form of a $S^{q}$  round sphere.
In order to write explicitly the form of the $(7-p)$-form potential $C^{(7-p)}$, let us represent the line element of the $S^{(8-p)}$ sphere in terms of a polar angle $\theta$ and the line element of a $S^{(7-p)}$ sphere, namely
\begin{equation}\eqlabel{S8mp}
 d\Omega^2_{8-p}\:=\:d\theta^2+\sin^2{\theta}\,d\Omega^2_{7-p}\,\,,
\end{equation}
with $\theta$ taking values in the range $0\le\theta\le \pi$ (see figure \ref{spheres}). 
Then,  the $(7-p)$-form potential can be taken to be
\begin{equation}\eqlabel{C7mp}
 C_{7-p}\:=\:-N d_p (\alpha')^{(7-p)/2}C_{p}(\theta)\,\,{\rm Vol}\,(S^{7-p})\,\,,
\end{equation}
 where $C_{p}(\theta)$ is the function uniquely defined by the conditions:
 \begin{equation}\frac{dC_p(\theta)}{d\theta}\:=\:-(7-p)\sin^{7-p}{\theta},\quad
 \quad\quad
 C_p(0)\:=\:0.
 \eqlabel{C-theta}
\end{equation}
In the case of black hole geometries, the metric \eqref{FGcoords} becomes
\begin{equation}\eqlabel{FGcoordsBH}
d\tilde{s}_{\mbox{\tiny $10$}}^2\:=\:
  \frac{-\mathtt{h}_{p}(\rho)dt^2+d\vec{x}^{\,2}}{\rho^2}+[\mathtt{h}_{p}(\rho)]^{-1}\frac{d\rho^2}{\rho^2}+
   u_p^2 d\Omega_{8-p}^2,
\end{equation}
with 
\begin{equation}\eqlabel{BlackF}
 \mathtt{h}_p(\rho)\:=\:1-\left(\frac{\rho}{\rho_{\mbox{\tiny $h$}}}\right)^{2\frac{7-p}{5-p}}.
\end{equation}
Parametrising the position of the event horizon by $\rho_{\mbox{\tiny $h$}}$, the background temperature takes the form
\begin{equation}\eqlabel{Tp}
 T\:=\:\frac{7-p}{2\pi(5-p)\rho_{\mbox{\tiny $h$}}}.
\end{equation}

\subsection{Generalised Conformal Symmetry}\label{subsec:GenConfSym}

The ``near-horizon'' geometry generated by a stack of $N$ D$p$-branes discussed in the previous section is dual to an $SU(N)$
supersymmetric Yang-Mills theory in $(p+1)$-dimensions, with coupling constant $g_{\mbox{\tiny YM}}$ which is generally dimensionful, except
for $p\,=\,3$, and with a trivial RG flow which is just due to the dimensionfulness of the coupling constant.

Following \cite{Kanitscheider:2008kd}, let us consider a generalised version of the $(p+1)$-dimensional Supersymmetric Yang-Mills Euclidean action
\begin{equation}\eqlabel{GenSYMact}
 \begin{split}
  S_{\mbox{\tiny $p+1$}}\:=\:-\int d^{p+1}x\:\sqrt{g_{\mbox{\tiny $(0)$}}}
   &\left\{
    -\Phi_{\mbox{\tiny $(0)$}}\frac{1}{4}\mbox{ Tr}\left\{F_{\mu\nu}F^{\mu\nu}\right\}+
    \frac{1}{2}\mbox{ Tr}\left\{X\left[D^2-\frac{p-1}{4p}R^{\mbox{\tiny $(p+1)$}}_{\mbox{\tiny $(0)$}}\right]X\right\}+
   \right.\\
   &+\left.
    \frac{\Phi_{\mbox{\tiny $(0)$}}^{-1}}{4}\mbox{ Tr}\left\{\left[X,X\right]^2\right\}
   \right\},
 \end{split}
\end{equation}
where $X^{\mbox{\tiny $I$}}$ are scalar fields, with $I\,=\,1,\ldots,\,9-p$, $g^{\mbox{\tiny $(0)$}}_{\mbox{\tiny $\mu\nu$}}$ is the background metric and
$\Phi_{\mbox{\tiny $(0)$}}$ is a background scalar field. The usual Supersymmetric Yang-Mills action is recovered for 
$g^{\mbox{\tiny $(0)$}}_{\mu\nu}\,=\,\delta_{\mu\nu}$ and $\Phi_{\mbox{\tiny $(0)$}}\,=\,g_{\mbox{\tiny YM}}^{-2}$. The action \eqref{GenSYMact} turns out to be
invariant under  a Weyl transformation of the following form
\begin{equation}\eqlabel{GenSYMWeyl}
 g_{\mbox{\tiny $(0)$}}\:\longrightarrow\:e^{2\omega}g_{\mbox{\tiny $(0)$}},\qquad
 X\:\longrightarrow\:e^{-\frac{p-1}{2}\omega}X,\qquad
 A_{\mu}\:\longrightarrow\:A_{\mu},\qquad
 \Phi_{\mbox{\tiny $(0)$}}\:\longrightarrow\:e^{-(p-3)\omega}\Phi_{\mbox{\tiny $(0)$}}.
\end{equation}
Defining the stress-energy tensor $T_{\mu\nu}$ and the scalar operator $\mathcal{O}_{\mbox{\tiny $\Phi$}}$ as
\begin{equation}\eqlabel{GenSYMTO}
 T_{\mu\nu}\:\overset{\mbox{\tiny def}}{=}\:\frac{2}{\sqrt{g_{\mbox{\tiny $(0)$}}}}\frac{\delta S_{\mbox{\tiny $(p+1)$}}}{\delta g_{\mbox{\tiny $(0)$}}^{\mu\nu}},
 \qquad
 \mathcal{O}_{\mbox{\tiny $\Phi$}}\:\overset{\mbox{\tiny def}}{=}\:
  \frac{1}{\sqrt{g_{\mbox{\tiny $(0)$}}}}\frac{\delta S_{\mbox{\tiny $(p+1)$}}}{\delta\Phi_{\mbox{\tiny $(0)$}}},
\end{equation}
one can obtain diffeomorphism and Ward identities
\begin{equation}\eqlabel{GenSYMId}
 \nabla^{\nu}\langle T_{\mu\nu}\rangle_{\mbox{\tiny $J$}}+\langle\mathcal{O}_{\mbox{\tiny $\Phi$}}\rangle_{\mbox{\tiny $J$}}\partial_{\mu}\Phi_{\mbox{\tiny $(0)$}}
  \,=\,0,
 \qquad
 \langle T_{\mu}^{\phantom{\mu}\mu}\rangle_{\mbox{\tiny $J$}}+(p-3)\Phi_{\mbox{\tiny $(0)$}}\langle\mathcal{O}_{\mbox{\tiny $\Phi$}}\rangle_{\mbox{\tiny $J$}}\:=\:0,
\end{equation}
with $\langle\cdot\rangle_{\mbox{\tiny $J$}}$ indicates the expectation value with respect to a source $J$. For 
$g^{\mbox{\tiny $(0)$}}_{\mu\nu}\,=\,\delta_{\mu\nu}$ and $\Phi_{\mbox{\tiny $(0)$}}\,=\,g_{\mbox{\tiny YM}}^{-2}$
the first equation in \eqref{GenSYMId} provides the stress-energy conservation, while the second one the tracelessness of the stress-energy tensor and
therefore the restoration of conformal symmetry. In particular, notice from the second equation of \eqref{GenSYMId} that conformal symmetry is broken because
of the dimensionfulness of the coupling constant.

In a theory of this type, the entropy $\mathcal{S}$ at finite temperature $T$ has to scale as
\begin{equation}\eqlabel{GenSYMEntr}
 \mathcal{S}\:=\:\mathtt{c}\left(g_{\mbox{\tiny eff}}^2(T),\,N,\,\ldots\right)\Omega_{p}T^{p},
\end{equation}
with $\Omega_{p}$ being the spacial volume, $g_{\mbox{\tiny eff}}^2(T)$ as defined in \eqref{effcc} and 
$\mathtt{c}\left(g_{\mbox{\tiny eff}}^2(T),\,N,\,\ldots\right)$ is a generic function of dimensionless parameters.

Furthermore, the two-point functions of an operator $\mathcal{O}$ need to have the following form
\begin{equation}\eqlabel{GenSYM2pt}
 \langle\mathcal{O}(x)\mathcal{O}(0)\rangle\:=\:\mathcal{R}
  \left(
   \mathtt{f}\left(g_{\mbox{\tiny eff}}^2(x),\,N,\,\ldots\right)\frac{1}{|x|^{2\Delta}}
  \right),
\end{equation}
with $g_{\mbox{\tiny eff}}^2(x)\,=\,g^2_{\mbox{\tiny YM}}Nx^{3-p}$, $\mathtt{f}\left(g_{\mbox{\tiny eff}}^2(x),\,N,\,\ldots\right)$ being a function of dimensionless 
parameters, and $\mathcal{R}$ provides the renormalised version of its argument. More precisely, these two objects  are the same for $x\,\neq\,0$, while at 
$x\,=\,0$ they differ by infinite renormalisation.

\subsection{Holographic impurities as probe D$(8-p)$-branes}\label{subsec:HolImpBr}

We would like now to introduce localised degrees of freedom in the $(p+1)$-dimensional $U(N)$ SYM theory. Before looking at general values of
$p\,<\,5$, let us review the $p\,=\,3$ case following \cite{Gomis:2006sb}.

\subsubsection{Wilson loops and D$5$-branes}\label{subsubsec:WLD5}

Let us start with considering a single probe D$5$-brane in $AdS_5\times S^5$ in such a way that it wraps an $S^{3}\,\subset\,S^5$ and extends
along the time and radial direction, so that the induced metric on the world-volume is $AdS_2\times S^3$. This system is the decoupling limit
of a system with $N$ D$3$-branes in flat space and a single probe D$5$-brane, whose degrees of freedom are carried by the $3-3$ strings 
(the ``ambient'' gauge theory), the $3-5$/$5-3$ strings which are localised in the codimension-$3$ defect identified by the intersection of the
two types of D-branes, and finally the non-dynamical $5-5$ strings. The action for the defect (point-like) degrees of freedom can be obtained by
performed T-duality on the D$0$/D$8$ system, obtaining
\begin{equation}\eqlabel{ConfImpAct}
 S_{\mbox{\tiny imp}}\:=\:\int dt\:
  \left[i\Psi^{\dagger}\partial_t\Psi + \Psi^{\dagger}\left(A_{t}+X_{\mbox{\tiny $I$}}v^{\mbox{\tiny $I$}}+a_{t}\right)\Psi-n a_t\right],
\end{equation}
where $A_t$ and $X_{\mbox{\tiny $I$}}$ are respectively the time-component of the gauge field and the scalars in $\mathcal{N}\,=\,4$ SYM, $v^{\mbox{\tiny $I$}}$
is a unit vector, $\Psi$ is the resulting fermionic impurity field, $a_t$ is the non-dynamical gauge field on the D$5$-brane, and $n$ is the unit of
background gauge field localised on the point-like defect. 

Integrating out the degrees of freedom associated to the D$5$-branes in the decoupling limit, it has been shown \cite{Gomis:2006sb} that the introduction
of the D$5$-branes corresponds to a $1/2$-BPS Wilson loop operator in $\mathcal{N}\,=\,4$ SYM in the anti-symmetric representation of the gauge group $U(N)$.

\subsubsection{General case}\label{subsubsec:WLgen}

As a natural generalisation of the D-brane construction reviewed above\footnote{A comment is now in order. While indeed the construction we are going to consider
introduces point-like degrees of freedom, a neat correspondence between the D$(8-p)$-branes and Wilson-loop operators has not been shown yet.}, one may think to 
introduce extra degrees of freedom in the boundary theory by considering in the bulk probe D$(8-p)$-branes wrapping an $S^{7-p}\,\subset\,S^{8-p}$ and extending along 
the radial direction. 
The embedding of such branes in the D$p$-brane background can be described by two functions $x^p\,\equiv\,z(\rho)$ and $\theta\,\equiv\,\theta(\rho)$, where 
$\theta$ is the angular coordinate introduced in  (\ref{S8mp}). 
For the time being, let us keep both the two embedding functions. The induced metric on the world-volume of the
D$(8-p)$-brane is
\bear
  &&d\tilde{s}^{2}_{\mbox{\tiny $9-p$}}\:=\:
  \mathtt{g}_{\alpha\beta}\,d\zeta^{\alpha}\,d\zeta^{\beta}\,=\,\rc\rc
 &&\qquad\qquad=\,
  \frac{-\mathtt{h}_p(\rho)\,dt^2}{\rho^2}+\left[1+\mathtt{h}_p(\rho)(z')^2+
    u_p^2\rho^2\mathtt{h}_p(\rho)(\theta')^2\right]\frac{d\rho^2}{\mathtt{h}_p(\rho)\rho^2}+ u_p^2\sin^{2}{\theta}\,\,d\Omega^2_{7-p}\,\,,\rc
 \label{IndMetr}
\eear
which is substantially the $AdS_2\times S^{7-p}$ geometry with the two functions $z(\rho)$ and $\theta(\rho)$ controlling,
respectively, the embedding of branes in the $(p+2)$-dimensional (conformal)-$AdS$ manifold, where they wrap a 
(conformally)-$AdS_2$ subspace, and the embedding in the transverse space. The action for the probe branes is the sum of a Dirac-Born-Infeld and a Wess-Zumino term
\begin{equation}\eqlabel{D8mpAct}
  S_{\mbox{\tiny D$(8-p)$}}\:=\:
   -T_{\mbox{\tiny D$(8-p)$}}\int d^{9-p}\zeta\:e^{-\phi}
    \sqrt{-\mbox{det}\left\{\mathtt{g}_{\alpha\beta}+(2\pi\alpha')F_{\alpha\beta}\right\}}+
   T_{\mbox{\tiny D$(8-p)$}}(2\pi\alpha')\int F\,\wedge\,C_{7-p}\,\,,
   \end{equation}
where $\mathtt{g}_{\alpha\beta}$ is the pullback of the near-horizon  string frame metric and $F$ is the world-volume abelian gauge field strength. The Wess-Zumino term in the action (\ref{D8mpAct}) acts as a source of the electric component $F_{t\rho}$ of the world-volume gauge field and, therefore, one cannot put consistently  $F_{t\rho}=0$. Let us take $\zeta^{\alpha}\,=\,(t,\rho,\theta^i)$ as world-volume coordinates, where the 
$\theta^i$ ($i=1,\cdots, 7-p$) parametrise the $S^{7-p}$ sphere in (\ref{S8mp}). Assuming that the electric field is independent of the angles $\theta^i$, one can write the action as:

\begin{equation}\eqlabel{D8mpAct2}
 \begin{split}
   &S_{\mbox{\tiny D$(8-p)$}}\,=\:-T_{\mbox{\tiny D$(8-p)$}}N B_p^{\frac{9-p}{2}}\Omega_{7-p}\,u_{p}^{7-p}
      \int dt\,d\rho\:
      \left\{
       \left(N\,e^{\phi}\right)^{\frac{2}{7-p}}\rho^{-2}\sin^{7-p}{\theta}\times
      \right.\\
  &\hspace{1cm}\times
      \left.
       \left[
        1+\mathtt{h}_p(z')^2+u_p^2\rho^2\mathtt{h}_p(\theta')^2-
         \frac{(2\pi\alpha')^2}{B_p^2}\frac{\rho^4 F_{t\rho}^2}{(N\,e^{\phi})^{\frac{4}{7-p}}}
       \right]^{1/2}
       +\frac{2\pi\alpha'}{B_p}F_{t\rho}C_p(\theta)
      \right\},
 \end{split} 
\end{equation}
where $\Omega_{7-p}=2\,\pi^{{8-p\over 2}}/\Gamma\big({8-p\over 2}\big)$ is the volume of the $S^{7-p}$. One notices that $S_{\mbox{\tiny D$(8-p)$}}$ depends on the embedding function $z$ and on the
world-volume gauge field just through their derivatives. Therefore, there are two first integrals of motion
\begin{equation}\eqlabel{IntMot}
 \frac{\partial\mathcal{L}_{\mbox{\tiny D$(8-p)$}}}{\partial F_{t\rho}}\:=\:\mbox{ const.}, \qquad\qquad
 \frac{\partial\mathcal{L}_{\mbox{\tiny D$(8-p)$}}}{\partial z'}\:=\:c_z\,\,,
\end{equation}
where ${\cal L}_{\mbox{\tiny D$(8-p)$}}$ is the Lagrangian density of the D$(8-p)$-brane, whose expression can be read from 
(\ref{D8mpAct2}). 
The first constant can be fixed through the quantisation condition \cite{Camino:2001at} 
\begin{equation}\eqlabel{QuantCond}
\int_{S^{7-p}}d^{7-p}\theta\,\,\frac{\partial\mathcal{L}_{\mbox{\tiny D$(8-p)$}}}{\partial F_{t\rho}}\:=\:n\left(2\pi\alpha'\right)T_{\mbox{\tiny F$1$}},
 \qquad n\,\in\,\mathbb{Z}.
\end{equation}
The integer $n$ in (\ref{QuantCond}) represents the number of fundamental strings (quarks) dissolved in the probe brane.  Computing the left-hand-side of \eqref{QuantCond}, the first integral of motion related to the world-volume gauge-field 
becomes
\begin{equation}\eqlabel{GaugeFI}
 c_f\frac{n}{N}\:=\:{\sin^{7-p}{\theta}\,\,\,{\cal F}_{t\rho}\over 
  \left[
   1+\mathtt{h}_p(z')^2+u_p^2\,\rho^2\,\mathtt{h}_p(\theta')^2- \,{\cal F}_{t\rho}\,
  \right]^{1/2}}-C_p(\theta),
\end{equation}
where we have introduced the rescaled world-volume gauge field ${\cal F}_{\alpha\beta}$, defined as
\beq
{\cal F}_{\alpha\beta}\:\overset{\mbox{\tiny def}}{=}\:
 \frac{2\pi\alpha'}{B_p}\frac{\rho^2 F_{\alpha\beta}}{(N\,e^{\phi})^{\frac{2}{7-p}}}\,\,,
\eeq
and the  constant $c_f$ is defined as
\begin{equation}\eqlabel{cf}
 c_f\:\overset{\mbox{\tiny def}}{=}\:\frac{T_{\mbox{\tiny F$1$}}B_p}{
   T_{\mbox{\tiny D$(8-p)$}}B_p^{\frac{9-p}{2}}\Omega_{7-p}u_p^{7-p}}\:=\:
   2\sqrt{\pi}\frac{\Gamma\left(\frac{8-p}{2}\right)}{\Gamma\left(\frac{7-p}{2}\right)}.
\end{equation}
The equation of motion \eqref{GaugeFI} can be easily solved for the gauge field-strength ${\cal F}_{t\rho}$ to get
\begin{equation}\eqlabel{Frt}
 {\cal F}_{t\rho}\:=\:C_{(p,n)}(\theta)
 \left[
  \frac{1+\mathtt{h}_p(z')^2+u_p^2\rho^2\mathtt{h}_p(\theta')^2}{\sin^{2(7-p)}{\theta}+C_{(p,n)}^2(\theta)}
 \right]^{1/2},
\end{equation}
where the function $C_{(p,n)}(\theta)$ has been defined as
\begin{equation}\eqlabel{Cpn}
 C_{(p,n)}(\theta)\:\overset{\mbox{\tiny def}}{=}\:C_p(\theta)+c_f\frac{n}{N}\,\,.
\end{equation}
As far as the first integral of motion related to the embedding function $z(\rho)$ is concerned, it acquires the form
\begin{equation}\eqlabel{zFI}
 \tilde{c}_z\:=\:-\frac{\rho^{-2}\sin^{7-p}{\theta}\left(N\,e^{\phi}\right)^{\frac{2}{7-p}}\,\,\mathtt{h}_p\,z'}{
  \left[
   1+\mathtt{h}_p\,(z')^2+u_p^2\,\rho^2\,\mathtt{h}_p\,(\theta')^2\,-\,{\cal F}_{t\rho}
  \right]^{1/2}}\,\,,
\end{equation}
where  $\tilde{c}_z$ is a new constant related to  $c_z$ by a rescaling
\begin{equation}\eqlabel{cz2}
\tilde{c}_z\:\overset{\mbox{\tiny def}}{=}\:
 \frac{c_z}{T_{\mbox{\tiny D$(8-p)$}}\,N\, B_p^{\frac{9-p}{2}}\Omega_{7-p}\,u_p^{7-p}}
 \:=\:
 \frac{2\sqrt{\pi}}{N\,T_{\mbox{\tiny F$1$}}}\frac{\Gamma\left(\frac{8-p}{2}\right)}{\Gamma\left(\frac{7-p}{2}\right)}
  \frac{c_z}{B_p}.
\end{equation}
Let us now evaluate the energy of the system. By performing a Legendre transform, the Hamiltonian $H$ of the
D$(8-p)$-brane is given by
\beq
H\,=\,\int_{S^{7-p}}\,d^{7-p}\,\theta\,\sqrt{g_{\mbox{\tiny $S^{7-p}$}}}\,\int d\rho
\,\Big[\,F_{t\rho}\,{\partial {\cal L}_{\mbox{\tiny D$(8-p)$}}\over \partial F_{t\rho}}\,-\,{\cal L}_{\mbox{\tiny D$(8-p)$}}\,\Big]\,\,.
\eeq
By using the explicit expression of the Lagrange density written in (\ref{D8mpAct2}), one gets
\begin{equation}\eqlabel{Hamilt}
 \begin{split}
  H\:=\:T_{\mbox{\tiny D$(8-p)$}}N B_p^{\frac{9-p}{2}}\Omega_{7-p} u_p^{7-p}&\int d\rho
   \left(N\,e^{\phi}\right)^{\frac{2}{7-p}}\rho^{-2}
   \left[
    1+\mathtt{h}_p(z')^2+\rho^2\mathtt{h}_p(\theta')^2
   \right]^{1/2}\times\\
  &\times
   \left[
    \sin^{2(7-p)}{\theta}+C_{(p,n)}^2(\theta)
   \right]^{1/2}.
 \end{split}
\end{equation}
Let us now discuss some possible configurations for the D$(8-p)$-brane. In particular, we will focus on the case
in which the position of the probe branes in both the (conformally) $AdS$ space and in the transverse space is fixed
(straight flux tubes with $\theta'=z'=0$), and on the  configurations in which the probe branes are allowed to bend either in the extended directions
or in the transverse space.

\section{Straight Flux Tube Configurations}\label{sec:Straight}

Let us start with the analysis of the configurations in which both the angular and linear coordinates $\theta$ and $z$ are
constants. Notice that these configurations introduce a localised defect on the boundary theory.  Following \cite{Camino:2001at}, the stable configurations are the ones minimising the energy \eqref{Hamilt},\ie\ the ones that satisfy the condition:
\begin{equation}\eqlabel{StableConf}
 \begin{split}
 0\:=\:\left.\frac{dH}{d\theta}\right|_{\mbox{\tiny
   $\left\{
     \begin{array}{l}
      \theta\,=\,\mbox{const}\\
      z\,=\,\mbox{const}
     \end{array}
   \right\}$
  }}
  \:=\:&T_{\mbox{\tiny D$(8-p)$}}N B_p^{\frac{9-p}{2}}\Omega_{7-p} u_p^{7-p}\int d\rho
  \left(N\,e^{\phi}\right)^{\frac{2}{7-p}}\rho^{-2}
%  \left[
%   1+\mathtt{h}_p(z')^2+\rho^2\mathtt{h}_p(\theta')^2
%  \right]^{1/2}
   \times\\
  &\times(7-p)\sin^{7-p}{\theta}
  \frac{\Lambda_{p,n}(\theta)}{\left[
   \sin^{2(7-p)}{\theta}+C_{p,n}^2(\theta)
  \right]^{1/2}},
 \end{split}
\end{equation}
where the function $\Lambda_{p,n}(\theta)$ is defined as
\begin{equation}\eqlabel{Lambda}
 \Lambda_{p,n}(\theta)\:\overset{\mbox{\tiny def}}{=}\:\sin^{6-p}{\theta}\cos{\theta}-C_{p,n}(\theta).
\end{equation}
The least energy condition \eqref{StableConf} can be satisfied if and only if
\begin{equation}\eqlabel{StableConf2}
 \sin{\theta}\:=\:0,\quad\mbox{ or }\quad \Lambda_{p,n}(\theta)\:=\:0.
\end{equation}
In the first case, such configurations occur  at $\theta\,=\,0,\,\pi$ which are points in which the sphere $S^{7-p}$
shrinks to zero size. In the second case, instead, the branes are located at $\theta\,=\,\bar{\theta}_{(p,n)}$ with
$\bar \theta_{(p,n)}$ defined by the condition itself: $\Lambda_{p,n}(\bar{\theta}_{(p,n)})\:=\:0$. The functions $\Lambda_{p,n}(\theta)$ for different values of $p$  are listed in appendix \ref{app:angles}. As it is clear from (\ref{Cpn}) they depend on the quantisation integer $n$ and on the rank of the gauge group through the combination
\beq
\nu\,=\,{n\over N}\,\,,
\label{filling_fraction}
\eeq
which we will refer to as the filling fraction. The reason for this name is the fact that, for a given value of $p$,  only for $1<\,n\,<N$ there exists a unique solution for the angles $\bar{\theta}_{(p,n)}$ in the range $0<\bar{\theta}_{(p,n)}<\pi$. The integer $n$ represents the number of impurity fermions introduced in the boundary theory and the ratio $\nu$ takes values in the range $0<\nu<1$. The upper bound of $n$ is a manifestation of the so-called stringy exclusion principle and is a piece of evidence supporting the identification of these brane configurations as the holographic duals of Wilson lines in the antisymmetric representation of the gauge group. Indeed, in the conformal case $p=3$ this identification was explicitly checked in refs. \cite{Yamaguchi:2006tq, Gomis:2006sb}. 

From the explicit formulas of the $\Lambda_{p,n}(\theta)$ functions displayed in appendix \ref{app:angles}, one can verify that the angles $\bar{\theta}_{(p,n)}$ satisfy the relation:
\beq
\bar{\theta}_{(p,N-n)}\,=\,\pi\,-\bar{\theta}_{(p,n)}\,\,.
\label{particle-hole}
\eeq
It follows that changing the polar angle of the embedding as $\theta\to\pi-\theta$ is equivalent to the particle-hole transformation $\nu\to1-\nu$.  The energy density (tension) of these configurations was derived in \cite{Camino:2001at} and it turns out to be
\begin{equation}\eqlabel{EnDens}
 \mathcal{E}_{(p,n)}\:=\:\frac{N\,T_{\mbox{\tiny F$1$}}}{2\sqrt{\pi}}\,
  \frac{\Gamma\left(\frac{7-p}{2}\right)}{\Gamma\left(\frac{8-p}{2}\right)}\sin^{6-p}{\bar{\theta}_{(p,n)}}.
\end{equation}
It follows from (\ref{particle-hole}) and (\ref{EnDens}) that the tension satisfies the relation:
\beq
\mathcal{E}_{(p,n)}\:=\:\mathcal{E}_{(p,N-n)}\,\,,
\eeq
\ie\ it is invariant under the transformation  $\nu\to 1-\nu$. 

The electric world-volume field $ \bar{\cal F}_{t\rho}$ for the flux-tube configuration $\theta=\bar\theta_{(p,n)}$ can be obtained from (\ref{Frt}) by using that 
$C_{(p,n)} ( \bar\theta_{(p,n)})=(\sin \bar\theta_{(p,n)})^{6-p}\,\cos \bar\theta_{(p,n)}$. One gets:
\begin{equation}\eqlabel{bar-Frt}
\bar{\cal F}_{t\rho}\:=\:
 \cos\bar\theta_{(p,n)}\,\,.
\end{equation}
Notice that $ \bar{\cal F}_{t\rho}$  changes its sign under the transformation $\nu\to1-\nu$. 

As is was proven in \cite{Camino:2001at}, in the case of the zero temperature background, the flux-tube configurations described above are the solution of a first-order BPS  equation for the embedding and the world-volume gauge field of the probe brane. One can show that this equation is the one that is obtained by imposing kappa symmetry to the D$(8-p)$-brane in such a way that it preserves 1/4 of the supersymmetry.

\subsection{Impurity Entropy}\label{subsec:ImpEntr}

Let us investigate some aspects of the thermodynamics of this class of systems. First of all, for $p\,<\,5$ the
relation \eqref{Tp} between the temperature $T$ and the position of the horizon $\rho_{\mbox{\tiny}}$ can be inverted
to obtain
\begin{equation}\eqlabel{rhT}
 \rho_{\mbox{\tiny h}}\:=\:\frac{7-p}{2\pi(5-p)T}.
\end{equation}
In order to compute the free energy and entropy of the flux tube configuration, let us evaluate the Euclidean action of one 
of such configurations that extends from the horizon at $\rho\,=\,\rho_{\mbox{\tiny h}}$ until a cutoff value of the radial
coordinate $\rho\,=\,\epsilon$. If we define $\beta_p$ as
\beq
\beta_{p}\:\overset{\mbox{\tiny def}}{=}
   \rho_{\mbox{\tiny h}}^{-\frac{2}{5-p}}
   \left(\frac{g_{\mbox{\tiny YM}}^2 N}{2(2\pi)^{p-2}}\right)^{\frac{2}{7-p}}
   \left(\frac{U_p}{u_p}\right)^{\frac{p-3}{5-p}},
   \label{beta_p}
\eeq
we get
\begin{equation}\eqlabel{EuclAct}
 I_{\mbox{\tiny D$(8-p)$}}^{\mbox{\tiny Eucl}}\Big|_{\mbox{\tiny on-shell}}\:=\:
 \frac{\mathcal{E}_{\mbox{\tiny $(p,n)$}}}{T}
\,\,\beta_{p}\,\,B_p\,\,\rho_{\mbox{\tiny h}}^{\frac{2}{5-p}}\,
 \int_{\epsilon}^{\rho_{\mbox{\tiny h}}}d\rho\:\rho^{-\frac{7-p}{5-p}}\,\,.
\end{equation}
To arrive at the expression (\ref{EuclAct}) we have integrated over a periodic Euclidean time circle of period $1/T$. 
Notice that the integral \eqref{EuclAct} diverges as $\epsilon\,\rightarrow\,0$ and therefore it needs to get
renormalised in a neighbourhood of the boundary. It easy to see that it is renormalised just by a term proportional to
the volume of the boundary, regularised with $\rho\,=\,\epsilon$
\begin{equation}\eqlabel{EuclActCt}
  I_{\mbox{\tiny D$(8-p)$}}^{\mbox{\tiny Eucl}}\Big|_{\mbox{\tiny ct}}\:=\:
  -{5-p\over 2}\,\,
 \frac{\mathcal{E}_{\mbox{\tiny $(p,n)$}}}{T}\,\,B_p\,
 \left(N\,e^{\phi}\big|_{\epsilon}\right)^{\frac{2}{7-p}}\,\,
\sqrt{\left.\mathtt{g}_{tt}\right|_{\mbox{\tiny $\epsilon$}}}\,,
\end{equation}
where $\mathtt{g}_{tt}\,|_{\mbox{\tiny $\epsilon$}}=\epsilon^{-2}$ is the $tt$ component of the induced metric (\ref{IndMetr}) at $\rho=\epsilon$.  Therefore, the renormalised action is given by
\begin{equation}\eqlabel{EuclActRen}
  I_{\mbox{\tiny D$(8-p)$}}^{\mbox{\tiny Eucl}}\Big|_{\mbox{\tiny ren}}\:=\:
  -{5-p\over 2}\,\,
  \frac{\mathcal{E}_{\mbox{\tiny $(p,n)$}}}{T}\,
  \,\,\beta_{p}\,\,B_p\,\,.
 \end{equation}
Let us rewrite (\ref{EuclActRen}) in a more convenient form. By using the definitions of $\beta_p$ and $B_p$ written in (\ref{beta_p}) and (\ref{DualMetric}), one easily proves that:
\beq
\beta_{p}\,\,B_p\,=\,{2\sqrt{\pi}\over (5-p)\,T_{\mbox{\tiny F$1$}}
}\,
\Bigg[{\Gamma\big({7-p\over 2}\big)\over (7-p)^2\,\pi}\,\Bigg]^{{1\over 5-p}}\,\,
 \left[g_{\mbox{\tiny eff}}^2(T)\right]^{\frac{1}{5-p}}T\,\,,
\eeq
where $g_{\mbox{\tiny eff}}(T)$ is the gauge theory effective coupling constant at the temperature $T$, defined as
\begin{equation}\eqlabel{geffT}
 g_{\mbox{\tiny eff}}^{2}(T)\:=\:g_{\mbox{\tiny YM}}^2 \,N \,T^{p-3}\,\,.
\end{equation}
By using the explicit expression of $\mathcal{E}_{\mbox{\tiny $(p,n)$}}$  written in (\ref{EnDens}) one can recast the renormalised action (\ref{EuclActRen}) as
\beq
 I_{\mbox{\tiny D$(8-p)$}}^{\mbox{\tiny Eucl}}\Big|_{\mbox{\tiny ren}}\,=\,-c_p\,N\,
 \big(\,\sin\bar \theta_{(p,n)}\,\big)^{6-p}\,\, 
 \left[g_{\mbox{\tiny eff}}^2(T)\right]^{\frac{1}{5-p}}\,\,,
\label{EuclActRen2}
 \eeq
where the coefficient $c_p$ is given by:
\beq
c_p\,=\,
{\Big[\,
\Gamma\big({7-p\over 2}\big)\,\,
\Big]^{{6-p\over 5-p}}\over 
\Gamma\big({8-p\over 2}\big)\,\,
\Big(\,(7-p)^2\,\pi\,\Big)^{{1\over 5-p}}
}\,\,.
\label{c_p}
\eeq
From the renormalised action \eqref{EuclActRen2} it is immediate to extract the free energy for the straight flux tubes, namely
\begin{equation}\eqlabel{FreeEn}
 F_{\mbox{\tiny D$(8-p)$}}^{\mbox{\tiny str}}\:=\:T\,I_{\mbox{\tiny D$(8-p)$}}^{\mbox{\tiny Eucl}}\Big|_{\mbox{\tiny ren}}
 \:=\:
 -c_p\,N\,
 \big(\,\sin\bar \theta_{(p,n)}\,\big)^{6-p}\,\, 
 \left[g_{\mbox{\tiny eff}}^2(T)\right]^{\frac{1}{5-p}}\,\,T
 \,\,.
\end{equation}
If we ignore non-abelian interactions among the D$(8-p)$-branes, 
the impurity entropy generated by $M$ D$(8-p)$-branes is just $M$ times the result obtained for $M=1$, namely
 \begin{equation}\eqlabel{ImpEntr}
 \mathcal{S}_{\mbox{\tiny D$(8-p)$}}^{\mbox{\tiny str}}\:=\:
  -M\frac{\partial F_{\mbox{\tiny D$(8-p)$}}^{\mbox{\tiny str}}}{\partial T}
  \:=\:{2c_p\over 5-p}\,\,M\,N\, \big(\,\sin\bar \theta_{(p,n)}\,\big)^{6-p}\,\, 
\left[g_{\mbox{\tiny eff}}^2(T)\right]^{\frac{1}{5-p}}\,\,.
 \end{equation}
It is interesting to study what happens in case of small filling-fractions $\nu\,=\,n/N$. The impurity entropy
\eqref{ImpEntr} depends on the filling-fraction $\nu$ through the factor $\sin^{6-p}{\bar{\theta}_{\mbox{\tiny $(p,n)$}}}$
contained in the energy density $\mathcal{E}_{\mbox{\tiny $(p,n)$}}$. In the small filling-fraction limit the angle $\bar\theta_{(p,n)}$ is also small and one can check \cite{Camino:2001at} that, at leading order in $n/N$, one has:
\beq
\Big(\bar\theta_{(p,n)}\Big)^{6-p}\,\approx\,
2\sqrt{\pi}\,\,\,
{\Gamma\Big({8-p\over 2}\Big)\over \Gamma\Big({7-p\over 2}\Big)}\,\,
{n\over N}\,+\,\cdots\,\,.
\label{theta-small-nu}
\eeq
Then,  the expansion of $\mathcal{E}_{\mbox{\tiny $(p,n)$}}$ is given by
\begin{equation}\eqlabel{EnDenSmall}
 \lim_{\frac{n}{N}\rightarrow0}\mathcal{E}_{\mbox{\tiny $(p,n)$}}\:=\:n\,T_{\mbox{\tiny F$1$}} \,
  \left[
   1-b_p\left(\frac{n}{N}\right)^{\frac{2}{6-p}}+\ldots
  \right],
 \qquad
 b_p\:\overset{\mbox{\tiny def}}{=}\:\frac{6-p}{2(8-p)}
  \left[
   2\sqrt{\pi}\frac{\Gamma\left(\frac{8-p}{2}\right)}{\Gamma\left(\frac{7-p}{2}\right)}
  \right]^{\frac{2}{6-p}},
\end{equation}
and, as a consequence, the impurity entropy can be written as
\begin{equation}\eqlabel{ImpEntrSmall}
 \begin{split}
  &\lim_{\frac{n}{N}\rightarrow0}\mathcal{S}_{\mbox{\tiny D$(8-p)$}}^{\mbox{\tiny str}}\:=\:
    n\,M\, a_p\left[g_{\mbox{\tiny eff}}^2(T)\right]^{\frac{1}{5-p}}
    \left[
     1-b_p\left(\frac{n}{N}\right)^{\frac{2}{6-p}}+\ldots
    \right],\\
  &\hspace{2cm}\mbox{ with }\quad a_p\:\overset{\mbox{\tiny def}}{=}\:
  {4\sqrt{\pi}\over 5-p}\,\,
\Bigg[\,{\Gamma\Big({7-p\over 2}\Big)\over
(7-p)^2\,\pi}\,\Bigg]^{{1\over 5-p}}\,\,,
 \end{split}
\end{equation}
One can verify that (\ref{ImpEntr}) and (\ref{ImpEntrSmall}) for $p=3$ coincide with the impurity entropy computed in \cite{Mueck:2010ja, Harrison:2011fs} for the 
holographic dual of the maximally supersymmetric Kondo model. Our results generalise those in \cite{Mueck:2010ja, Harrison:2011fs} for non-conformal D$p$-brane 
backgrounds. 

From \eqref{ImpEntrSmall} we notice that the dependence of $\mathcal{S}$ on the filling fraction is, in general, 
non-analytic. However,  in the particular case of the D$4$/D$4$ system, the impurity entropy turns out to be  analytic in 
the filling fraction. Actually, it is possible to obtain a simple closed expression of 
$\mathcal{S}_{\mbox{\tiny D$4$}}^{\mbox{\tiny str}}$ as a function of $\nu$. In particular, for such a case
the system is in the minimal-energy configuration if the probe D$4$-brane is located at 
$\theta\,=\,\bar{\theta}_{\mbox{\tiny $(4,n)$}}$ such that
\begin{equation}\eqlabel{D4D4angle}
 \cos{\bar{\theta}_{\mbox{\tiny $(4,n)$}}}\:=\:1-2\nu,
\end{equation}
satisfying the condition \eqref{StableConf2}. Using \eqref{D4D4angle} the expression for the entropy density for $p\,=\,4$
can be written as
\begin{equation}\eqlabel{D4D4ImpEntr}
 \mathcal{S}_{\mbox{\tiny D$4$}}^{\mbox{\tiny str}}\:=\:\frac{2n\,M}{9}\,g^2_{\mbox{\tiny eff}}(T)\left(1-\nu\right).
\end{equation}
where $g^2_{\mbox{\tiny eff}}(T)\,=\,g^2_{YM}\,N\,T$ in this $p=4$ case.

The internal energy ${\cal E}$ is obtained from the free energy $F$ by means of the standard thermodynamic formula
\beq
{\cal E}\,=\,F+T {\mathcal S}\,\,.
\eeq
For the straight brane configuration one can immediately  obtain ${\cal E}$ by combining (\ref{FreeEn}) and (\ref{ImpEntr}), namely
\beq
{\cal E}_{\mbox{\tiny D$(8-p)$}}^{\mbox{\tiny str}}\,=\,{p-3\over 2}\,\,T\,
\mathcal{S}_{\mbox{\tiny D$(8-p)$}}^{\mbox{\tiny str}}\,\,.
\eeq
Notice that ${\cal E}_{\mbox{\tiny D$5$}}^{\mbox{\tiny str}}$ vanishes, while 
${\cal E}_{\mbox{\tiny D$(8-p)$}}^{\mbox{\tiny str}}$ becomes negative for $p<3$.

In order to complete our thermodynamics analysis, we can compute the impurity specific heat from the impurity entropy, namely
\eqref{ImpEntr}
\begin{equation}\eqlabel{ImpSpecHeat}
 C_{\mbox{\tiny D$(8-p)$}}^{\mbox{\tiny str}}\:=\:
  T\frac{\partial\mathcal{S}_{\mbox{\tiny D$(8-p)$}}^{\mbox{\tiny str}}}{\partial T}\:=\:
  \frac{2(p-3)c_p}{(5-p)^2}\,\,M\,N\,({\rm sin}\,\bar\theta_{(p,n)})^{6-p}\,
  \left[g^2_{\mbox{\tiny eff}}(T)\right]^{\frac{1}{5-p}},
\end{equation}
which vanishes for $p\,=\,3$ and is negative for $p\,<\,3$, giving a signature of a thermodynamic instability. 
From \eqref{ImpSpecHeat}, it is easy to compute the impurity susceptibility
\begin{equation}\eqlabel{ImpSusc}
 \chi^{\mbox{\tiny str}}_{\mbox{\tiny D$(8-p)$}}\:\propto\:\frac{\partial C_{\mbox{\tiny D$(8-p)$}}^{\mbox{\tiny str}}}{\partial T}\:\propto\:
 (p-3)^2 \frac{\left[g^2_{\mbox{\tiny eff}}(T)\right]^{\frac{1}{5-p}}}{T}\:=\:(p-3)^2\frac{\left[g^2_{\mbox{\tiny YM}} N\right]^{\frac{1}{5-p}}}{T^{2\frac{4-p}{5-p}}},
\end{equation}
which is constant for $p\,=\,4$.

\section{Hanging Flux Tube Configurations}\label{sec:Hanging}

Let us now consider the configuration in which $\theta$ is a constant and the embedding of the probe brane in the
$(p+2)$-dimensional (conformally)-$AdS$ manifold is described through the scalar $z(\rho)$. The probe branes are
located at $\theta\,=\,\bar{\theta}_{(p,n)}$.

Analysing the form of the first integral of motion \eqref{zFI}, one can notice that, in the case of black hole 
embedding phase, the regularity condition at the horizon fixes the (rescaled) first integral of motion $\tilde{c}_z$ to be
zero, and, as a consequence, the embedding function $z$ must have a trivial profile. 

Focusing instead on the configuration in which the probe brane lies completely outside the black hole, \eqref{zFI} implies
the existence of a turning point. Inverting \eqref{zFI} with respect to $z'$ and integrating the obtained equation, one gets
\begin{equation}\eqlabel{zInt}
 z(\rho)\:=\:\pm\int_{\rho}^{\rho_{\mbox{\tiny t}}}\frac{c_z^{\mbox{\tiny $(p,n)$}}d\rho'}{
  \sqrt{
   \mathtt{h}_p
  \left[\left(N\,e^{\phi}\right)^{\frac{4}{7-p}}(\rho')^{-4}\,\mathtt{h}_p-\left(c_z^{\mbox{\tiny $(p,n)$}}\right)^2\right]}},
 \quad\qquad
 c_z^{\mbox{\tiny $(p,n)$}}\:\overset{\mbox{\tiny def}}{=}\:\frac{\tilde{c}_z}{\sin^{6-p}{\bar{\theta}_{p,n}}}.
\end{equation}
It is easy to see from \eqref{zInt} and \eqref{cz2} that the constant $c_z^{\mbox{\tiny $(p,n)$}}$ is actually related to 
the energy density \eqref{EnDens}
\begin{equation}\eqlabel{cpn}
 c_z^{\mbox{\tiny $(p,n)$}}\:=\:\frac{c_z}{\mathcal{E}_{(p,n)}\,B_p},
\end{equation}
where $c_z$ is the constant introduced in (\ref{IntMot}) and the position $\rho_{\mbox{\tiny t}}$ of the  turning point is given by
\begin{equation}\eqlabel{TurnPoint}
 \rho_{\mbox{\tiny t}}\:=\:
  \frac{\rho_{\mbox{\tiny h}}}{\left[1+
  \big({\rho_h\over \beta_p}\big)^2\,
   \left(c_z^{\mbox{\tiny $(p,n)$}}\right)^2\right]^{\frac{5-p}{2(7-p)}}},
 \end{equation}
 Let us next introduce a new radial variable $\sigma$, related to $\rho$ by means of the expression
 \beq
 \sigma\,=\,{\rho\over \rho_h}\,\,.
 \label{sigma-def}
 \eeq
 Clearly, the turning point in this new variable is just
 \beq
 \sigma_{\mbox{\tiny t}}\,=\,
 \left[1+
  \Big({\rho_h\over \beta_p}\Big)^2\,
   \left(c_z^{\mbox{\tiny $(p,n)$}}\right)^2\right]^{\frac{p-5}{2(7-p)}}\,\,.
   \label{sigma_t}
 \eeq
 Moreover, the separation on the boundary  between the $n$-quark and the $n$-antiquark is given by
\begin{equation}\eqlabel{Sep}
 L\:=\:{2\rho_{\mbox{\tiny h}}^{2}\,c_z^{\mbox{\tiny $(p,n)$}}\over \beta_p}
 \,\,
  \sigma_{\mbox{\tiny t}}^{\frac{7-p}{5-p}}
  \,\,
  {\cal I}(\sigma_{\mbox{\tiny t}})\,\,,
  \end{equation}
where ${\cal I}(\sigma_{\mbox{\tiny t}})$ is the integral
\beq
{\cal I}(\sigma_{\mbox{\tiny t}})\,=\,
\int_{0}^{\sigma_{\mbox{\tiny t}}}d\sigma\:\frac{\sigma^{\frac{7-p}{5-p}}}{
  \left[
   \left(1-\sigma^{2\frac{7-p}{5-p}}\right)
   \left(\sigma_{\mbox{\tiny t}}^{2\frac{7-p}{5-p}}-\sigma^{2\frac{7-p}{5-p}}\right)
  \right]^{1/2}}\,\,.
 \label{I_sigma_t}
\eeq
By using (\ref{sigma_t}) one can eliminate in (\ref{Sep}) the constant $c_z^{\mbox{\tiny $(p,n)$}}$ in favour of $\sigma_{\mbox{\tiny t}}$, with the result
\beq
L\,=\,2\rho_{h}\,\,
\sqrt{1-\big(\,\sigma_{\mbox{\tiny t}}\,\big)^{{2(7-p)\over 5-p}}}\,\,\,\,
{\cal I}(\sigma_{\mbox{\tiny t}})\,\,.
\eeq
By applying the same techniques as in appendix \ref{app:dimerintegrals}, the integral (\ref{I_sigma_t}) can be obtained in terms of the hypergeometric function. One gets:
\beq
{\cal I}(\sigma_{\mbox{\tiny t}})\,=\,
{\Gamma\Big({6-p\over 7-p}\Big)\over \Gamma\Big({5-p\over2(7-p)}\Big)}\,\,\sqrt{\pi}\,\,
\sigma_t\,\,
F\Big(\,{1\over 2}\,,\,1-{1\over 7-p}\,;\,{3\over 2}-{1\over 7-p}\,;\,
\sigma_{\mbox{\tiny t}}^{2\frac{7-p}{5-p}}\,\Big)\,\,.
\eeq
Let us use these results to write $L$ in a closed form. Actually, it is very convenient to define the parameter $\gamma$ as
\beq
\gamma\,\overset{\mbox{\tiny def}}{=}\,
\sigma_{\mbox{\tiny t}}^{2\frac{7-p}{5-p}}\,\,.
 \label{gamma_def}
\eeq
It follows from this definition  that $0\le \gamma\le 1$. Moreover, $L$ can be written as
\beq
L\,=\,{7-p\over (5-p)\,\sqrt{\pi}\,T}\,\,
{\Gamma\Big({6-p\over 7-p}\Big)\over \Gamma\Big({5-p\over2(7-p)}\Big)}\,\,
\gamma^{{5-p\over 2(7-p)}}\,\,\sqrt{1-\gamma}\,\,
F\Big(\,{1\over 2}\,,\,1-{1\over 7-p}\,;\,{3\over 2}-{1\over 7-p}\,;\,
\gamma\,\Big)\,\,,\qquad
\label{L_hyper}
\eeq
where we used (\ref{Tp}) to eliminate $\rho_h$ and  we wrote the result in terms of the temperature $T$.  The energy for this hanging configuration is
\begin{equation}\eqlabel{EnHang}
 E\:=\:2\,\mathcal{E}_{\mbox{\tiny $(p,n)$}}\,\beta_p\,B_p\,
  \sigma_{\mbox{\tiny t}}^{\frac{7-p}{5-p}}\,
             \int_{\epsilon}^{\sigma_{\mbox{\tiny t}}}\frac{d\sigma}{\sigma^{\frac{7-p}{5-p}}}\:
       \frac{\sqrt{1-\sigma^{2\frac{7-p}{5-p}}}}{\sqrt{\sigma_{\mbox{\tiny t}}^{2\frac{7-p}{5-p}}-
        \sigma^{2\frac{7-p}{5-p}}}}\,\,,
 \end{equation}
where the cut-off $\epsilon$ has been introduced to regularise the Hamiltonian near the boundary $\sigma=0$, where the integral 
diverges. It can be regulated by minimal subtraction, {\it i.e} by
the energy corresponding to a disconnected configuration. Therefore, let us look at the divergences near the boundary
$\sigma\,=\,0$, where the embedding function $z(\rho)$ and its first derivative have the following asymptotic
expansion
\begin{equation}\eqlabel{zasymp}
 \begin{split}
  &z'(\sigma)\:=\:\pm
  {\rho_h^2\over \beta_p}\,\,c_z^{\mbox{\tiny $(p,n)$}}\,
  \sigma^{\frac{7-p}{5-p}}
    \left[
     1+
     \frac{1}{2}\left(1+\sigma_{\mbox{\tiny t}}^{-2\frac{7-p}{5-p}}\right)\sigma^{2\frac{7-p}{5-p}}+
     \mathcal{O}\left(\sigma^{4\frac{7-p}{5-p}}\right)
    \right],\\
  &z(\sigma)\:=\:\mp {L\over 2}\,\pm\,
  {5-p\over 2(6-p)}\, {\rho_h^2\over \beta_p}\,\,c_z^{\mbox{\tiny $(p,n)$}}\,
  \sigma^{2\frac{6-p}{5-p}}
   +\mathcal{O}\left(\sigma^{2\frac{13-2p}{5-p}}\right).
 \end{split}
\end{equation}
It is easy to read off the divergent boundary Hamiltonian
\begin{equation}\eqlabel{Hdiv}
 H\Big|_{\mbox{\tiny div}}\:=\:(5-p)\,\mathcal{E}_{\mbox{\tiny $(p,n)$}}\beta_p \,B_p\,\,\epsilon^{-\frac{2}{5-p}}.
\end{equation}
As a consequence, one needs to add just a counterterm which is proportional to the volume of the boundary, regularised 
with $\sigma\,=\,\epsilon$
\begin{equation}\eqlabel{Hdiv2}
 H\Big|_{\mbox{\tiny ct}}\:=\:
 -(5-p)
 \,\mathcal{E}_{\mbox{\tiny $(p,n)$}}\,B_p\:
 \left(N\,e^{\phi}\big|_{\epsilon}\right)^{\frac{2}{7-p}}\sqrt{-\left.\mathtt{g}_{tt}\right|_{\mbox{\tiny $\epsilon$}}}\,\,.
\end{equation}
As noticed in \cite{Benincasa:2009ze}, the divergent Hamiltonian \eqref{Hdiv} is exactly the divergent boundary Hamiltonian 
that one would obtain from probe D-branes in an $AdS$-space of dimensions $2/(5-p)+1$, and the counterterm \eqref{Hdiv2} is 
actually given in terms of the volume of the boundary of this $AdS$-space.

Therefore, the renormalised Hamiltonian is given by
\begin{equation}\eqlabel{Hren}
 H\Big|_{\mbox{\tiny ren}}\:=\:\lim_{\epsilon\rightarrow0}
 \left[
  H\Big|_{\mbox{\tiny on-shell}}+H\Big|_{\mbox{\tiny ct}}
 \right]
\end{equation}
and the free energy can be written as
\begin{equation}\eqlabel{FreeEnHang}
 F^{\cup}_{\mbox{\tiny D$(8-p)$}}
 \:=\:2\,\mathcal{E}_{\mbox{\tiny $(p,n)$}}\,\,\beta_p\,B_p\,
      \left[
       \int_{0}^{\sigma_{\mbox{\tiny t}}}\frac{d\sigma}{
       \sigma^{\frac{7-p}{5-p}}}\:
       \left(
        \frac{\sigma_{\mbox{\tiny t}}^{\frac{7-p}{5-p}}\,\,
        \sqrt{1-\sigma^{2\frac{7-p}{5-p}}}}
        {\sqrt{\sigma_{\mbox{\tiny t}}^{2\frac{7-p}{5-p}}-
        \sigma^{2\frac{7-p}{5-p}}}}-1
       \right)-\frac{5-p}{2}\,\sigma_{\mbox{\tiny t}}^{-\frac{2}{5-p}}
     \right].
\end{equation}
Let us now change to a new variable $\xi$ in the integral (\ref{FreeEnHang}), defined as
\beq
\xi\,=\,\Big(\,{\sigma_{\mbox{\tiny t}}\over \sigma}\,\Big)^{{2\over 5-p}}\,\,.
\eeq
After this change of variable, it is easy to verify that $ F^{\cup}_{\mbox{\tiny D$(8-p)$}}$ becomes
\beq
 F^{\cup}_{\mbox{\tiny D$(8-p)$}}\,=\,(5-p)\,
 \mathcal{E}_{\mbox{\tiny $(p,n)$}}\,\,\beta_p\,B_p\,
 \gamma^{-{1\over 7-p}}\,\,
  \left[
       \int_{1}^{\infty}d\xi \:
       \left(
        \frac{
        \sqrt{\xi^{7-p}-\gamma}}
        {\sqrt{\xi^{7-p}-1}}-1
       \right)\,-\,1
     \right]\,\,,
     \label{F_cup}
 \eeq
where $\gamma$ is related to $\sigma_{\mbox{\tiny t}}$ as in (\ref{gamma_def}). The term inside the square brackets in (\ref{F_cup}) is just the integral $J(7-p, \gamma)$ calculated in appendix \ref{app:dimerintegrals}. Thus, we can write
\beq
F^{\cup}_{\mbox{\tiny D$(8-p)$}}\,=\,(5-p)\,
 \mathcal{E}_{\mbox{\tiny $(p,n)$}}\,\,\beta_p\,B_p\,
 \gamma^{-{1\over 7-p}}\,\,
J(7-p, \gamma)\,\,.
\label{Fcup-J}
\eeq
The prefactor in (\ref{Fcup-J}) can be related to the free energy of the straight brane. Indeed, according to (\ref{EuclActRen}) and (\ref{FreeEn}) one has
$(5-p)\, \mathcal{E}_{\mbox{\tiny $(p,n)$}}\,\,\beta_p\,B_p\,=\,
F^{\mbox{\tiny str}}_{\mbox{\tiny D$(8-p)$}}$. Moreover, by using the explicit value of the integral 
$J(7-p, \gamma)$ (see eq. (\ref{J-int-value})), one gets
\beq
F^{\cup}_{\mbox{\tiny D$(8-p)$}}\,=\,2\,F^{\mbox{\tiny str}}_{\mbox{\tiny D$(8-p)$}}\,
\gamma^{-{1\over 7-p}}\,\,
{\Gamma\Big({6-p\over 7-p}\Big)\over \Gamma\Big({5-p\over2(7-p)}\Big)}\,\,
\sqrt{\pi}\,\,
F\Big(\,-{1\over 2}\,,\,-{1\over 7-p}\,;\,{5-p\over 2(7-p)}\,;\,
\gamma\,\Big)\,\,.
\label{Fcup-hyper}
\eeq
The constant $\gamma$ appearing on the expressions of the length $L$ and free energy $F^{\cup}_{\mbox{\tiny D$(8-p)$}}$ (eqs. (\ref{L_hyper}) and (\ref{Fcup-hyper})) parametrises the turning point of the hanging brane configuration, \ie\ how much the brane penetrates into the bulk. The configuration with $\gamma\to 0$ corresponds to $\rho_{\mbox{\tiny t}}\to 0$, which means that the brane lies completely on the boundary in this limit. On the contrary, when $\gamma=1$  one has $\rho_{\mbox{\tiny t}}=\rho_h$ and, thus, this configuration corresponds to two disconnected straight branes ending on the black hole horizon. It is easy to check that our formula (\ref{Fcup-hyper}) is consistent with this interpretation. Indeed, as
\beq
F\Big(\,-{1\over 2}\,,\,-{1\over 7-p}\,;\,{5-p\over 2(7-p)}\,;\,
1\,\Big)\,=\,{1\over \sqrt{\pi}}\,\,
{\Gamma\Big({5-p\over2(7-p)}\Big)\over \Gamma\Big({6-p\over 7-p}\Big)}
\,\,,
\eeq
one gets
\beq
\lim_{\gamma\to 1}\,F^{\cup}_{\mbox{\tiny D$(8-p)$}}\,=\,2\,
F^{\mbox{\tiny str}}_{\mbox{\tiny D$(8-p)$}}\,\,.
\label{Fcup-gamma1}
\eeq

Finally, we can now compute the one-point correlator of the boundary operator $\mathcal{O}_{\mbox{\tiny $z$}}$
associated with the scalar function $z$. In order to perform this computation, we consider the Euclidean version of the action \eqref{D8mpAct2} at 
$\theta\,=\,\bar{\theta}_{\mbox{\tiny $(p,n)$}}$
\begin{equation}\eqlabel{EuclActConstAngl}
 \begin{split}
  &I_{\mbox{\tiny D$(8-p)$}}^{\mbox{\tiny Eucl}}\:=\:T_{\mbox{\tiny D$(8-p)$}}N B_{p}^{\frac{9-p}{2}}\Omega_{\mbox{\tiny $7-p$}}u_p^{7-p}\int dt d\rho\:
   \left(N\,e^{\phi}\right)^{\frac{2}{7-p}}\rho^{-2}%\sqrt{g_{\mbox{\tiny $AdS_2^{\mbox{\tiny BH}}$}}}
   \Big\{
    \sin^{7-p}{\bar{\theta}_{\mbox{\tiny $(p,n)$}}}\times
    \\
  &\left.
    \hspace{1cm}\times
    \left[
     1+\mathtt{h}_{p}\left(z'\right)^2-\mathcal{F}_{t\rho}^2
    \right]^{1/2}+
    \mathcal{F}_{t\rho}C_p\left(\bar{\theta}_{\mbox{\tiny $(p,n)$}}\right)
   \right\}\:=\\
  &\phantom{I_{\mbox{\tiny D$(8-p)$}}^{\mbox{\tiny Eucl}}}\:=\:\frac{\mathcal{E}_{\mbox{\tiny $(p,n)$}}B_p}{\sin^{6-p}{\bar{\theta}}_{\mbox{\tiny $(p,n)$}}}
   \left[\frac{g_{\mbox{\tiny YM}}^2 N}{2(2\pi)^{p-2}}\right]^{\frac{1}{5-p}}\left(\frac{d_p}{u_p}\right)^{\frac{p-3}{(5-p)(7-p)}}\int dt d\rho\:
   \rho^{-\frac{7-p}{5-p}}
   \Big\{
    \sin^{7-p}{\bar{\theta}_{\mbox{\tiny $(p,n)$}}}\times
    \\
  &\left.
    \hspace{1cm}\times
    \left[
     1+\mathtt{h}_{p}\left(z'\right)^2-\mathcal{F}_{t\rho}^2
    \right]^{1/2}+
    \mathcal{F}_{t\rho}C_p\left(\bar{\theta}_{\mbox{\tiny $(p,n)$}}\right)
   \right\},
 \end{split}
\end{equation}
where the second equality has been written using the explicit expression for the dilaton \eqref{FGcoords2} and $U_p$ \eqref{Up} as well as the relation \eqref{cf}
and the expression for the energy density \eqref{EnDens}. It is interesting to notice that the overall term $\rho^{-\frac{7-p}{5-p}}$ in \eqref{EuclActConstAngl}
can be seen as the determinant of an $AdS$ black-hole metric from two-dimension to $1+q$, with $q\,=\,2/(5-p)$. Therefore, we can think that our probe D$(8-p)$-brane configuration is actually in a higher dimensional $AdS$ black-hole background (times $S^{8-p}$) and that the induced metric on the brane has an 
(asymptotically) $AdS_{1+q}$ factor, with $q$ integer -- then we can perform the analytic continuation to the value $2/(5-p)$. This action is renormalised
exactly by the same term \eqref{Hdiv2} used to renormalised the Hamiltonian -- notice that if one writes the explicit expression for the dilaton in
\eqref{Hdiv2}, the counterterm can be seen as proportional to the volume of the boundary of $AdS_{1+q}$.
In this way, we can compute the one-point function for the operator $\mathcal{O}_{\mbox{\tiny $z$}}$ dual to the embedding function $z(\rho)$ as
\begin{equation}\eqlabel{1ptCorrZ}
 \langle\mathcal{O}_{\mbox{\tiny $z$}}\rangle\:=\:
 \lim_{\epsilon\rightarrow0}\frac{1}{\epsilon^{\Delta_{z}}}\frac{1}{\sqrt{\left.g_{\mbox{\tiny $\partial AdS_{1+q}$}}\right|_{\epsilon}}}
  \frac{\delta \left.I_{\mbox{\tiny D$(8-p)$}}^{\mbox{\tiny Eucl}}\right|_{\mbox{\tiny ren}}}{\delta z(\epsilon)}.
\end{equation}
where $\left.g_{\mbox{\tiny $\partial AdS_{1+q}$}}\right|_{\epsilon}$ is the determinant of the metric at the boundary of $AdS_{\mbox{\tiny $1+q$}}$ regularized at 
$\rho\,=\,\epsilon$, and $\Delta_{z}\:=\:2/(5-p)$. Using equation \eqref{1ptCorrZ}, the explicit expression for the one-point correlator is
\begin{equation}\eqlabel{1ptCorrZb}
 \begin{split}
  \langle\mathcal{O}_{\mbox{\tiny $z$}}\rangle\:=\:&
   \frac{\mathcal{E}_{\mbox{\tiny $(p,n)$}}B_p}{\sin^{6-p}{\bar{\theta}}_{\mbox{\tiny $(p,n)$}}}
   \left[\frac{g_{\mbox{\tiny YM}}^2 N}{2(2\pi)^{p-2}}\right]^{\frac{1}{5-p}}\left(\frac{d_p}{u_p}\right)^{\frac{p-3}{(5-p)(7-p)}}\times\\
   &\times\lim_{\epsilon\rightarrow0}\frac{1}{\epsilon^{\frac{2}{5-p}}}\frac{1}{\sqrt{\left.g_{\mbox{\tiny $\partial AdS_{1+q}$}}\right|_{\epsilon}}}
   \sin{\bar{\theta}_{\mbox{\tiny $(p,n)$}}}\left.\frac{\mathfrak{h}_p\,z'}{\left[1+\mathfrak{h}_p\left(z'\right)^2\right]^{1/2}}\right|_{\epsilon}\:=\:-c_{z},
 \end{split}
\end{equation}
where the last equality has been obtained by using the equation of motion \eqref{zFI} for $z$\footnote{One has to remember that this computation has been performed by
using the Euclidean action, as usual. If one repeats the same calculation using the Minkowski signature, the result would be the same up to a sign.}, with
$c_{z}$ has been defined in eq \eqref{IntMot}.

\section{Dimer Thermodynamics}\label{sec:dimer_thermo}

\begin{figure}
 \centering%
 {\scalebox{.70}{\includegraphics{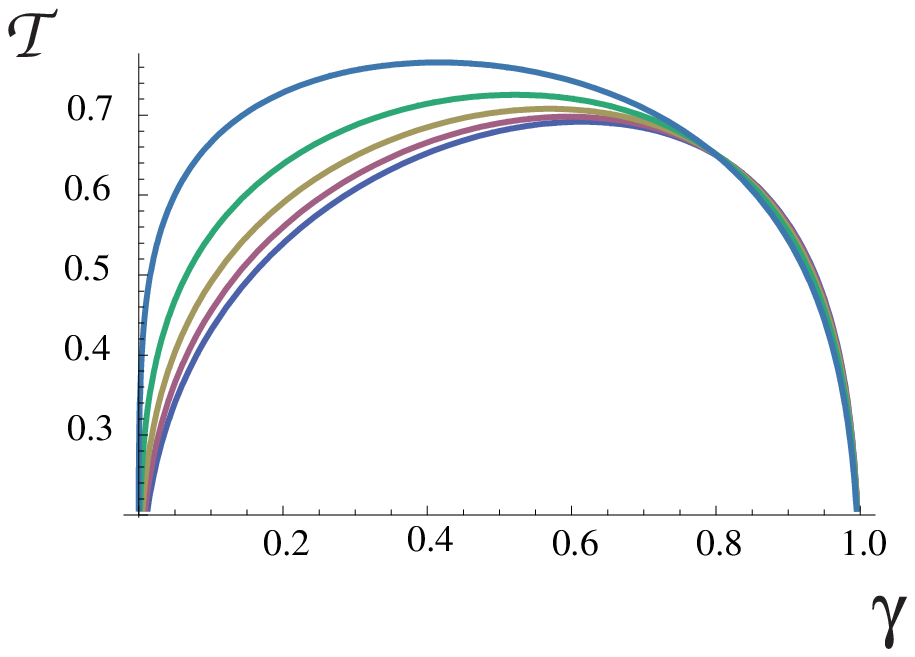}}
\qquad\qquad\qquad
  {\scalebox{.70}{\includegraphics{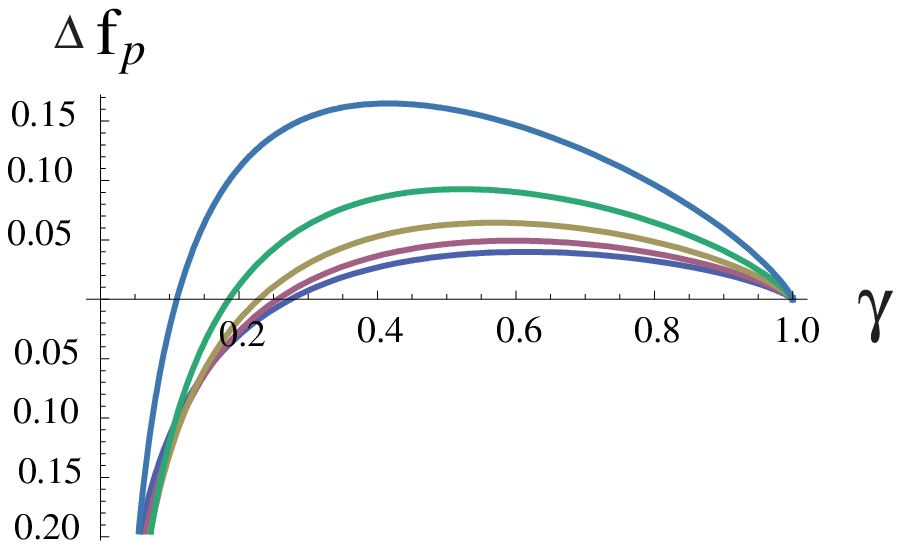}}}}
 \caption{On the left we plot the reduced temperature ${\cal T}$ versus the parameter $\gamma$ for $0\le p\le 4$. As $p$ is increased the maximum of the curve is shifted towards lower values of $\gamma$. On the right we plot $\Delta f_p$ versus $\gamma$ for $0\le p\le 4$. The points where the different curves cut the horizontal axis define the $\gamma$ parameter $\gamma_p^*$ of the  dimerisation transition . }
 \label{Tgamma}
\end{figure}

In this section we will study in detail the thermodynamic properties of the connected configuration of the D$(8-p)$-brane. First of all, we will rewrite the free energy (\ref{Fcup-hyper}) in  a more convenient way. With this purpose, let us define two functions $h(\gamma)$ and $g(\gamma)$ as follows:
\bear
&&h(\gamma)\,\,\overset{\mbox{\tiny def}}{=}\,\,
\gamma^{-{1\over 7-p}}\,\,
F\Big(\,-{1\over 2}\,,\,-{1\over 7-p}\,;\,{1\over 2}\,-\,{1\over 7-p}\,;\,
\gamma\,\Big)\,\,,
\rc\rc
&&g(\gamma)\,\,\overset{\mbox{\tiny def}}{=}\,\,
\gamma^{{1\over 2}-{1\over 7-p}}\,\,\sqrt{1-\gamma}\,\,
F\Big(\,{1\over 2}\,,\,1-{1\over 7-p}\,;\,{3\over 2}-{1\over 7-p}\,;\,
\gamma\,\Big)\,\,,
\label{h_g_def}
\eear
as well as the constant $\Delta_p$, given by
\beq
\Delta_p\,=\,
{\Gamma\Big({6-p\over 7-p}\Big)\over \Gamma\Big({5-p\over2(7-p)}\Big)}\,\,
\sqrt{\pi}\,\,.
\eeq
Then, the free energy (\ref{Fcup-hyper}) can be written as
\beq
F^{\cup}_{\mbox{\tiny D$(8-p)$}}(T)\,=\,2\,\Delta_p\,\,
F^{str}_{\mbox{\tiny D$(8-p)$}}(T)\,\,h(\gamma)\,\,,
\label{F_cup_h}
\eeq
where we have stressed the fact that both $F^{\cup}_{\mbox{\tiny D$(8-p)$}}$ and
$F^{str}_{\mbox{\tiny D$(8-p)$}}$ depend on $T$. The dependence of $F^{str}_{\mbox{\tiny D$(8-p)$}}$ on $T$ is shown explicitly in (\ref{FreeEn}). Moreover,  $F^{\cup}_{\mbox{\tiny D$(8-p)$}}$ depends implicitly on $T$ through its dependence on the parameter  $\gamma$. Actually, by rewriting (\ref{L_hyper}) as
\beq
T\,=\,{7-p\over (5-p)\,\pi\,L}\,\,\Delta_p\,g(\gamma)\,\,,
\label{T-gamma}
\eeq
we get the explicit relation between the temperature $T$ and the parameter $\gamma$ (for fixed length $L$). In order to study the dependence of the temperature on $\gamma$, let us define the dimensionless reduced temperature ${\cal T}$ as
\beq
{\cal T}\,\,\,\overset{\mbox{\tiny def}}{=}\,\,{(5-p)\pi\over (7-p)\Delta_{p}}\,\,L\,T\,\,.
\label{reducedT}
\eeq
It follows from (\ref{T-gamma})  that ${\cal T}(\gamma)$ is just given by the function 
$g(\gamma)$, namely
\beq
{\cal T}(\gamma)\,=\,g(\gamma)\,\,.
\label{reducedT-gamma}
\eeq
Let us now analyse the behaviour of ${\cal T}(\gamma)$ near $\gamma\approx 0, 1$. From the explicit expression of $g(\gamma)$ in terms of the hypergeometric function (see the second equation in (\ref{h_g_def})), one can verify that
\bear
&&{\cal T}(\gamma)\,\approx\, \gamma^{{1\over 2}-{1\over 7-p}}\,\,,
\hspace{6.55cm}\gamma\approx 0\,\,,\rc\rc
&&{\cal T}(\gamma)\,\approx\, -\,
{\Gamma\Big({3\over 2}-{1\over 7-p}\Big)\over
 \sqrt{\pi}\,\Gamma\Big(1-{1\over 7-p}\Big)}\,\,\sqrt{1-\gamma}\,\,
 \log (1-\gamma)\,\,,\qquad\qquad \gamma\approx 1\,\,,
\eear
and, therefore,  ${\cal T}(\gamma)\to 0$ as $\gamma\to 0,1$. Actually, by plotting the function ${\cal T}(\gamma)$  for different values of $p$ one discovers that it reaches a maximum at some value  $\gamma=\gamma_{p}^{\mbox{\tiny max}}$ with $0<\gamma_{p}^{\mbox{\tiny max}}<1$.  These  plots are shown in figure \ref{Tgamma} for $0\le p\le 4$. 

The fact that ${\cal T}$ has an upper bound implies that above a certain temperature 
${\cal T}_p^{\mbox{\tiny max}}={\cal T} (\gamma=\gamma_{p}^{\mbox{\tiny max}})$ (which depends on $p$), only the disconnected solution exists. The values of $\gamma_{p}^{\mbox{\tiny max}}$ and ${\cal T}_p^{\mbox{\tiny max}}$ can be obtained by requiring the vanishing of the first derivative of $g(\gamma)$. 
We can calculate $g'(\gamma)$  by differentiating the right-hand-side of the second equation in (\ref{h_g_def}). By doing so we get a term which contains the derivative on the hypergeometric function, which we  simplify by using the relation
\beq
\gamma\,(\,1-\gamma)\,{d\over d\gamma}\,\,F(a,b;c;\gamma)\,=\,
(c-a)\,F(a-1,b;c;\gamma)\,+\,(a-c+b\gamma)\,F(a,b;c;\gamma)\,\,.
\eeq
Then, $g'(\gamma)$ can be written as a combination of two hypergeometric functions which, remarkably,  can be simplified to be:
\beq
g'(\gamma)\,=\,
{5-p\over 2(7-p)}\,\,
\gamma^{-{1\over 2}-{1\over 7-p}}\,\,
(1-\gamma)^{-{1\over 2}}\,
F\Big(\,-{1\over 2}\,,\,1-{1\over 7-p}\,;\,{1\over 2}\,-\,{1\over 7-p}\,;\,
\gamma\,\Big)\,\,.
\label{gprime}
\eeq
Therefore,  the maximum of the reduced temperature ${\cal T}(\gamma)$ occurs at $\gamma=\gamma^{\mbox{\tiny max}}_p$, where $\gamma^{\mbox{\tiny max}}_p$ is the solution of the equation:
\beq
F\Big(\,-{1\over 2}\,,\,1-{1\over 7-p}\,;\,{1\over 2}\,-\,{1\over 7-p}\,;\,
\gamma^{\mbox{\tiny max}}_p\,\Big)\,=\,0\,\,.
\eeq
The numerical values of $\gamma^{\mbox{\tiny max}}_p$ for $0\le p\le 4$ and of the corresponding reduced temperatures ${\cal T}^{\mbox{\tiny max}}_p$ have been written in table \ref{critical}. One notices that $\gamma^{\mbox{\tiny max}}_p$ becomes smaller as $p$ is increased, while ${\cal T}^{\mbox{\tiny max}}_p$ grows with $p$. 

For ${\cal T}< {\cal T}^{\mbox{\tiny max}}_p$
it is clear from figure \ref{Tgamma} that there are two connected solutions with different $\gamma$ for a given value of  ${\cal T}$. Recall that $\gamma$ parametrizes the holographic coordinate of the turning point of the connected configurations. In addition, we have the disconnected configuration,  which is the only one that exists for  ${\cal T}> {\cal T}^{\mbox{\tiny max}}_p$ and is competing with the two connected ones when  the temperature  is lowered below ${\cal T}^{\mbox{\tiny max}}_p$. In order to determine which one of these three configurations is thermodynamically more stable we have to find out which one has the lowest free energy. One can establish numerically that the configuration with higher $\gamma$  (\ie\ extending deeper in the bulk)  has higher free energy than the connected one with lower $\gamma$. To compare the latter with the disconnected one, let us define the reduced difference of free energies $\Delta f_p$ as:
\beq
\Delta f_p\,\,\overset{\mbox{\tiny def}}{=}\,\,
{F^{\cup}_{\mbox{\tiny D$(8-p)$}}\,-\,2\,F^{\mbox{\tiny str}}_{\mbox{\tiny D$(8-p)$}}\over 
\big|\,2\,F^{\mbox{\tiny str}}_{\mbox{\tiny D$(8-p)$}}\,\big|}\,\,.
\label{Deltafp_definition}
\eeq
From (\ref{F_cup_h}) we get that, as a function of $\gamma$,   $\Delta f_p$ is given by:
\beq
\Delta f_p\,=\,1\,-\,\Delta_p\,h(\gamma)\,\,.
\label{Deltafp_gamma}
\eeq
Clearly, by its  definition (\ref{Deltafp_definition}) and  by (\ref{Fcup-gamma1}) it follows that $\Delta f_p\to 0$ as $\gamma\to 1$. Moreover, near $\gamma\approx 0$ one can check that $\Delta f_p$ diverges as $\Delta f_p\approx -\Delta_p\,\gamma^{-{1\over 7-p}}\to-\infty$.  The functions $\Delta f_p$ have been plotted in figure \ref{Tgamma}. It turns out that 
$\Delta f_p(\gamma)$ vanishes for an intermediate value $\gamma=\gamma^*_p$ which, according to (\ref{Deltafp_gamma}), satisfies:
\beq
h(\gamma^*_p)\,=\,{1\over \Delta_p}\,\,,
\qquad\qquad \gamma^*_p<1\,\,.
\eeq
\begin{table}
\centering
\begin{tabular}[b]{|c|c|c|c|c|c|}   
\hline
 $p$  & $\gamma^{max}_p$ & ${\cal T}^{max}_p$&
 $\gamma^*_p$& ${\cal T}^{*}_p$& $\Delta \epsilon_p^*$
  \\ 
\hline
\  0 & $0.6141$ & $0.6916$ & $0.2745$& $0.5927$& $1.5150$
\\    \hline
\  1 & $0.5965$ & $0.6979$ & $0.2557$& $0.5995$& $1.5274$
\\    \hline
\  2 & $0.5693$ & $0.7078$ & $0.2285$& $0.6101$& $1.5465$
\\ \hline
\  3 & $0.5215$ & $0.7254$ & $0.1855$& $0.6293$& $1.5799$
\\ \hline
\  4 & $0.4155$ & $0.7658$ & $0.1102$& $0.6752$& $1.6541$
\\ \hline
\end{tabular}
\caption{Numerical values of $\gamma^{\mbox{\tiny max}}_p$,  ${\cal T}^{\mbox{\tiny max}}_p$,  $\gamma^*_p$ and ${\cal T}^{*}_p$ for $0\le p\le 4$.  On the last column we have displayed the reduced latent heat of the dimerisation transition, as defined in (\ref{latent-heat-transition}). 
}
\label{critical}
\end{table}

The different numerical  values of $\gamma^*_p$ for $0\le p\le 4$ have been collected in table \ref{critical}, together with the corresponding values of the temperature ${\cal T}^{*}_p={\cal T} (\gamma^*_p)$.  As shown in table \ref{critical}, $\gamma_p^*< \gamma_p^{max}$, which means that the connected configuration with lower $\gamma$ is the more stable one when ${\cal T} <{\cal T}_p^*$.  At ${\cal T} ={\cal T}_p^*$ it takes place a dimerisation transition of the type advocated in  \cite{Kachru:2009xf}, in which  two disconnected branes are recombined to form a hanging brane and the impurities of the boundary theory condense to form bonds or dimers.  In order to complete the thermodynamic description of this dimerised phase, let us calculate the entropy of the connected configuration. By computing the derivative   of (\ref{F_cup_h}) with respect to $T$, we obtain:
\beq
{\mathcal S}^{\cup}_{\mbox{\tiny D$(8-p)$}}(T)\,=\,-{\partial \over \partial T}\,\,
F^{\cup}_{\mbox{\tiny D$(8-p)$}}(T)\,=\,2\Delta_p\,\,
\Big[\,{\mathcal S}^{\mbox{\tiny str}}_{\mbox{\tiny D$(8-p)$}}\,h(\gamma)\,-\,
F^{\mbox{\tiny str}}_{\mbox{\tiny D$(8-p)$}}\,\,{dh\over d\gamma}\,\,
\Big({dT\over d\gamma}\Big)^{-1}\,\,\Big]\,\,.
\label{Scup}
\eeq
By using  that (see (\ref{FreeEn}) and (\ref{ImpEntr}))
\beq
F^{\mbox{\tiny str}}_{\mbox{\tiny D$(8-p)$}}\,=\,-{5-p\over 2}\,\,T\,{\mathcal S}^{\mbox{\tiny str}}_{\mbox{\tiny D$(8-p)$}}\,\,,
\label{Fstr-Sstr}
\eeq
together with (\ref{T-gamma}), the relation (\ref{Scup}) can be written as
\beq
{\mathcal S}^{\cup}_{\mbox{\tiny D$(8-p)$}}(T)\,=\,2\Delta_p\,\,
{\mathcal S}^{\mbox{\tiny str}}_{\mbox{\tiny D$(8-p)$}}\,
\Big[\,h(\gamma)\,+\,{5-p\over 2}\,\,g(\gamma)\,\,{h'(\gamma)\over g'(\gamma)}\,
\Big]\,\,.
\eeq
The derivative of $g(\gamma)$ has been calculated above (see (\ref{gprime})). We
now compute  $h'(\gamma)$. To carry out this calculation we will differentiate the right-hand-side of the first equation in (\ref{h_g_def}) and we will  use the relation:
\beq
{d\over d\gamma}\,\,F(a,b;c;\gamma)\,=\,{ab\over c}\,\,
F(a+1,b+1;c+1;\gamma)\,\,.
\label{d-hyper}
\eeq
Proceeding in this way,  one can represent $h'(\gamma)$ as a combination of two hypergeometric functions which, remarkably, can be written  as a single hypergeometric function. One gets
\beq
h'(\gamma)\,=\,-{1\over 7-p}\,\,\gamma^{-1-{1\over 7-p}}\,\,
F\Big(\,-{1\over 2}\,,\,1-{1\over 7-p}\,;\,{1\over 2}\,-\,{1\over 7-p}\,;\,
\gamma\,\Big)\,\,.
\label{hprime}
\eeq
Amazingly, the hypergeometric functions on the right-hand-sides of (\ref{hprime}) and (\ref{gprime}) are the same and, therefore,  the ratio $h'(\gamma)/ g'(\gamma)$ is remarkably simple, namely
\beq
{h'(\gamma)\over g'(\gamma)}\,=\,-{2\over 5-p}\,\,
\sqrt{{1-\gamma\over \gamma}}\,\,.
\eeq
Therefore, the entropy of the connected configuration can be written as
\beq
{\mathcal S}^{\cup}_{\mbox{\tiny D$(8-p)$}}(T)\,=\,2\Delta_p\,\,
{\mathcal S}^{\mbox{\tiny str}}_{\mbox{\tiny D$(8-p)$}}\,
\Big[\,h(\gamma)\,-\,\sqrt{{1-\gamma\over \gamma}}\,g(\gamma)\,\Big]\,\,.  
\label{Scup-h-g}
\eeq
Again, the magic of the hypergeometric functions allows us to write the term in brackets in (\ref{Scup-h-g}) in a simplified form. One can actually check that
\beq
h(\gamma)\,-\,\sqrt{{1-\gamma\over \gamma}}\,g(\gamma)\,=\,
{2(6-p)(7-p)\over (19-3p)(5-p)}\,\,\sigma(\gamma)\,\,,
\label{h-g-sigma}
\eeq
where $\sigma(\gamma)$ is a new function defined as
\beq
\sigma(\gamma)\,\overset{\mbox{\tiny def}}{=}\,
\gamma^{1-{1\over 7-p}}\,\,
F\Big(\,{1\over 2}\,,\,1-{1\over 7-p}\,;\,{5\over 2}\,-\,{1\over 7-p}\,;\,
\gamma\,\Big)\,\,.
\label{sigma-def}
\eeq
By using this result, one can easily show that the entropy can be written as
\beq
{\mathcal S}^{\cup}_{\mbox{\tiny D$(8-p)$}}(T)\,=\,\hat\Delta_p\,
{\mathcal S}^{\mbox{\tiny str}}_{\mbox{\tiny D$(8-p)$}}(T)\,
\sigma(\gamma)\,\,,
\label{Scup-simply}
\eeq
where the coefficient $\hat\Delta_p$ is given by:
\beq
\hat\Delta_p\,\overset{\mbox{\tiny def}}{=}\,
{\Gamma\Big(2-{1\over 7-p}\Big)\over \Gamma\Big({5\over 2}-{1\over 7-p}\Big)}\,\,
\sqrt{\pi}\,\,.
\eeq
As a check of the equation (\ref{Scup-simply}), let us take the limit $\gamma\to 1$. One can  easily show that
\beq
\sigma (\gamma=1)\,=\,{2\over \hat\Delta_p}\,\,,
\label{sigma-gamma_1}
\eeq
and, therefore, one has
\beq
\lim_{\gamma\to 1}\,\,{\mathcal S}^{\cup}_{\mbox{\tiny D$(8-p)$}}(T)\,=\,
2\,{\mathcal S}^{\mbox{\tiny str}}_{\mbox{\tiny D$(8-p)$}}(T)\,\,,
\eeq
as expected. In order to study in more detail the entropy of the dimers, let us define the reduced entropy $\mathtt{s}_p(T)$ as the ratio
\beq
\mathtt{s}_p(T)\,\overset{\mbox{\tiny def}}{=}\,{{\mathcal S}^{\cup}_{\mbox{\tiny D$(8-p)$}}(T)\over
2\,{\mathcal S}^{\mbox{\tiny str}}_{\mbox{\tiny D$(8-p)$}}(T)}\,\,.
\eeq
It follows from (\ref{Scup-simply}) that $s_p(T)$ is given by
\beq
\mathtt{s}_p(T)\,=\,{\hat\Delta_p\over 2}\,\,\sigma(\gamma)\,\,.
\label{reduced-S}
\eeq
By plotting the right-hand-side of (\ref{reduced-S}) versus $\gamma$ one realises that $\mathtt{s}_p$ is a monotonic function of $\gamma$ which grows from zero to one as $\gamma$ varies in the interval $[0,1]$.  Let us study the behaviour of the entropy of the dimer near  $T\sim 0$, which corresponds to $\gamma\approx 0$. From the definition of $\sigma(\gamma)$ in (\ref{sigma-def}) it follows that  $\sigma(\gamma)\approx \gamma^{1-{1\over 7-p}}\to 0$ for small $\gamma$. Moreover, it is straightforward to prove from  the expression of the reduced temperature in (\ref{reducedT-gamma}) that 
$\gamma\approx \big(\,{\cal T}\,\big)^{{2(7-p)\over 5-p}}$ for small $\gamma$. Thus, we have for small ${\cal T}$ the reduced entropy of the dimer is given by
\beq
\mathtt{s}_p\approx {\hat\Delta_p\over 2} \,
\big(\,{\cal T}\,\big)^{{2(6-p)\over 5-p}}\,\,.
\eeq
Let us now use this result to obtain the  low temperature behaviour of the entropy ${\mathcal S}^{\cup}_{\mbox{\tiny D$(8-p)$}}$. 
By using  (\ref{ImpEntr}) for $M=1$ and (\ref{reducedT}),  it follows that, for low $T$, the entropy of the dimer of length $L$ behaves as
\beq
{\mathcal S}^{\cup}_{\mbox{\tiny D$(8-p)$}}\approx \alpha_p\,N\,
\big(\,\sin\bar \theta_{(p,n)}\,\big)^{6-p}\,\, 
\left(g_{\mbox{\tiny YM}}^2 \,N \,L^{3-p} \right)^{\frac{1}{5-p}}\,\, 
\left(LT \right)^{{9-p\over 5-p}}\,\,,
\label{S-low -T}
\eeq
where $\alpha_p$  is a numerical coefficient given by:
\beq
\alpha_p\,=2\pi^2\,\,
\Bigg[{(5-p)\,\sqrt{\pi}\over (7-p)^2}\Bigg]^{{7-p\over 5-p}}\,
{\Big[\,
\Gamma\big({7-p\over 2}\big)\,\Big]^{{6-p\over 5-p}}\,\Gamma\Big(2-{1\over 7-p}\Big)
\over 
\Gamma\big({8-p\over 2}\big)\,\Gamma\Big({5\over 2}-{1\over 7-p}\Big)
}\,\,.
\eeq
Notice that we arranged the right-hand-side of (\ref{S-low -T}) in such a way that all factors are dimensionless. The first two factors in parenthesis contain the dependence on the filling faction and on the Yang-Mills coupling respectively, while the last factor contains the dependence on the temperature.  Notice that the power of $T$ in (\ref{S-low -T}) is positive for all values of $p$ (contrary to what happens with the straight-brane configuration) which means that the entropy of the dimer vanishes at $T=0$. In particular, for $p=3$, the entropy of the dimer configuration of D5-branes in the $AdS_5\times S^5$ background is given by
\beq
{\mathcal S}^{\cup}_{\mbox{\tiny D$5$}}\sim \,N\,\big(\,\sin\bar \theta_{(3,n)}\,\big)^{3}\,\, \sqrt{\lambda}\,(L\,T\,)^3\,\,,
\eeq
where the relation between the angle $\bar \theta_{(3,n)}$ and the filling fraction has been written in  (\ref{theta-3}), $\lambda$ is the 't Hooft coupling of the bulk ${\cal N}=4$ theory  and the numerical coefficient is just $\alpha_3$.

Let us next compute the internal energy for the connected configuration, which is given by:
\beq
{\cal E}^{\cup}_{\mbox{\tiny D$(8-p)$}}\,=\,F^{\cup}_{\mbox{\tiny D$(8-p)$}}\,+\,T
{\mathcal S}^{\cup}_{\mbox{\tiny D$(8-p)$}}\,\,.
\eeq
By combining (\ref{F_cup_h}), (\ref{Fstr-Sstr}) and (\ref{Scup-h-g}), one gets 
\beq
{\cal E}^{\cup}_{\mbox{\tiny D$(8-p)$}}\,=\,\Delta_p\,T\,{\mathcal S}^{\mbox{\tiny str}}_{\mbox{\tiny D$(8-p)$}}(T)\,\Big[\,(p-3)\,h(\gamma)\,-\,2\sqrt{{1-\gamma\over \gamma}}\,g(\gamma)\,\Big] \,\,.
\label{Ecup_hg}
\eeq
Let us rewrite (\ref{Ecup_hg}) by using (\ref{h-g-sigma}) to eliminate $h$.  After some simplifications one has
\beq
{\cal E}^{\cup}_{\mbox{\tiny D$(8-p)$}}(T)\,=\,
T\,{\mathcal S}^{\mbox{\tiny str}}_{\mbox{\tiny D$(8-p)$}}(T)\,\Big[\,{p-3\over 2}\,\,\hat\Delta_p\,\sigma(\gamma)\,+\,(p-5)\,\Delta_p\,\sqrt{{1-\gamma\over \gamma}}\,g(\gamma)\,\Big]\,\,.
\label{Ecup_sigmag}
\eeq
Notice that, due to (\ref{sigma-gamma_1}), the right-hand-side of (\ref{Ecup_sigmag}) reduces to $2\, {\cal E}^{\mbox{\tiny str}}_{\mbox{\tiny D$(8-p)$}}(T)$ as $\gamma\to 1$, as it should. 
In order to characterise the dimerisation transition, let us define the reduced latent heat
$\Delta\epsilon_p$ as
\beq
\Delta\epsilon_p\,\overset{\mbox{\tiny def}}{=}\,
{2{\cal E}^{\mbox{\tiny str}}_{\mbox{\tiny D$(8-p)$}}\,-\,{\cal E}^{\cup}_{\mbox{\tiny D$(8-p)$}}
\over
T\,{\mathcal S}^{\mbox{\tiny str}}_{\mbox{\tiny D$(8-p)$}}}\,\,.
\eeq
Taking into account that $2\,{\cal E}^{\mbox{\tiny str}}_{\mbox{\tiny D$(8-p)$}}=(p-3)\,
T\,{\mathcal S}^{\mbox{\tiny str}}_{\mbox{\tiny D$(8-p)$}}\,$,  it follows that
\beq
\Delta\epsilon_p(\gamma)\,=\,(3-p)\,\Big[\,{\hat\Delta_p\over 2}\,\sigma(\gamma)\,-\,1
\Big]\,+\,(5-p)\,\Delta_p\,\sqrt{{1-\gamma\over \gamma}}\,g(\gamma)\,\,.
\eeq
By construction $\Delta\epsilon_p(\gamma)\to 1$ when $\gamma\to 1$. Moreover, 
$\Delta\epsilon_p(\gamma)\to +\infty$ as $\gamma\to 0$ since
$\Delta\epsilon_p(\gamma)\approx (5-p)\,\Delta_p\,\gamma^{-{1\over 7-p}}$ for small $\gamma$. By plotting 
$\Delta\epsilon_p(\gamma)$ as a function of $\gamma$ for different values of $p$ one can verify that $\Delta\epsilon_p(\gamma)\ge 0$ for $\gamma\in [0,1]$. It follows that the reduced latent heat at the dimerisation transition, given by
\beq
\Delta\epsilon_p^*\,=\,\Delta\epsilon_p(\gamma=\gamma_p^*)
\,\,,
\label{latent-heat-transition}
\eeq
is always positive and therefore the dimerisation transition is a first-order phase transition.  The values of $\Delta\epsilon_p^*$ for different $p$'s have been written in the last column of table \ref{critical}. 

\begin{figure}
 \centering%
 {\scalebox{.75}{\includegraphics{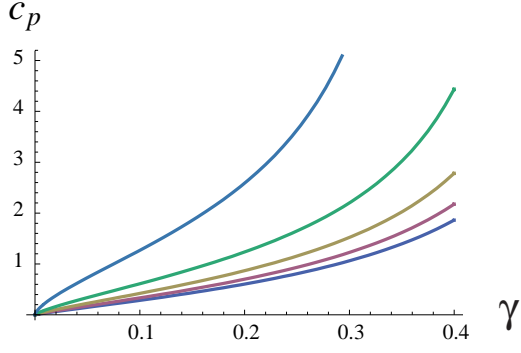}}}
 \caption{Reduced specific heats $\mathtt{c}_p(\gamma)$ for $0\le p\le 4$. Only a portion of the region $0\le \gamma\le \gamma_p^{\mbox{\tiny max}}$, where $\mathtt{c}_p\ge 0$, has been represented in the plot. The curve that grows faster (slower) corresponds to $p=4$ ($p=0$). }
 \label{cp-gamma}
\end{figure}

We now compute the specific heat for the dimer configuration. Proceeding as in the calculation of the entropy, we arrive at
\beq
C^{\cup}_{\mbox{\tiny D$(8-p)$}}(T)\,=\,T\,{\partial\over \partial T}\,\,
{\mathcal S}^{\cup}_{\mbox{\tiny D$(8-p)$}}(T)\,=\,
\hat\Delta_p\,
{\mathcal S}^{\mbox{\tiny str}}_{\mbox{\tiny D$(8-p)$}}(T)\,\Big[\,
{p-3\over 5-p}\,\sigma(\gamma)\,+\,T\,{d\sigma\over d\gamma}\,\,
\Big({dT\over d\gamma}\Big)^{-1}\,\,\Big]\,\,.
\label{Ccup_Tprime}
\eeq
By using identities satisfied by the hypergeometric functions, the derivative of $\sigma(\gamma)$ with respect to $\gamma$ can be simply written as
\beq
\sigma'(\gamma)\,=\,{6-p\over 7-p}\,\,\gamma^{-{1\over 7-p}}\,\,
F\Big(\,{1\over 2}\,,\,2-{1\over 7-p}\,;\,{5\over 2}\,-\,{1\over 7-p}\,;\,
\gamma\,\Big)\,\,.
\label{sigma_prime}
\eeq
Moreover, since $T$ can be written in terms of the function $g$ (see (\ref{T-gamma})), equation (\ref{Ccup_Tprime})  can be recast as
\beq
C^{\cup}_{\mbox{\tiny D$(8-p)$}}(T)\,=\,\hat \Delta_p\,
{\mathcal S}^{\mbox{\tiny str}}_{\mbox{\tiny D$(8-p)$}}(T)\,
\Big[\,{p-3\over 5-p}\,\sigma(\gamma)\,+\,g(\gamma)\,{\sigma'(\gamma)\over g'(\gamma)}
\,\Big]\,\,,
\eeq
where $g'(\gamma)$ and $\sigma'(\gamma)$ are written in (\ref{gprime}) and (\ref{sigma_prime}) respectively. In figure \ref{cp-gamma} we plot the reduced specific heat, defined as
\beq
\mathtt{c}_p\,=\,{C^{\cup}_{\mbox{\tiny D$(8-p)$}}\over 
\hat \Delta_p \,{\mathcal S}^{\mbox{\tiny str}}_{\mbox{\tiny D$(8-p)$}}}\,=\,
{p-3\over 5-p}\,\sigma(\gamma)\,+\,g(\gamma)\,{\sigma'(\gamma)\over g'(\gamma)}
\,\,,
\label{reduced-cp}
\eeq
as a function of $\gamma$ for different values of $p$. Notice that $\mathtt{c}_p\to\infty $ as 
$\gamma\to \gamma_p^{\mbox{\tiny max}}$, since the derivative of $g(\gamma)$ vanishes at this point. Actually, one can check that $\mathtt{c}_p(\gamma)$ is positive in the region $0\le\gamma\le \gamma_{p}^{\mbox{\tiny max}}$  for all values of $p\le 4$. Notice that the denominator $\hat \Delta_p \,{\mathcal S}^{\mbox{\tiny str}}_{\mbox{\tiny D$(8-p)$}}$ in (\ref{reduced-cp}) is always positive and, thus, $C^{\cup}_{\mbox{\tiny D$(8-p)$}}\ge 0$ for $\gamma\in [0,\gamma_p^{\mbox{\tiny max}})$.  The behaviour of $C^{\cup}_{\mbox{\tiny D$(8-p)$}}(T)$ for low $T$ can be obtained directly from (\ref{S-low -T}). It follows that 
$C^{\cup}_{\mbox{\tiny D$(8-p)$}}(T)\sim T^{{9-p\over 5-p}}\to 0$ as $T\to 0$.

\section{Impurity Fluctuations}\label{sec:FluctRev}

In this section, following \cite{Camino:2001at, Faraggi:2011ge} we discuss the fluctuations on the probe branes around the straight flux tube configurations in the 
zero temperature background. The goal is to check the stability of these configurations and to carry out the holographic renormalisation program for the corresponding
action of the fluctuations and to compute the one-point and two-point functions for the fluctuation operators. 

The action that governs the fluctuations of the D$(8-p)$-brane probe can be obtained by expanding its DBI action around the configuration with $z'=0$ and 
$\theta=\bar{\theta}_{\mbox{\tiny $(p,n)$}}$. The induced metric on the world-volume for this configuration at zero temperature is just obtained by taking $z'=\theta'=0$ and 
$\mathtt{h}_p=1$ in (\ref{IndMetr}), namely
\beq
 \tilde{g}^{\mbox{\tiny $(0)$}}_{\alpha\,\beta}\,\,d\zeta^{\alpha}\,d\zeta^{\beta}\,=\,
 {-dt^2\,+\,d\rho^2\over \rho^2}\,+\,u_p^2\,
 \sin^2{\bar{\theta}_{\mbox{\tiny $(p,n)$}}}\,d\Omega_{7-p}^{2}\,\,.
 \label{zero-order-metric}
\eeq
Notice that this metric is of the type $AdS_2\times S^{7-p}$, with 
$u_p^2\,\sin^2{\bar{\theta}_{\mbox{\tiny $(p,n)$}}}$ being the radius of the $S^{7-p}$ sphere. 

Recall that our probe branes have a non-trivial world-volume gauge field 
$\bar{\cal F}_{t\rho}\:=\: \cos\bar\theta_{(p,n)}$ (see (\ref{bar-Frt})). Due to this, the metric that is relevant for the fluctuations is not the induced metric $ \tilde{g}^{\mbox{\tiny $(0)$}}$ but the so-called open string metric, which includes the effects of the world-volume gauge field. In order to define this metric let us consider the matrix 
$\big(\, \tilde{g}^{\mbox{\tiny $(0)$}}\,+\,\bar{\cal F}\,\big)^{-1}$, which is the inverse of the matrix appearing in the DBI Lagrangian for the unperturbed configuration. Let us write this inverse as the sum of a symmetric and an antisymmetric matrix
\beq
\Big(\, \tilde{g}^{\mbox{\tiny $(0)$}}\,+\,\bar{\cal F}\,\Big)^{-1}\,=\,
{\cal G}^{-1}\,+\,{\cal J}\,\,,
\eeq
with ${\cal J}$ being the antisymmetric component. Then, the symmetric matrix 
$ \mathcal{G}_{\alpha\beta}$ is defined as the open string metric. In our case it is straightforward to obtain the expression of $\mathcal{G}_{\alpha\beta}$. One gets
\begin{equation}\eqlabel{OpenStrMetr}
 \mathcal{G}_{\alpha\beta}d\zeta^{\alpha}d\zeta^{\beta}\:=\:\sin^2{\bar{\theta}_{\mbox{\tiny $(p,n)$}}}
 \left[
  {-dt^2\,+\,d\rho^2\over \rho^2}\,+\,u_p^2\,\,d\Omega_{7-p}^{2}
  \right]\,\,. 
\end{equation}

By analysing the possible fluctuations of the D$(8-p)$-brane  probe one can identify
two decoupled channels:
\begin{itemize}
 \item fluctuations $\chi$ of the Cartesian coordinates;
 \item fluctuations $(\xi,\,f)$ of the embedding function $\theta$ and of the world-volume gauge field.
\end{itemize}

These two channels will be studied separately in the next two subsections.

\subsection{Fluctuations of the Cartesian Coordinates}\label{subsec:CartFluct}

Let us assume that the Cartesian coordinates are not fixed but, instead, they fluctuate, and let $\chi$ be the
corresponding fluctuation. In this fluctuating configuration, the induced metric on the probe D$(8-p)$-branes becomes
\begin{equation}\eqlabel{IndMetrChi}
 d\tilde{s}^2_{\mbox{\tiny $9-p$}}\:=\:\tilde{g}^{\mbox{\tiny $(0)$}}_{\alpha\beta}d\zeta^{\alpha}d\zeta^{\beta}+
 \tilde{g}^{\mbox{\tiny $(1)$}}_{\alpha\beta}d\zeta^{\alpha}d\zeta^{\beta},
 \qquad
 \tilde{g}^{\mbox{\tiny $(1)$}}_{\alpha\beta}\:\overset{\mbox{\tiny def}}{=}\:
 \rho^{-2}\left(\partial_{\alpha}\chi\right)\left(\partial_{\beta}\chi\right),
\end{equation}
with $\tilde{g}^{\mbox{\tiny $(0)$}}_{\alpha\beta}$ being the  $AdS_2\times S^{7-p}$ zero-order metric \eqref{zero-order-metric}. By plugging the metric (\ref{IndMetrChi}) into the DBI action and by expanding to second order in the fluctuation, one gets that
 the action for  $\chi$ can be written as
\begin{equation}\eqlabel{ChiAct}
 \begin{split}
  S^{\mbox{\tiny $(\chi)$}}\:=\:-{\mathcal{E}_{\mbox{\tiny $(p,n)$}}B_p\sin^2{\bar{\theta}_{\mbox{\tiny $(p,n)$}}}\over \Omega_{7-p}}\,
 \int d^{9-p}\zeta&\left(N\,e^{\phi}\right)^{\frac{2}{7-p}}\sqrt{-g_{\mbox{\tiny $AdS_2$}}}\sqrt{g_{\mbox{\tiny $S^{7-p}$}}}
  \times\\
  &\times\left[
   \frac{1}{2}\rho^{-2}\mathcal{G}^{\alpha\beta}\left(\partial_{\alpha}\chi\right)\left(\partial_{\beta}\chi\right)+
   \mathcal{O}\left(\chi^4\right)
  \right],
 \end{split}
\end{equation}
where $g_{\mbox{\tiny $AdS_2$}}$ and $g_{\mbox{\tiny $S^{7-p}$}}$ are respectively the determinant of the $AdS_2$ metric
and of the unit sphere $S^{7-p}$, while $\mathcal{G}_{\alpha\beta}$ is the 
open string metric (\ref{OpenStrMetr}).  In (\ref{ChiAct}) we have written the global coefficient in terms of the flux-tube tension ${\mathcal{E}_{\mbox{\tiny $(p,n)$}}}$ and of the constant  $B_p$ defined in (\ref{DualMetric}). 

It is convenient to rewrite the dilaton by  factorising the constant contained in its definition \eqref{FGcoords2} and to introduce  a 
rescaled dilaton $\varphi$ as
\begin{equation}\eqlabel{DilResc}
 N\,e^{\phi}\:
 =
 \:\left(\frac{g_{\mbox{\tiny YM}}^2 N}{2(2\pi)^{p-2}}\right)^{\frac{7-p}{2(5-p)}}
  \left(\frac{d_p}{u_p^{7-p}}\right)^{\frac{p-3}{2(5-p)}}\,e^{\frac{(p-3)(7-p)}{2(5-p)}\varphi}
  \,
  =\,
  N\,e^{\phi_*}\,e^{\frac{(p-3)(7-p)}{2(5-p)}\varphi}\,,
\end{equation}
so that the new dilaton $\varphi$ is just $\varphi\,=\,\log{\rho^{-1}}$ and, in the second step of (\ref{DilResc}), we have defined a new constant $\phi_*$. In terms of $\varphi$ the action acquires the form
\begin{equation}\eqlabel{ChiActDilRes}
 S^{\mbox{\tiny $(\chi)$}}\:=\:-
 \tau_{\mbox{\tiny $(p,n)$}}
\int d^{9-p}\zeta\:
  e^{\frac{p-3}{5-p}\varphi}\,\sqrt{-g_{\mbox{\tiny $AdS_2$}}}\sqrt{g_{\mbox{\tiny $S^{7-p}$}}}
  \left[
   \frac{1}{2}\rho^{-2}\mathcal{G}^{\alpha\beta}(\partial_{\alpha}\chi)(\partial_{\beta}\chi)+
   \mathcal{O}(\chi^4)
  \right]\,\,,
\end{equation}
where the effective tension $\tau_{\mbox{\tiny $(p,n)$}}$ is given by:
\beq
\tau_{\mbox{\tiny $(p,n)$}}\,\overset{\mbox{\tiny def}}{=}\,
{B_p\,\big(\,N\,e^{\phi_*}\,\big)^{{2\over  7-p}}\over \Omega_{7-p}}\,\,
\mathcal{E}_{\mbox{\tiny $(p,n)$}}\,\,\sin^2{\bar{\theta}_{\mbox{\tiny $(p,n)$}}}
\,\,.
\eeq
By using the explicit values of $B_p$, $\phi_*$ and $\mathcal{E}_{\mbox{\tiny $(p,n)$}}$ it is straightforward to verify that $\tau_{\mbox{\tiny $(p,n)$}}$ can be written in terms of field theory quantities as
\beq
\tau_{\mbox{\tiny $(p,n)$}}\,=\,
{\Gamma\big({5-p\over 2}\big)\over 4\,\pi^{{10-p\over 2}}}\,\,
\Bigg[\,{\Gamma\big({5-p\over 2}\big) \over (5-p)\,(2\pi)^{p-2}}\,\,g^2_{YM}\,N\,\Bigg]
^{{1\over 5-p}}\,N\,\sin^{8-p}{\bar{\theta}_{\mbox{\tiny $(p,n)$}}}\,
\overset{\mbox{\tiny def}}{=}\,
\tilde\tau_{\mbox{\tiny $(p,n)$}}\,\,\Big[\,g^2_{YM}\,N\,\Big]^{{1\over 5-p}}
\,\,.
\label{tau(p,n)}
\eeq

It is interesting to notice that the action \eqref{ChiActDilRes} is equivalent to the action of the Cartesian fluctuations
for a brane wrapping an $AdS_{1+\sigma}$-space rather than $AdS_2$
\begin{equation}\eqlabel{ChiAct4}
 S_{\mbox{\tiny  $7+\sigma-p$}}^{\mbox{\tiny $(\chi)$}}\:=\:-T_{\mbox{\tiny D$(7+\sigma-p)$}}\int d^{8+\sigma-p}\zeta\,
  \sqrt{-g_{\mbox{\tiny $AdS_{\sigma+1}$}}}\sqrt{g_{\mbox{\tiny $S^{7-p}$}}}
  \left[
   \frac{1}{2}\rho^{-2}\mathcal{G}^{\alpha\beta}\left(\partial_{\alpha}\chi\right)\left(\partial_{\beta}\chi\right)+
   \mathcal{O}\left(\chi^4\right)
  \right],
\end{equation}
up to a compactification on a $T^{\sigma-1}$ torus in such a way that the metric is
\begin{equation}\eqlabel{AdSq1}
 d\tilde{s}^2_{\mbox{\tiny $AdS_{1+\sigma}$}}\:=\:\frac{-dt^2+d\rho^2}{\rho^2}+
  e^{2\frac{p-3}{(5-p)(\sigma-1)} \varphi}\delta_{\bar{a}\bar{b}}d\zeta^{\bar{a}} d\zeta^{\bar{b}},
\end{equation}
where $\bar{a},\,\bar{b}$ run on the $\sigma-1$ extra directions, and
the the extra coordinates $\zeta^{\bar{a}}$ take values in the interval $[0, 2\pi R_{\mbox{\tiny T}}]$. Notice that the metric (\ref{AdSq1}) corresponds to the one of the $AdS_{1+\sigma}$ space only when the prefactor multiplying the line element of the extra coordinates is $e^{2\varphi}=\rho^{-2}$. This happens when $\sigma$ is analytically continued to take the factional value $\sigma\,=\,2/(5-p)$. Moreover, the compactification radius $R_{\mbox{\tiny T}}$ is related to the tensions $T_{\mbox{\tiny D$(7+\sigma-p)$}}$ and $\tau_{\mbox{\tiny $(p,n)$}}$ by means of the equation
\begin{equation}\eqlabel{ChiTens}
 T_{\mbox{\tiny D$(7+\sigma-p)$}}\left(2\pi R_{\mbox{\tiny T}}\right)^{\sigma-1}\:=\:
\tau_{\mbox{\tiny $(p,n)$}}.
\end{equation}
This action can be considered as the action of a  D$(7+\sigma-p)$-brane in an $AdS_{1+p+\sigma}\,\times\,S^{8-p}$ wrapping an 
$AdS_{1+\sigma}\,\times\,S^{7-p}$ submanifold. 

%From this aspect, the ``conformal'' dimension of the operator
%$\hat{\mathcal{O}}_{\mbox{\tiny $\chi$}}$ dual to the fluctuations $\chi$ can be easily computed following 
%\cite{Arean:2006pk}
%\begin{equation}\eqlabel{ChiDim}
% \Delta_{\mbox{\tiny $\chi$}}\:=\:\frac{1+(\sigma-1)}{2}+\alpha_{\mbox{\tiny $+$}}-\alpha_{\mbox{\tiny $-$}}\:=\:
%  \frac{2l+7-p}{5-p},
%\end{equation}
%where $\alpha_{\pm}$ are related to the asymptotic behaviour of the fluctuating field $\chi$ as the boundary is approached.
In order to easily compute the asymptotic behaviour of the fluctuating field $\chi$, one can consider the factor $\rho^{-2}$
in the square brackets of \eqref{ChiAct4} as a further dilaton factor and then the action \eqref{ChiAct4} can be
conveniently written as
\begin{equation}\eqlabel{ChiAct2}
% \begin{split}
  S_{\mbox{\tiny $7+\sigma-p$}}^{\mbox{\tiny $(\chi)$}}\:=\:-T_{\mbox{\tiny D$(7+\sigma-p)$}}
   \int d^{8-p+\sigma}\zeta\:e^{2 \varphi}\sqrt{-g_{\mbox{\tiny $AdS_{1+\sigma}$}}}\sqrt{g_{\mbox{\tiny $S^{7-p}$}}}
  \left[
   \frac{1}{2}\mathcal{G}^{\alpha\beta}\left(\partial_{\alpha}\chi\right)\left(\partial_{\beta}\chi\right)+
   \mathcal{O}\left(\chi^4\right)
  \right].
% \end{split}
\end{equation}
As before, one can consider the dilaton factor as further extra-dimensions in the world-volume
\begin{equation}\eqlabel{ChiAct5}
  S_{\mbox{\tiny $7+q-p$}}^{\mbox{\tiny $(\chi)$}}\:=\:-T_{\mbox{\tiny D$(7+q-p)$}}\int d^{8+q-p}\zeta\,
  \sqrt{-g_{\mbox{\tiny $AdS_{1+q}$}}}\sqrt{g_{\mbox{\tiny $S^{7-p}$}}}
  \left[
   \frac{1}{2}\mathcal{G}^{\alpha\beta}\left(\partial_{\alpha}\chi\right)\left(\partial_{\beta}\chi\right)+
   \mathcal{O}\left(\chi^4\right)
  \right],
\end{equation}
where $q$ is given by
\beq
q\,=\,2(6-p)/(5-p)\,\,,
\label{q-def}
\eeq
and the compactification on the $T^{q-p}$ torus is determined by the relation between the tensions of the D$(7+q-p)$-brane and $\tau_{\mbox{\tiny $(p,n)$}}$, namely
\begin{equation}\eqlabel{ChiTens2}
 T_{\mbox{\tiny D$(7+q-p)$}}\left(2\pi R_{\mbox{\tiny T}}\right)^{q-1}\:=\:
 \tau_{\mbox{\tiny $(p,n)$}} .
\end{equation}

Let us next consider an ansatz for the fluctuation field $\chi$ such that is factorises as the product of functions of the AdS space and of the $S^{7-p}$ sphere, 
namely:
\beq
\chi\,=\, Y_{l} (S^{7-p})\,\tilde \chi (t,\rho)\,=\,e^{iEt}\,Y_{l} (S^{7-p})\,\hat \chi (E,\rho)\,\,.
\label{chi-separated}
\eeq
Notice that we have assumed that $\chi$ does not depend on the coordinates of the extra dimensions.  Moreover, in (\ref{chi-separated}) the functions $Y_{l} (S^{7-p})$ are the scalar spherical harmonics on the $S^{7-p}$ sphere which, among other quantum numbers, depend on an integer $l$.  The  $Y_{l} (S^{7-p})$  are eigenfunctions 
of the Laplacian operator on $S^{7-p}$, with an eigenvalue which depends on $l$ and $p$:
\begin{equation}\eqlabel{ChiAnsatz}
 \Box_{\mbox{\tiny $S^{7-p}$}}Y_{l}\:=\:-l(l+6-p)Y_{l}.
\end{equation}
By plugging this ansatz for $\chi$,
the action \eqref{ChiAct5} becomes the action of a free massive particle propagating in $AdS_{1+q}$
\begin{equation}\eqlabel{ChiAct3}
 S_{\mbox{\tiny $7+q-p$}}^{\mbox{\tiny $(\chi)$}}\:=\:-T_{\mbox{\tiny D$(7+q-p)$}}\,\,
  \mathcal{N}_{\mbox{\tiny $l$}}\,\,
   \int dt\,d^{q-1}\zeta\,d\rho\:\sqrt{-g_{\mbox{\tiny $AdS_{1+q}$}}}\,
  \left[\frac{1}{2}\,
   \left(\partial\tilde{\chi}\right)^2+{M^2\over 2}\,\tilde{\chi}^2\,
 \right]\,\,,
\end{equation}
where $\mathcal{N}_{\mbox{\tiny $l$}}$ is the integral
\beq
\mathcal{N}_{\mbox{\tiny $l$}}\,=\,
\int d^{7-p}\zeta\:\,\sqrt{g_{\mbox{\tiny $S^{7-p}$}}}\:\,\left(Y_{l}(S^{7-p})\right)^2\,\,,
\eeq
and the mass $M^2$  of the $l^{{\rm th}}$ Kaluza-Klein mode  is given by
\begin{equation}\eqlabel{ChiMass}
 M^2\:=\:\frac{l(l+6-p)}{u_p^2}\:=\:\frac{4l(l+6-p)}{(5-p)^2}.
\end{equation}
The equation of motion of $\tilde\chi$ can be simply written as the equation of motion of a free massive particle in $AdS_{q+1}$
\begin{equation}\eqlabel{EoMq1}
 \left(-\Box_{q+1}+M^2\right)\tilde{\chi}\:=\:0,
\end{equation}
with $\Box_{q+1}$ being the d'Alembertian in $AdS_{q+1}$. In Euclidean signature, the regular solution of (\ref{EoMq1}) can be 
written in terms of the modified Bessel function of the second type
\begin{equation}\eqlabel{EoMq2}
 \tilde{\chi}(t,\,\rho)\:=\:e^{iEt}\,\hat{\chi}(E,\rho)\,=\,e^{iEt}\,
 (E\rho)^{q/2}K_{\tilde{\alpha}}(E\rho),\qquad
 \tilde{\alpha}\:\overset{\mbox{\tiny def}}{=}\:\sqrt{M^2+\frac{q^2}{4}}\:\equiv\:\frac{2l+6-p}{5-p},
\end{equation}
and the asymptotic behaviour  of $\tilde{\chi}$ as the boundary is approached turns out to be 
\begin{equation}\eqlabel{NormNonNorm}
\tilde{\chi}\:\sim\:\rho^{2\alpha_{-}}\left(\tilde{\chi}_{-}^{\mbox{\tiny $(0)$}}+\ldots\right)+
 \rho^{2\alpha_{+}}\left(\tilde{\chi}_{+}^{\mbox{\tiny $(0)$}}+\ldots\right),
\end{equation}
with the exponents $\alpha_{\mp}$ obtained by solving 
\begin{equation}\eqlabel{amp}
 M^2\:=\:2\alpha\left(2\alpha-q\right),\quad\Longrightarrow\quad
 \left\{
  \begin{array}{l}
   \alpha_{-}\:=\:-\frac{l}{5-p},\\
   \phantom{\ldots}\\
   \alpha_{+}\:=\:\frac{l+6-p}{5-p}\,\,.
  \end{array}
 \right. 
\end{equation}
Notice that the parameter $\tilde{\alpha}$ in \eqref{EoMq2} is nothing but the difference between $\alpha_{+}$ and 
$\alpha_{-}$ (\ie\ $ \tilde{\alpha}=\alpha_{+}-\alpha_{-}$).

In the framework of holographic renormalisation, the non-normalisable fluctuation 
$\hat{\chi}_{-}^{\mbox{\tiny $(0)$}}$ acts as a source  of the 
operator $\hat{\mathcal{O}}_{\mbox{\tiny $\chi$}}$ dual to the fluctuation $\chi$, whereas  $\hat{\chi}_{+}^{\mbox{\tiny $(0)$}}$ as a vev.  In this respect some comments are now in order. Sourcing the mode $\chi$ in the bulk theory is equivalent to sourcing an irrelevant operator 
$\hat{\mathcal{O}}_{\mbox{\tiny $\chi$}}$ at the boundary. We are working perturbatively in the mode $\chi$, $\chi$
being a fluctuation. This is equivalent to consider parametrically small sources of the irrelevant operator 
$\hat{\mathcal{O}}_{\mbox{\tiny $\chi$}}$. As first noticed in \cite{vanRees:2011fr} (see \cite{vanRees:2011ir} for a 
further reference on holographic renormalisation of irrelevant operators), one can consistently holographic
renormalise the theory order by order in $\chi$, making the $n$-point correlator of the dual irrelevant operator
normalised up to some fixed $n$\footnote{We thank Balt van Rees for discussion about this point.}. We are interested in 
computing the two point function, and therefore it will be enough to consider and renormalise the part of the
action which is quadratic in the fluctuations. 

\subsubsection{Holographic Renormalisation}\label{subsubsec:HR}

As a first step, let us consider the asymptotic expansion of $\hat{\chi}$ in a neighbourhood of the boundary
$\rho\,=\,0$. This expression can be written as
\begin{equation}\eqlabel{BoundChi}
 \begin{split}
 \tilde{\chi}\:\overset{\mbox{\tiny $\rho\,\rightarrow\,0$}}{\sim}\:
  \rho^{2\alpha_{-}}
  &\left[
   \sum_{k=0}^{\infty} \tilde{\chi}_{-}^{\mbox{\tiny $(2k)$}}\rho^{2k}+
   \rho^{2\tilde{\alpha}}
    \left[
     \sum_{k=0}^{\infty} \tilde{\chi}_{+}^{\mbox{\tiny $(2k)$}}\rho^{2k}+
     \delta_{\beta,\tilde{\alpha}-1}\sum_{k=0}^{\infty}\vartheta_{+}^{\mbox{\tiny $(2k)$}}\rho^{2k}\log{\rho}\,+
    \right.
   \right.\\
   &\left.
    \left.\hspace{.5cm}+
     \delta_{\beta,\tilde{\alpha}-1}\sum_{k=0}^{\infty}\sum_{l=2}^{s}\sigma_{+}^{\mbox{\tiny $(2k,l)$}}\rho^{2k}\log^l(\rho)
   \right]
  \right],
 \quad
 \beta\:=\:
 \left\{
  \begin{array}{l}
   \tilde{\alpha}-1,\,\; \tilde{\alpha}\,\in\,\mathbb{Z}_{+}\\
   \mbox{$[\tilde{\alpha}]$}\,\; \mbox{otherwise}
  \end{array}
 \right.
 \end{split}
\end{equation}
where $[a]$ denotes the integer part of $a$ and
the logarithmic terms can be present if and only if $\tilde{\alpha}$ is an integer (implying that $p$ must be even), which 
can never occur at leading order as it can be easily seen from the explicit expression for $\tilde{\alpha}$ in \eqref{EoMq2}.
More precisely, the logarithmic terms can start to appear at order $\mathcal{O}(\rho^{2\alpha_{-}+n})$, with $n$ being
a (non-zero) multiple of $4$. Therefore, the modes characterised by logarithmic terms in the asymptotic expansion close to
the boundary have quantum number $l$ given by
\begin{equation}\eqlabel{Chil}
 l\:=\:(5-2\tilde{p})\tilde{n}-(3-\tilde{p}),\qquad \tilde{n}\:\ge\:\frac{3-\tilde{p}}{5-2\tilde{p}}\,\in\,\mathbb{Z}_{+},
  \quad
 \tilde{p}\:=\:0,\,1,\,2,
\end{equation}
where $\tilde{p}$ and $\tilde{n}$ are respectively related to $p$ and $n$ by $n\,=\,4\tilde{n}$ and $p\,=\,2\tilde{p}$.
Furthermore, in absence of the logarithmic terms, the coefficients in such an expansion can be obtained recursively to
be
\begin{equation}\eqlabel{BoundChiCoeff}
 \tilde{\chi}_{\mp}^{{\mbox{\tiny $(2k)$}}}\:=\:
  \frac{\Box_{\mbox{\tiny $\mathtt{g}$}} \tilde{\chi}_{\mp}^{\mbox{\tiny $(2k-2)$}}}{4k(\tilde{\alpha}-k)}\:=\:
  \frac{\Gamma(\tilde{\alpha}-k)}{4^k \Gamma(k+1)\Gamma(\tilde{\alpha})}
 \Box_{\mbox{\tiny $\mathtt{g}$}}^k \tilde{\chi}_{\mp}^{\mbox{\tiny $(0)$}},\qquad 
 k\,\in\,[1,\,+\infty[,
\end{equation}
with $\Box_{\mbox{\tiny $\mathtt{g}$}}$ being the d'Alembertian with respect to the boundary metric $\mathtt{g}$. In presence of the
logarithmic terms, the recursive relation \eqref{BoundChiCoeff} still holds for the coefficients 
$\tilde{\chi}_{-}^{\mbox{\tiny $(2k)$}}$, $k\,<\,\tilde{\alpha}$, and the coefficient $\vartheta_{+}^{\mbox{\tiny $(0)$}}$
of the first logarithmic term is
\begin{equation}\eqlabel{BoundChiLog}
 \vartheta_{+}^{\mbox{\tiny $(0)$}}\:=\:
  \frac{\Box_{\mbox{\tiny $\mathtt{g}$}} \tilde{\chi}_{-}^{\mbox{\tiny $(2\tilde{a}-2)$}}}{2\tilde{\alpha}}\:=\:
  \frac{2\Box_{\mbox{\tiny $\mathtt{g}$}}^{\tilde{\alpha}} \tilde{\chi}_{\mbox{\tiny $-$}}^{\mbox{\tiny $(0)$}}}{
   4^{\tilde{\alpha}}\Gamma(\tilde{\alpha}+1)\Gamma(\tilde{\alpha})}\,\,.
\end{equation}

The on-shell action for the fluctuation is
\begin{equation}\eqlabel{ChiActOnShell}
 S^{\mbox{\tiny $(\chi)$}}\:\sim\:\frac{1}{2}\int_{\mbox{\tiny $\mathcal{M}_{\varepsilon}$}}dt d^{q-1}\zeta\:
  \left.\sqrt{\mathtt{g}}\, \tilde{\chi}\,\rho\,\partial_{\rho}\, \tilde{\chi}\right|_{\mbox{\tiny $\varepsilon$}},
\end{equation}
with $\varepsilon$ being a cut-off for small $\rho$ and $\mathcal{M}_{\varepsilon}$ is the boundary of $AdS_{1+q}$ at $\rho=\varepsilon$.  Using the asymptotic expansion \eqref{BoundChi}, the divergent part
of the action turns out to be
\begin{equation}\eqlabel{ChiActDiv}
 \begin{split}
 \left. S^{\mbox{\tiny $(\chi)$}} \right|_{\mbox{\tiny div}}\:\sim\:
  \frac{1}{2}\int_{\mbox{\tiny $\mathcal{M}_{\varepsilon}$}}dt d^{q-1}\zeta\:\varepsilon^{-2\tilde{\alpha}}
  &\left[
   2\sum_{k=0}^{\beta}\varepsilon^{2k}\sum_{m=0}^{k}(\alpha_{-}+k-m)
    \tilde{\chi}_{-}^{\mbox{\tiny $(2m)$}} \tilde{\chi}_{-}^{\mbox{\tiny $(2k-2m)$}}+
  \right.\\
  &\left.\hspace{.5cm}+
   2(\alpha_{+}+\alpha_{-}) \tilde{\chi}_{-}^{\mbox{\tiny $(0)$}}\vartheta_{+}^{\mbox{\tiny $(0)$}}
   \varepsilon^{2\tilde{\alpha}}\log{\varepsilon}
  \right],
 \end{split}
\end{equation}
and by re-expressing the action in terms of the original field $\tilde{\chi}$, one gets the following counterterm action
\begin{equation}\eqlabel{ChiActCount}
 S_{\mbox{\tiny ct}}^{\mbox{\tiny $(\chi)$}}\:\sim\:\int_{\mbox{\tiny $\mathcal{M}_{\mbox{\tiny $\varepsilon$}}$}}
  dt\, d^{q-1}\zeta\:\sqrt{\left.\mathtt{g}\right|_{\mbox{\tiny $\varepsilon$}}}
  \left[
   \sum_{k=0}^{\beta}a^{\mbox{\tiny $(2k)$}} \tilde{\chi}\,\Box_{\mbox{\tiny $\mathtt{g}$}}^k\, \tilde{\chi}+
   a^{\mbox{\tiny $(2\tilde{\alpha})$}} \tilde{\chi}\,\Box_{\mbox{\tiny $\mathtt{g}$}}^{\mbox{\tiny $2\tilde{\alpha}$}}\, \tilde{\chi}
   \log{\varepsilon}
  \right],
\end{equation}
with $a^{\mbox{\tiny $(0)$}}\,=\,2\alpha_{-}$. Thus, the renormalised action becomes
\begin{equation}\eqlabel{ChiActRen}
 \begin{split}
  \left. S^{\mbox{\tiny $(\chi)$}}\right|_{\mbox{\tiny ren}}\:&=\:\lim_{\varepsilon\rightarrow0}
  \left[
   \left.S^{\mbox{\tiny $(\chi)$}}\right|_{\varepsilon}+S^{\mbox{\tiny $(\chi)$}}_{\mbox{\tiny ct}}
  \right]\:=\\
  &=\:\lim_{\varepsilon\rightarrow0}
  \left[
  \frac{T_{\mbox{\tiny D$(7-q+p)$}}}{2}\,\mathcal{N}_{\mbox{\tiny $l$}}
  \int_{\mbox{\tiny $\mathcal{M}_{\mbox{\tiny $\varepsilon$}}$}}dt\, d^{q-1}\zeta
  \sqrt{\left.\mathtt{g}\right|_{\mbox{\tiny $\varepsilon$}}}\,\,
  \Big(\,\tilde{\chi}\,\varepsilon\,\partial_{\varepsilon}\, \tilde{\chi}\,\Big)_{
 \big |_{\mbox{\tiny $\varepsilon$}}}\,
  +\right.\\
  &\left.\phantom{=\:}+
   \frac{T_{\mbox{\tiny D$(7-q+p)$}}}{2}\,\,\mathcal{N}_{\mbox{\tiny $l$}}
  \int_{\mbox{\tiny $\mathcal{M}_{\mbox{\tiny $\varepsilon$}}$}}dt\, d^{q-1}\zeta
  \sqrt{\left.\mathtt{g}\right|_{\mbox{\tiny $\varepsilon$}}}
   \left[\,
    \sum_{k=0}^{\beta}a^{\mbox{\tiny $(2k)$}} \tilde{\chi}\,\Box_{\mbox{\tiny $\mathtt{g}$}}^k \,\tilde{\chi}+
    a^{\mbox{\tiny $(2\tilde{\alpha})$}} \tilde{\chi}\,\Box_{\mbox{\tiny $\mathtt{g}$}}^{\mbox{\tiny $2\tilde{\alpha}$}}\, \tilde{\chi}
    \log{\varepsilon}
   \right]_{
 \big |_{\mbox{\tiny $\varepsilon$}}}\,
 \right],
 \end{split}
\end{equation}
where $\mathcal{N}_{\mbox{\tiny $l$}}$ has been defined as the integral of the spherical harmonics in \eqref{ChiAct3}

\subsubsection{One-Point and Two-Point Correlation Function}\label{subsubsec:1pt}

Let us now use the renormalised action to compute the one-point correlation function, which is defined as
\begin{equation}\eqlabel{1ptCF}
 \langle\hat{\mathcal{O}}_{\mbox{\tiny $\chi$}}\rangle\:=\:\lim_{\varepsilon\rightarrow0}
  \frac{1}{\varepsilon^{2\alpha_{+}}}\frac{1}{\sqrt{\left.\mathtt{g}\right|_{\mbox{\tiny $\varepsilon$}}}}
  \frac{\left.\delta S^{\mbox{\tiny $(\chi)$}}\right|_{\mbox{\tiny ren}}}{\delta\tilde{\chi}(t,\,\varepsilon)}.
\end{equation}
The first variation of the renormalised action turns out to be
\begin{equation}\eqlabel{1varAct}
 \delta \left.S^{\mbox{\tiny $(\chi)$}}\right|_{\mbox{\tiny ren}}\:=\:
 T_{\mbox{\tiny D$(7-q+p)$}}\,\,\mathcal{N}_{\mbox{\tiny $l$}}
 \int_{\mbox{\tiny $\mathcal{M}_{\mbox{\tiny $\varepsilon$}}$}}dt\, d^{q-1}\zeta\:
 \varepsilon^{-2\alpha_{-}}
  \left[2(\alpha_{-}-\alpha_{+})\tilde{\chi}_{+}^{\mbox{\tiny $(0)$}}+\vartheta_{+}^{\mbox{\tiny $(0)$}}\right]
 \delta\tilde{\chi}(t,\varepsilon),
\end{equation}
and therefore the one-point function is given by
\begin{equation}\eqlabel{1ptCF2}
 \begin{split}
 \langle\hat{\mathcal{O}}_{\mbox{\tiny $\chi$}}\rangle\:&=\:
\tau_{\mbox{\tiny $(p,n)$}}\,\mathcal{N}_{\mbox{\tiny $l$}}
  \left[
   2(\alpha_{-}-\alpha_{+})\,{\tilde \chi_{+}}^{\mbox{\tiny $(0)$}}+
   \delta_{\beta,\tilde{\alpha}-1}\vartheta_{+}^{\mbox{\tiny $(0)$}}
  \right]\:=\\
 &=\: 
 \tau_{\mbox{\tiny $(p,n)$}}\,\mathcal{N}_{\mbox{\tiny $l$}}
  \left[
   -2\frac{2l+6-p}{5-p}\tilde{\chi}_{+}^{\mbox{\tiny $(0)$}}+
   \delta_{\beta,\tilde{\alpha}-1}\,
   \frac{2\,\Box_{\mbox{\tiny $\mathtt{g}$}}^{\tilde{\alpha}}\,\tilde{\chi}_{\mbox{\tiny $-$}}^{\mbox{\tiny $(0)$}}}{
    4^{\tilde{\alpha}}\Gamma(\tilde{\alpha}+1)\Gamma(\tilde{\alpha})}
  \right]\,\,.
 \end{split}
\end{equation}
Typically, the second term in \eqref{1ptCF2} can be removed completely by adding a finite counter-term, corresponding
to the matter conformal anomaly in $AdS_{q+1}$ \cite{Petkou:1999fv}.

The coefficient $\tilde{\chi}_{+}^{\mbox{\tiny $(0)$}}$ can be read off from the full-solution \eqref{EoMq1}.  In order to perform this calculation, let us extract the time dependence of $\tilde\chi (t,\rho)$ and work with $\hat\chi (E,\rho)$ (see eq. (\ref{chi-separated})).
Expanding
the latter in a neighbourhood of the boundary we have to distinguish two cases, namely
\begin{equation}\eqlabel{FullSolExp}
 \begin{split}
  &\tilde{\alpha}\mbox{ non-integer:}\\
  &\qquad\hat{\chi}(E,\rho)\:=\:\rho^{2\alpha_{-}}\hat{\chi}_{-}^{\mbox{\tiny $(0)$}}(E)
    \left[
     1+\frac{(E\rho)^2}{4(\tilde{\alpha}+1)}+\ldots-\frac{\left(E\rho\right)^{2\tilde{\alpha}}}{2^{2\tilde{\alpha}}}
     \frac{\Gamma(1-\tilde{\alpha})}{\Gamma(1+\tilde{\alpha})}+\ldots
    \right],\\
  &\tilde{\alpha}\mbox{ integer:}\\
  &\qquad\hat{\chi}(E,\rho)\:=\:\rho^{2\alpha_{-}}\hat{\chi}_{-}^{\mbox{\tiny $(0)$}}(E)
    \left[
     1+\frac{\left(E\rho\right)^2}{4(1-\tilde{\alpha})}+\ldots+
     \frac{(-1)^{\tilde{\alpha}}\left(E\rho\right)^{2\tilde{\alpha}}}{2^{2\tilde{\alpha}}\Gamma(\tilde{\alpha})
      \Gamma(\tilde{\alpha}-1)}\times
    \right.\\
  &\hspace{2.5cm}
     \left.\times
      \left[
       H_{\tilde{\alpha}}-2\gamma_{\mbox{\tiny EM}}-\frac{2}{\tilde{\alpha}(\tilde{\alpha}-1)}\log{\frac{E}{2}}
      \right]+
     \frac{(-1)^{\tilde{\alpha}+1}2\left(E\rho\right)^{2\tilde{\alpha}}}{2^{2\tilde{\alpha}}\Gamma(\tilde{\alpha}+1)
      \Gamma(\tilde{\alpha})}\log{\rho}+\ldots
    \right],
 \end{split}
\end{equation}
where $\gamma_{\mbox{\tiny EM}}=.577$ is the Euler-Mascheroni constant and, for a given integer $n$, $H_n$ is the harmonic number ($H_n=\sum_{k=1}^{n}\,{1\over k}$). By using these results the coefficient $\hat{\chi}_{+}^{\mbox{\tiny $(0)$}}$ turns out to be
\begin{equation}\eqlabel{Xp0}
 \hat{\chi}_{+}^{\mbox{\tiny $(0)$}}\:=\:
 \left\{
  \begin{array}{l}
  - \frac{E^{2\tilde{\alpha}}}{2^{2\tilde{\alpha}}}\frac{\Gamma(1-\tilde{\alpha})}{\Gamma(1+\tilde{\alpha})}
    \hat{\chi}_{-}^{\mbox{\tiny $(0)$}}(E),\hspace{5.2cm}
    \tilde{\alpha}\mbox{ non-integer}\\
   \phantom{\ldots}\\
   \frac{(-1)^{\tilde{\alpha}}E^{2\tilde{\alpha}}}{2^{2\tilde{\alpha}}\Gamma(\tilde{\alpha})\Gamma(\tilde{\alpha}-1)}
    \left[
     H_{\tilde{\alpha}}-2\gamma-\frac{2}{\tilde{\alpha}(\tilde{\alpha}-1)}\log{\frac{E}{2}}
    \right]
    \hat{\chi}_{-}^{\mbox{\tiny $(0)$}}(E),\quad
    \tilde{\alpha}\mbox{ integer}
  \end{array}
 \right.
\end{equation}
The correlator \eqref{1ptCF2} therefore acquires the following form
\begin{equation}\eqlabel{1ptCF3}
 \langle\hat{\mathcal{O}}_{\mbox{\tiny $\chi$}}(E)\rangle\:=\:-\mathcal{N}_{\mbox{\tiny $l$}}^{\mbox{\tiny $(p,n)$}}
 2\frac{2l+6-p}{5-p}\left[g^2_{\mbox{\tiny eff}}(E)\right]^{\frac{1}{5-p}}E^{\frac{4l+3(5-p)}{5-p}}\Theta(E),
\end{equation}
where
$\mathcal{N}_{\mbox{\tiny $l$}}^{\mbox{\tiny $(p,n)$}}=\mathcal{N}_{\mbox{\tiny $l$}}\,
\tilde\tau_{\mbox{\tiny $(p,n)$}}$ ($\tilde\tau_{\mbox{\tiny $(p,n)$}}$ has been defined in (\ref{tau(p,n)})), $g_{\mbox{\tiny eff}}^2(E)\,\overset{\mbox{\tiny def}}{=}\,g_{\mbox{\tiny YM}}^2 N E^{p-3}$ and 
$\Theta(E)\,=\,E^{-2\tilde{\alpha}}\hat{\chi}_{+}^{\mbox{\tiny $(0)$}}$. Notice that 
$\Theta(E)$ can be written in terms of the source $ \hat{\chi}_{-}^{\mbox{\tiny $(0)$}}(E)$ by using  (\ref{Xp0}). This relation depends on whether $\tilde\alpha$ is integer or not. As 
$\tilde\alpha\,=\,1+\,{2l+1\over 5-p}$, one has that $\tilde\alpha$ is integer iff $2l+1=0\,\,\,
({\rm mod} (5-p))$.  This never occurs if $p$ is odd and it happens in all cases if $p=4$. In the remaining cases $p=0,2$ it only occurs for some particular values of the Kaluza-Klein mode $l$. Furthermore, for $\tilde{\alpha}$ integer, the only relevant term is the one containing $\log{E}$, while the others are scheme dependent and
we will omit them.

Finally, the two-point correlator can be obtained from \eqref{1ptCF3} by differentiating with respect to the source
$\hat{\chi}_{-}^{\mbox{\tiny $(0)$}}$. It is easy to see that for $\tilde{\alpha}$ non integer, it is just given by
contact terms
\begin{equation}\eqlabel{2ptCFa}
 \langle\hat{\mathcal{O}}_{\mbox{\tiny $\chi$}}\hat{\mathcal{O}}_{\mbox{\tiny $\chi$}}\rangle\:=\:-
 \mathcal{N}_{\mbox{\tiny $l$}}^{\mbox{\tiny $(p,n)$}} 
 \frac{2\tilde{\alpha}}{4^{\tilde{\alpha}}}\left[g^2_{\mbox{\tiny eff}}(E)\right]^{\frac{1}{5-p}}E^{\frac{4l+3(5-p)}{5-p}}
 \frac{\Gamma(1-\tilde{\alpha})}{\Gamma(1+\tilde{\alpha})}.
\end{equation}
while for $\tilde{\alpha}$ integer it has a logarithmic dependence
\begin{equation}\eqlabel{2ptCFb}
 \langle\hat{\mathcal{O}}_{\mbox{\tiny $\chi$}}\hat{\mathcal{O}}_{\mbox{\tiny $\chi$}}\rangle\:=\:
 \frac{2(-1)^{\mbox{\tiny $\tilde{\alpha}$}}\mathcal{N}_{\mbox{\tiny $l$}}^{\mbox{\tiny $(p,n)$}}}{
 4^{\tilde{\alpha}}\Gamma(\tilde{\alpha})\Gamma(\tilde{\alpha}-1)}
 \left[g^2_{\mbox{\tiny eff}}(E)\right]^{\frac{1}{5-p}}E^{\frac{4l+3(5-p)}{5-p}}
 \log{\frac{E^{2\tilde{\alpha}}}{\mu^{2\alpha}}}.
\end{equation}
The two-point correlator in position space can be obtained by Fourier transforming \eqref{2ptCFa} and \eqref{2ptCFb}. With this purpose, we will use
 the relations
\begin{equation}\eqlabel{FourTrans}
 \begin{split}
  &\mathtt{I}_{\mbox{\tiny a}}(t)\:\overset{\mbox{\tiny def}}{=}\:\int dE\,e^{-itE}E^{2\tilde{\alpha}}\:=\:2^{2\tilde{\alpha}+1}\,\sqrt{\pi}\,
   \frac{\Gamma\left(1/2-\tilde{\alpha}\right)}{\Gamma\left(-\tilde{\alpha}\right)}\frac{1}{|t|^{1+2\tilde{\alpha}}},
   \hspace{2.3cm} \tilde{\alpha}\mbox{ non-integer},\\
  &\mathtt{I}_{\mbox{\tiny b}}(t)\:\overset{\mbox{\tiny def}}{=}\:\int dE\,e^{-itE}E^{2\tilde{\alpha}}\log{E}\:=\:
   (-1)^{1+\tilde{\alpha}}\,{\left(2\tilde{\alpha}\right)!\over 2}\, \mathcal{R}\left( {1\over |t|^{1+2\tilde \alpha}}\right)\,\,,
  \quad\qquad \tilde{\alpha}\mbox{ integer},
 \end{split}
\end{equation}
where $\mathcal{R}\left(\cdot\right)$ indicates the renormalised version
of its argument and is defined as
\beq
\mathcal{R}\left( {1\over |t|^{1+2\tilde \alpha}}\right)\:\overset{\mbox{\tiny def}}{=}\:
{1\over |t|^{1+2\tilde \alpha}}\,-\,{2\gamma_{\mbox{\tiny EM}}\over 
\left(2\tilde{\alpha}\right)!}
\,\delta^{\mbox{\tiny $(2\tilde{\alpha})$}}(t)\,\,,
\eeq
with $\delta^{(\mbox{\tiny $m$})}(t)$ denoting the $m^{{\rm th}}$ derivative of the Dirac $\delta$-function. The two-point correlator in position space becomes
\begin{equation}\eqlabel{2ptCFc}
 \begin{split}
  &\langle\hat{\mathcal{O}}_{\mbox{\tiny $\chi$}}\hat{\mathcal{O}}_{\mbox{\tiny $\chi$}}\rangle\:=\:-
   \mathcal{N}_{\mbox{\tiny $l$}}^{\mbox{\tiny $(p,n)$}}4\tilde{\alpha}\sqrt{\pi}
    \frac{\Gamma\left(1/2-\tilde{\alpha}\right)\Gamma\left(1-\tilde{\alpha}\right)}{\Gamma\left(-\tilde{\alpha}\right)\Gamma\left(1+\tilde{\alpha}\right)}
    \frac{\left[g_{\mbox{\tiny eff}}^2(t)\right]^{\frac{1}{5-p}}}{|t|^{2\Delta_{\mbox{\tiny $\chi$}}}},\qquad
    \tilde{\alpha} \mbox{ non-integer},\\
  &\langle\hat{\mathcal{O}}_{\mbox{\tiny $\chi$}}\hat{\mathcal{O}}_{\mbox{\tiny $\chi$}}\rangle\:=\:-
    \frac{\left(2\tilde{\alpha}\right)!\,\mathcal{N}_{\mbox{\tiny $l$}}^{\mbox{\tiny $(p,n)$}}}{
    4^{\tilde{\alpha}}\Gamma(\tilde{\alpha})\Gamma(\tilde{\alpha}-1)}
     \mathcal{R}\left(\frac{\left[g_{\mbox{\tiny eff}}^2(t)\right]^{\frac{1}{5-p}}}{|t|^{2\Delta_{\mbox{\tiny $\chi$}}}}\right),
     \hspace{2.2cm}\tilde{\alpha}\mbox{ integer},
 \end{split}
\end{equation}
where the effective coupling constant is defined as 
$g_{\mbox{\tiny eff}}^2(t)\,\overset{\mbox{\tiny def}}{=}\,g_{\mbox{\tiny YM}}^2 N |t|^{3-p}$ and $\mathcal{N}_{l}^{\mbox{\tiny $(p,\,n)$}}\,\sim\,N$. Notice that
the exponent of $|t|$ in \eqref{2ptCFc} provides twice the generalised conformal dimension $\Delta_{\chi}$ 
of the operator $\hat{\mathcal{O}}_{\mbox{\tiny $\chi$}}$, which is given by
\beq
\Delta_{\chi}\,=\,{2\over 5-p}\,l\,+\,2\,\,.
\eeq
For $p\,=\,3$ this result coincides with the correct value $\Delta_{\chi}\,=\,\,l\,+\,2$ for the truly conformal case \cite{Camino:2001at,Faraggi:2011ge,Harrison:2011fs}. Notice that
$\Delta_{\chi}$ is fractional for $p<3$.

\subsection{Fluctuations of the Angular Embedding Function and World-volume Gauge Field}\label{subsec:EmbGfFluct}

Let us move on to the angular embedding, whose fluctuations are coupled to the ones of the world-volume gauge field.
We analyse the fluctuations around the configuration $\theta\:=\:\bar{\theta}_{\mbox{\tiny $(p,n)$}}$, namely
\begin{equation}\eqlabel{FluctAng}
 \theta\:=\:\bar{\theta}_{\mbox{\tiny $(p,n)$}}+\xi,\qquad\qquad
 \mathcal{F}_{\alpha\beta}
\:=\:
  \mathcal{F}^{\mbox{\tiny $(0)$}}_{\alpha\beta}\,+\,\mathcal{F}^{\mbox{\tiny $(1)$}}_{\alpha\beta}\,\,,
\end{equation}
where the only non-zero component of $\mathcal{F}^{\mbox{\tiny $(0)$}}_{\alpha\beta}$ is 
$\mathcal{F}^{\mbox{\tiny $(0)$}}_{t\rho}\,=\,\cos{\bar{\theta}_{\mbox{\tiny $(p,n)$}}}$ and, in terms of  the rescaled
dilaton $\varphi$ defined in \eqref{DilResc}, the fluctuation field can be rewritten as 
\beq
\mathcal{F}^{\mbox{\tiny $(1)$}}_{\alpha\beta}\,=\,e^{\frac{3-p}{5-p} \varphi}
\,f_{\alpha\beta}\,\,.
\eeq
When these fluctuations are switched  on the induced metric on the probe D$(8-p)$-branes acquires a further term given by
\begin{equation}\eqlabel{AngIndMetr}
 \tilde{g}_{\alpha\beta}^{\mbox{\tiny $(1)$}}\:=\:u_p^2
 \left[
  (\partial_{\alpha}\xi)(\partial_{\beta}\xi)+
  \left(
   2\sin{\bar{\theta}_{\mbox{\tiny $(p,n)$}}}\cos{\bar{\theta}_{\mbox{\tiny $(p,n)$}}}\xi+
   (\cos^{2}{\bar{\theta}_{\mbox{\tiny $(p,n)$}}}-\sin^{2}{\bar{\theta}_{\mbox{\tiny $(p,n)$}}})\, \right)\,\xi^2
   \delta_{\alpha}^{\phantom{\alpha}a}\delta_{\beta}^{\phantom{\beta}b}\,g_{ab}
 \,\,
 \right]\,\,,
\end{equation}
where $a,b$  are indices along the $S^{7-p}$ sphere and $g_{ab}$ denotes its round metric.  The D$(8-p)$-brane action for these fluctuations acquires the following form
\begin{equation}\eqlabel{AngAc}
 \begin{split}
  S_{\mbox{\tiny D$(8-p)$}}\:=\:-\tau_{\mbox{\tiny $(p,n)$}}\,\,  &\int d^{9-p}\zeta\:e^{\frac{p-3}{5-p}\varphi}\sqrt{-g_{\mbox{\tiny $AdS_2$}}}\sqrt{g_{\mbox{\tiny $S^{7-p}$}}}
   \times\\
  &\times
   \left[
    \sqrt{\mbox{det}\left\{\mathbb{I}+X\right\}}
     +\frac{(\mathcal{F}_{t\rho}^{\mbox{\tiny $(0)$}}+\mathcal{F}_{t\rho}^{\mbox{\tiny $(1)$}})\,\,
      C_{p}(\bar{\theta}_{\mbox{\tiny $(p,n)$}}+\xi)}{
     \sin^{8-p}{\bar{\theta}_{\mbox{\tiny $(p,n)$}}}}
   \right],
 \end{split}
\end{equation}
where $C_p(\theta)$ is the function defined in (\ref{C-theta}) and $X$ is a matrix defined as
\begin{equation}\eqlabel{Xdef}
 X\:\overset{\mbox{\tiny def}}{=}\:\left(\tilde{g}^{\mbox{\tiny $(0)$}}+\mathcal{F}^{\mbox{\tiny $(0)$}}\right)^{-1}
 \left(\tilde{g}^{\mbox{\tiny $(1)$}}+\mathcal{F}^{\mbox{\tiny $(1)$}}\right).
\end{equation}
Expanding the action \eqref{AngAc} and keeping the terms up to the second order in the fluctuations, one gets
\begin{equation}\eqlabel{AngfAct}
 \begin{split}
  S^{\mbox{\tiny $(\xi,f)$}}\:&=\:-\tau_{\mbox{\tiny $(p,n)$}}\,\, 
   \int d^{9-p}\zeta\:e^{\frac{p-3}{5-p} \varphi}\sqrt{-g_{\mbox{\tiny $AdS_2$}}}\sqrt{g_{\mbox{\tiny $S^{7-p}$}}}
   \times\\
  &\times
   \left[
    \frac{u_p^2}{2}\,\mathcal{G}^{\alpha\beta}(\partial_{\alpha}\xi)(\partial_{\beta}\xi)-
    \frac{7-p}{2}\frac{\xi^2}{\sin^2{\bar{\theta}_{\mbox{\tiny $(p,n)$}}}}+
    \frac{1}{4}\mathcal{G}^{\alpha\gamma}\mathcal{G}^{\beta\delta}\,
    e^{2\frac{3-p}{5-p} \varphi}\,
     f_{\alpha\beta}f_{\gamma\delta}-
    \frac{7-p}{\sqrt{-g_{\mbox{\tiny $AdS_2$}}}}
    \frac{\xi \,e^{\frac{3-p}{5-p} \varphi}\, f_{t\rho}}{\sin^3{\bar{\theta}_{\mbox{\tiny $(p,n)$}}}}
   \right],
 \end{split}
\end{equation}
where the open-string metric $\mathcal{G}_{\alpha\beta}$ is defined as in \eqref{OpenStrMetr}. Such a Lagrangian coincides
with eq $(2.61)$ in \cite{Camino:2001at} if the components $f_{ab}$ along the $S^{7-p}$ directions are taken to be zero. Let us represent $f_{\alpha\,\beta}$ in terms of a potential $a_{\alpha}$ as $f_{\alpha\,\beta}=\partial_{\alpha}a_{\beta}-\partial_{\beta}a_{\alpha}$. Then,  the equations of motion for the fluctuations $(\xi,\,a_{\beta})$ have the following form
\begin{equation}\eqlabel{AngfEom}
 \begin{split} 
  &0\:=\:\frac{1}{\sqrt{-\hat{g}}}\partial_{\alpha}
    \left[
     \sqrt{-\hat{g}}\,\mathcal{G}^{\alpha\beta}\,\partial_{\beta}\xi
    \right]+
   \frac{7-p}{u_p^2\sin^2{\bar{\theta}_{\mbox{\tiny $(p,n)$}}}}
   \left[
    \xi+\frac{e^{\frac{3-p}{5-p} \varphi}\,f_{t\rho}}{\sqrt{-g_{\mbox{\tiny $AdS_2$}}}\sin{\bar{\theta}_{\mbox{\tiny $(p,n)$}}}}
   \right],\\
  &0\:=\:\frac{1}{\sqrt{-\hat{g}}}\partial_{\alpha}
    \left[
     \sqrt{-\hat{g}}\,\mathcal{G}^{\alpha\gamma}\mathcal{G}^{\beta\delta}\,
    e^{2\frac{3-p}{5-p} \varphi}\, f_{\gamma\delta} +
     \frac{(7-p)\sqrt{g_{\mbox{\tiny $S^{7-p}$}}}}{\sin^{3}{\bar{\theta}_{\mbox{\tiny $(p,n)$}}}}\xi
     \left[
      \delta_{\rho}^{\alpha}\delta_{t}^{\beta}-\delta_{\rho}^{\beta}\delta_{t}^{\alpha}
     \right]
    \right],
 \end{split}
\end{equation}
where $\sqrt{-\hat{g}}\,\overset{\mbox{\tiny def}}{=}\,e^{\frac{p-3}{5-p} \varphi}\sqrt{-g_{\mbox{\tiny $AdS_2$}}}
\sqrt{g_{\mbox{\tiny $S^{7-p}$}}}$. Similarly to the case analysed in Section \ref{subsec:CartFluct}, the equations of motion \eqref{AngfEom} are
 equivalent to the ones in $AdS_{\hat q+1}$ compactified on a $T^{\hat q-1}$ and with $\hat q$ analytically continued to take the value
 \beq
 \hat q\,=\,{2\over 5-p}\,\,.
 \eeq
 Indeed, it is straightforward to check that one can rewrite the system (\ref{AngfEom}) as
 \begin{equation}\eqlabel{AngfEom3}
 \begin{split}
  &0\:=\:\Box_{\mbox{\tiny $\hat q+1$}}\,\xi+u_p^{-2}\Box_{\mbox{\tiny $S^{7-p}$}}\,\xi+
     \frac{7-p}{u_p^2}
     \left[\xi+\frac{f_{t\rho}}{\sqrt{-g_{\mbox{\tiny $AdS_{\hat q+1}$}}}\sin{\bar{\theta}_{\mbox{\tiny $(p,n)$}}}}\right],\\
  &0\:=\:\frac{1}{\sqrt{-g_{\mbox{\tiny $AdS_{\hat q+1}$}}}}\partial_{\bar{A}}
     \left[\sqrt{-g_{\mbox{\tiny $AdS_{\hat q+1}$}}}\,g^{\bar{A}\bar{C}}\,\mathcal{G}^{BD}\,
    e^{2\frac{3-p}{5-p} \varphi}\, f_{\bar{C}D}
     +\frac{7-p}{\sin{\bar{\theta}_{\mbox{\tiny $(p,n)$}}}}
     \xi
     \left[
      \delta_{\rho}^{\bar{A}}\delta_{t}^{B}-\delta_{\rho}^{B}\delta_{t}^{\bar{A}}
     \right]
     \right]+
      \\
  &\phantom{0\:=\:}
     +\frac{u_p^{-2}}{\sqrt{g_{\mbox{\tiny $S^{7-p}$}}}}\partial_{a}
     \left[\sqrt{g_{\mbox{\tiny $S^{7-p}$}}}\,g^{ac}\mathcal{G}^{BD}\,
      e^{2\frac{3-p}{5-p} \varphi}\, f_{cD}
     \right],
 \end{split}
\end{equation}
Let us now analyse in details the different modes, following the classification in \cite{Kruczenski:2003be}.

\subsubsection{Coupled Modes}\label{subsubsec:CoupMod}

Let us now choose the following ansatz for the scalar $\xi$ and the gauge field
\begin{equation}\eqlabel{IIImodAns}
 \xi\:=\:e^{iEt}Y_{l}(S^{7-p})\hat{\xi}(\rho),\qquad
 a_{a}\:=\:e^{iEt}\nabla_{a}Y_{l}(S^{7-p})\hat{a}(\rho),\qquad
 a_{\rho}\:=\:e^{iEt}Y_{l}(S^{7-p})\hat{a}_{\rho}(\rho),
\end{equation}
with all the other components of the gauge field set to zero. Notice that the field component $a_a$ is a gradient in the $S^{7-p}$ sphere. Therefore,
it is always possible to perform a gauge transformation
such that the components of the gauge field along the $S^{7-p}$ are set to zero while the non-zero components are the
ones along the $AdS_{2}$-directions. Accordingly, in the following we will adopt the following ansatz for the components of the gauge field
\begin{equation}\eqlabel{IIImodAns2}
 a_t\:=\:-iE\,e^{iEt}Y_{l}(S^{7-p})\hat{a}_t(\rho),\qquad
 a_{\rho}\:=\:e^{iEt}Y_{l}(S^{7-p})\left(\hat{a}_{\rho}-\partial_{\rho}\hat{a}_t\right).
\end{equation}
With such an ansatz, the equations of motion \eqref{AngfEom3} acquire the following form
\begin{equation}\eqlabel{IIImodEom}
 \begin{split}
  &0\:=\:\Box_{\mbox{\tiny $\hat q+1$}}\hat{\xi}-E^2 g^{tt}\hat{\xi}-\frac{4l(l+6-p)}{(5-p)^2}\hat{\xi}+
    \frac{4(7-p)}{(5-p)^2}\left[\hat{\xi}+iE\frac{\hat{a}_{\rho}}{\sqrt{-g_{\mbox{\tiny $AdS_{\hat q+1}$}}}
     \sin{\bar{\theta}}_{\mbox{\tiny $(p,n)$}}}\right],
  \\
  &0\:=\:\partial_{\rho}
    \left[
     \frac{\hat{a}_{\rho}}{\sqrt{-g_{\mbox{\tiny $AdS_{\hat  q+1}$}}}\sin{\theta}}-i\frac{7-p}{E}\hat{\xi}
    \right]-
    \frac{4l(l+6-p)}{(5-p)^2}\frac{g_{\rho\rho}\,\hat{a}_{t}}{\sqrt{-g_{\mbox{\tiny $AdS_{\hat  q+1}$}}}
     \sin{\bar{\theta}}_{\mbox{\tiny $(p,n)$}}},\\
  &0\:=\:E^2
    \left[
     \frac{\hat{a}_{\rho}}{\sqrt{-g_{\mbox{\tiny $AdS_{\hat q+1}$}}}\sin{\theta}}-i\frac{7-p}{E}\hat{\xi}
    \right]+
    \frac{4l(l+6-p)}{(5-p)^2}\frac{g_{tt}\left(\hat{a}_{\rho}-\partial_{\rho}\hat{a}_t\right)}{
     \sqrt{-g_{\mbox{\tiny $AdS_{\hat q+1}$}}}\sin{\bar{\theta}}_{\mbox{\tiny $(p,n)$}}},\\
  &0\:=\:E^2\frac{g_{\rho\rho}\,\hat{a}_{t}}{\sqrt{-g_{\mbox{\tiny $AdS_{\hat  q+1}$}}}\sin{\bar{\theta}}_{\mbox{\tiny $(p,n)$}}}+
    \partial_{\rho}
    \left[
     \frac{g_{tt}\left(\hat{a}_{\rho}-\partial_{\rho}\hat{a}_t\right)}{
     \sqrt{-g_{\mbox{\tiny $AdS_{\hat  q+1}$}}}\sin{\bar{\theta}}_{\mbox{\tiny $(p,n)$}}}
    \right].
 \end{split}
\end{equation}
where $\Box_{\mbox{\tiny $\hat q+1$}}$ is the d'Alembertian  operator for $AdS_{\hat q+1}$. This is a system of four equations in three unknowns $\left(\hat{\xi},\,\hat{a}_{t},\,\hat{a}_{\rho}\right)$, and therefore
one of the equations must be redundant. It is easy to see that solving the second and third equations in \eqref{IIImodEom} 
for their last terms and inserting such solutions in the last equation, the latter turns out to be an identity. So, we
can consider the last equation in \eqref{IIImodEom} as redundant and focus on the first three. Their form suggests that
it is convenient to define two new fields $\left(\tilde{\xi},\,\tilde\eta\right)$
\begin{equation}\eqlabel{IIImodXiEta}
 \tilde{\xi}\:\overset{\mbox{\tiny def}}{=}\:-\frac{i}{E}\,\hat{\xi}\,e^{iEt}\,,\qquad
\tilde \eta\:\overset{\mbox{\tiny def}}{=}\:\Bigg(\,
  \frac{\hat{a}_{\rho}}{\sqrt{-g_{\mbox{\tiny $AdS_{\hat q+1}$}}}\sin{\theta}}-i\frac{7-p}{E}\hat{\xi}\,\,\Bigg)\,e^{iEt}\,.
\end{equation}
Now, eliminating the field $\hat{a}_t$ by using the second equation in \eqref{IIImodEom}, the equations of motion can
be reduced to the following system of two equations
\begin{equation}\eqlabel{IIImodEom2}
 \begin{split}
  &0\:=\:\Box_{\mbox{\tiny $\hat q+1$}}\,\tilde{\xi}-
    \frac{4}{(5-p)^2}\left[l(l+6-p)+(7-p)(6-p)\right]\tilde{\xi}+\frac{4(7-p)}{(5-p)^2}\,
    \tilde\eta,\\\\
  &0\:=\:\Box_{\mbox{\tiny $\hat q+1$}}\,\tilde\eta-\frac{4(l+6-p)}{(5-p)^2}\tilde\eta+
   \frac{4(7-p)l(l+6-p)}{(5-p)^2}\,\tilde{\xi}.
 \end{split}
\end{equation}
It is convenient to rewrite the above system 
of two equations in matrix form
\begin{equation}\eqlabel{IIImodEom3}
 \Box_{\mbox{\tiny $\hat{q}+1$}}\psi\:=\:\mathcal{M}_{\mbox{\tiny $(l,p)$}}\,\psi,
 \qquad
 \psi\:=\:
 \begin{pmatrix}
  \tilde{\xi}\\
 \tilde \eta
 \end{pmatrix},
\end{equation}
and the mass matrix $\mathcal{M}_{\mbox{\tiny $(l,p)$}}$, as in \cite{Camino:2001at}, is 
\begin{equation}\eqlabel{IImodMassMatrix}
 \mathcal{M}_{\mbox{\tiny $(l,p)$}}\:\overset{\mbox{\tiny def}}{=}\:\frac{4}{(5-p)^2}
 \begin{pmatrix}
  l(l+6-p)+(7-p)(6-p) & p-7\\
  {} & {} \\
  (p-7)l(l+6-p) & l(l+6-p)
 \end{pmatrix}.
\end{equation}
The eigenvalues $\lambda^{\mbox{\tiny $(i)$}}$ and eigenvectors 
$\psi^{\mbox{\tiny $(i)$}}$  of $\mathcal{M}_{\mbox{\tiny $(l,p)$}}$ are
\begin{equation}\eqlabel{IIImodEigen}
 \begin{split}
  &\lambda^{\mbox{\tiny $(1)$}}\:=\:\frac{4(l+6-p)(l+7-p)}{(5-p)^2},\qquad
    \psi^{\mbox{\tiny $(1)$}}\:=\:(l+6-p)\tilde{\xi}-\tilde\eta,\qquad
    l\,\ge\,0,\\
  &\lambda^{\mbox{\tiny $(2)$}}\:=\:\frac{4l(l-1)}{(5-p)^2},\hspace{3.1cm}
    \psi^{\mbox{\tiny $(2)$}}\:=\:\tilde{\xi}+\tilde\eta,\hspace{2.5cm}
    l\,\ge\,1,
 \end{split}
\end{equation}
and they satisfy the differential equation
\begin{equation}\eqlabel{IIImodEom4}
 \Box_{\mbox{\tiny $\hat q+1$}}\psi^{\mbox{\tiny $(i)$}}\:=\:\lambda^{\mbox{\tiny $(i)$}}\psi^{\mbox{\tiny $(i)$}}.
\end{equation}
It is now straightforward to obtain the behaviour of the normalizable and non-normalizable modes for $\psi^{\mbox{\tiny $(i)$}}$
in a neighbourhood of the boundary $\rho\,\rightarrow\,0$. One gets
\begin{equation}\eqlabel{IIImodNnN}
 \begin{split}
  &\psi^{\mbox{\tiny $(1)$}}\:\overset{\mbox{\tiny $\rho\rightarrow0$}}{\sim}\:
    \psi^{\mbox{\tiny $(1)$}}_{-}\rho^{2\alpha_{-}^{(1)}}+\psi^{\mbox{\tiny $(1)$}}_{+}\rho^{2\alpha_{+}^{(1)}},\quad
   \left\{
    \begin{array}{l}
     \alpha_{-}^{(1)}\:=\:-\frac{l+6-p}{5-p},\\
     \phantom{\ldots}\\
     \alpha_{+}^{(1)}\:=\:\frac{l+7-p}{5-p}\,\,,
    \end{array}
   \right. \\\\
  &\psi^{\mbox{\tiny $(2)$}}\:\overset{\mbox{\tiny $\rho\rightarrow0$}}{\sim}\:
    \psi^{\mbox{\tiny $(2)$}}_{-}\rho^{2\alpha_{-}^{(2)}}+\psi^{\mbox{\tiny $(2)$}}_{+}\rho^{2\alpha_{+}^{(2)}},\quad
   \left\{
    \begin{array}{l}
     \alpha_{-}^{(2)}\:=\:-\frac{l-1}{5-p},\\
     \phantom{\ldots}\\
     \alpha_{+}^{(2)}\:=\:\frac{l}{5-p}.
    \end{array}
   \right.
 \end{split}
\end{equation}

In order to compute the correlation functions for the decoupled modes $\psi^{\mbox{\tiny $(i)$}}$, one can notice
that their equations of motion \eqref{IIImodEom4} are just the equation of motion for free massive scalars in $AdS_{\hat q+1}$ 
and, therefore, the dynamics of the fluctuations -- up to quadratic order -- can be described by the following
effective Euclidean action
\begin{equation}\eqlabel{IIImodEffAct}
 S_{\mbox{\tiny eff}}^{\mbox{\tiny $(\psi^{(i)})$}}\:=\:
 T_{\mbox{\tiny D$(7+\hat q-p)$}}\,\,
  \mathcal{N}_{\mbox{\tiny $l$}}\,\,
 \int dt\, d^{\hat q-1}\zeta\, d\rho\:\sqrt{g_{\mbox{\tiny $AdS_{\hat q+1}$}}}\,\frac{1}{2}
 \left[
  \left(\partial\psi^{\mbox{\tiny $(i)$}}\right)^2+\lambda^{\mbox{\tiny $(i)$}}
   \left(\psi^{\mbox{\tiny $(i)$}}\right)^2
 \right].
\end{equation}
The holographic renormalisation of this action goes exactly as in \ref{subsubsec:HR}, so that the one-point function
can be written as
\begin{equation}\eqlabel{IIImod1ptCF}
 \langle\hat{\mathcal{O}}_{\mbox{\tiny $\psi^{(i)}$}}\rangle\:=\:-2\tilde{\alpha}^{\mbox{\tiny $(i)$}}\, \mathcal{N}_{\mbox{\tiny $l $}}^{\mbox{\tiny $(p,n)$}}\,  \left[g^2_{\mbox{\tiny eff}}(E)\right]^{\frac{1}{5-p}}E^{2\tilde{\alpha}^{(i)}-\frac{3-p}{5-p}}
   \Theta^{\mbox{\tiny $(i)$}}(E),
\end{equation}
with $\tilde{\alpha}^{\mbox{\tiny $(i)$}}\,\overset{\mbox{\tiny def}}{=}\,\alpha^{(i)}_{+}-\alpha^{(i)}_{-}$ and
$\Theta(E)$ given by
\begin{equation}\eqlabel{IIImodTE}
 \Theta(E)\:=\:
 \left\{
  \begin{array}{l}
  - \frac{\Gamma(1-\tilde{\alpha}^{\mbox{\tiny $(i)$}})}{\Gamma(1+\tilde{\alpha}^{\mbox{\tiny $(i)$}})}
    \psi^{\mbox{\tiny $(i)$}}_{-},\hspace{7.2cm} \tilde{\alpha}^{\mbox{\tiny $(i)$}} \mbox{ non-integer}\\
   \phantom{\ldots}\\
   \frac{(-1)^{\tilde{\alpha}^{(i)}}}{4^{\tilde{\alpha}^{(i)}}\Gamma(\tilde{\alpha}^{\mbox{\tiny $(i)$}})
    \Gamma(\tilde{\alpha}^{\mbox{\tiny $(i)$}}-1)}
    \left[
     H_{\mbox{\tiny $\tilde{\alpha}^{\mbox{\tiny $(i)$}}$}}-2\gamma_{\mbox{\tiny EM}}-
      \frac{2}{\tilde{\alpha}^{\mbox{\tiny $(i)$}}(\tilde{\alpha}^{\mbox{\tiny $(i)$}}-1)}\log{\frac{E}{2}}
    \right]\psi^{\mbox{\tiny $(i)$}}_{-},\quad \tilde{\alpha}^{\mbox{\tiny $(i)$}} \mbox{ integer}.
  \end{array}
 \right.
\end{equation}
From the values of  $\alpha^{\mbox{\tiny $(i)$}}_{\pm}$ written in (\ref{IIImodNnN}), one immediately determines the values of  $\tilde{\alpha}^{\mbox{\tiny $(i)$}}$, namely
\beq
\tilde{\alpha}^{\mbox{\tiny $(1)$}}\,=\,{2l+3\over 5-p}\,+\,2\,\,,
\qquad\qquad
\tilde{\alpha}^{\mbox{\tiny $(2)$}}\,=\,{2l-1\over 5-p}\,\,.
\eeq
Then, it follows that $\tilde{\alpha}^{\mbox{\tiny $(1)$}}$ is integer iff $2l+3=0\,\,
({\rm mod} (5-p))$, while $\tilde{\alpha}^{\mbox{\tiny $(1)$}}\in {\mathbb Z}$ only when 
$2l-1=0\,\, ({\rm mod} (5-p))$. As in the case of the $\chi$ fluctuation, this can only happen when $p$ is even and always happens if $p=4$, whereas for $p=0,2$ it occurs for some values of $l$.

Similarly to the case of the fluctuations of the Cartesian coordinates, the two point function turns out to have the 
following functional form
\begin{equation}\eqlabel{IIImod2ptCF}
 \begin{split}
   &\langle\hat{\mathcal{O}}_{\mbox{\tiny $\psi^{(i)}$}}\hat{\mathcal{O}}_{\mbox{\tiny $\psi^{(i)}$}}\rangle\:\sim\:
    \left[g^2_{\mbox{\tiny eff}}(E)\right]^{\frac{1}{5-p}}E^{2\tilde{\alpha}^{(i)}+\frac{3-p}{5-p}},
    \hspace{2.4cm}\tilde{\alpha}^{\mbox{\tiny $(i)$}}\mbox{ non-integer},\\
   &\langle\hat{\mathcal{O}}_{\mbox{\tiny $\psi^{(i)}$}}\hat{\mathcal{O}}_{\mbox{\tiny $\psi^{(i)}$}}\rangle\:\sim\:
    \left[g^2_{\mbox{\tiny eff}}(E)\right]^{\frac{1}{5-p}}E^{2\tilde{\alpha}^{(i)}+\frac{3-p}{5-p}}
    \log{\frac{E^{\mbox{\tiny $2\tilde{\alpha}^{\mbox{\tiny $(i)$}}$}}}{
     \mu^{\mbox{\tiny $2\tilde{\alpha}^{\mbox{\tiny $(i)$}}$}}}},
    \qquad\tilde{\alpha}^{\mbox{\tiny $(i)$}}\mbox{ integer},
 \end{split}
\end{equation}
which in the coordinate space becomes
\begin{equation}\eqlabel{IIImod2ptCFb}
 \begin{split}
  &\langle\hat{\mathcal{O}}_{\mbox{\tiny $\psi^{(i)}$}}\hat{\mathcal{O}}_{\mbox{\tiny $\psi^{(i)}$}}\rangle\:\sim\:
   \frac{\left[g^2_{\mbox{\tiny eff}}(t)\right]^{\frac{1}{5-p}}}{|t|^{2\Delta_{\psi}^{\mbox{\tiny $(i)$}}}},
   \hspace{3cm}\tilde{\alpha}^{\mbox{\tiny $(i)$}}\mbox{ non-integer}\\
  &\langle\hat{\mathcal{O}}_{\mbox{\tiny $\psi^{(i)}$}}\hat{\mathcal{O}}_{\mbox{\tiny $\psi^{(i)}$}}\rangle\:\sim\:
   \mathcal{R}\left(\frac{\left[g^2_{\mbox{\tiny eff}}(t)\right]^{\frac{1}{5-p}}}{|t|^{2\Delta_{\psi}^{\mbox{\tiny $(i)$}}}}\right),
%   \log{|2\pi t|^{-2\tilde{\alpha}^{\mbox{\tiny $(i)$}}}},
   \hspace{1.9cm}\tilde{\alpha}^{\mbox{\tiny $(i)$}}\mbox{ integer}.
 \end{split}
\end{equation}
with the exponent $\Delta_{\psi}^{\mbox{\tiny $(i)$}}$ given by
\begin{equation}\eqlabel{IIImodConfDim}
 \Delta_{\psi}^{\mbox{\tiny $(i)$}}\:=\:\tilde{\alpha}^{\mbox{\tiny $(i)$}}+1-\frac{1}{5-p}\:=\:
 \left\{
  \begin{array}{l}
   \frac{2}{5-p}\,(l+1)\,+\,3,\qquad i\,=\,1,\\
    \phantom{\ldots}\\
   \frac{2}{5-p}\,(l-1)\,+\,1,\qquad i\,=\,2.
  \end{array}
 \right.
\end{equation}
Notice that the exponents $\Delta_{\psi}^{\mbox{\tiny $(i)$}}$ correctly reproduce the conformal dimensions in the
case $p\,=\,3$, where $\Delta_{\psi}^{\mbox{\tiny $(1)$}}\,=\,l+4$ and $\Delta_{\psi}^{\mbox{\tiny $(2)$}}\,=\,l$ \cite{Camino:2001at,Faraggi:2011ge,Harrison:2011fs}. Again the conformal dimensions are fractional for $p<3$. 

\subsubsection{Internal Gauge Field Modes}\label{subsubsec:Imod}

Finally, let us consider an ansatz for which the scalar $\xi$ is set to zero and the only non-zero component for the gauge
field are the ones along the $S^{7-p}$-directions
\begin{equation}\eqlabel{ImodAns}
 \xi\:=\:0,\qquad f_{ta}\:=\:\partial_{t}a_a,\qquad f_{\rho a}\:=\:\partial_{\rho}a_{a},\qquad 
  f_{ab}\:=\:\partial_{a}a_b-\partial_{b}a_a,
\end{equation}
with 
\begin{equation}\eqlabel{ImodAns2}
 a_a\:=\:Y^{l}_a\left(S^{7-p}\right)\tilde{a}(t,\rho)\:\equiv\:Y^{l}_a\left(S^{7-p}\right)e^{iEt}\hat{a}(\rho),
\end{equation}
where $Y^{l}_a\left(S^{7-p}\right)$ is a vector spherical harmonic on the $S^{7-p}$ sphere.
With such an ansatz, the equations of motion \eqref{AngfEom3} can be written as
\begin{equation}\eqlabel{ImodEom}
 \begin{split}
  &0\:=\:\partial_{t}\big(\nabla_{a}a^a\big)\,\,,\\
  &0\:=\:\partial_{\rho}\big(\nabla_{a}a^a\big)\,\,,\\
  &0\:=\:\frac{g_{\rho\rho}}{\sqrt{-g_{\mbox{\tiny $AdS_{\hat q+1}$}}}}\partial_{t}^2a^b-
    \partial_{\rho}
     \left[
      \frac{g_{tt}}{\sqrt{-g_{\mbox{\tiny $AdS_{\hat q+1}$}}}}\partial_{\rho}a^b
     \right]-\\
  &\phantom{0\:=\quad}
    -\frac{g_{tt}g_{\rho\rho}}{\sqrt{-g_{\mbox{\tiny $AdS_{\hat q+1}$}}}}u_p^{-2}
    \left(
     \Delta\,a^b - \nabla^b\nabla_c a^c
    \right),
 \end{split}
\end{equation}
where the operators $\Delta$  (the Hodge-de Rham  operator for one-forms on the $S^{7-p}$)  and $\nabla_{b}$ are defined as
\begin{equation}\eqlabel{ImodNabl}
 \Delta a^b\:=\:\Box_{\mbox{\tiny $S^{7-p}$}} \,a^b - R^{b}_c\, a^c,\qquad
 \nabla_{b}\,a^b\:=\:\frac{1}{\sqrt{g_{\mbox{\tiny $S^{7-p}$}}}}\,\partial_{b}
  \left[
   \sqrt{g_{\mbox{\tiny $S^{7-p}$}}}\,g^{bc}\,a_c
  \right],
\end{equation}
with $R^b_c$ being the Ricci tensor on the $S^{7-p}$-sphere. From the first two equations \eqref{ImodEom}, one can deduce 
the requirement
\begin{equation}\eqlabel{ImodCond}
 \nabla_{b}\,a^b\:=\:0,
\end{equation}
which implies that the last term in the third equation \eqref{ImodEom} vanishes. Furthermore, from the decomposition 
\eqref{ImodAns2} the condition \eqref{ImodCond} implies that the vector spherical harmonic $Y^{l}_a\left(S^{7-p}\right)$ has
to be such that
\begin{equation}\eqlabel{ImodCond2}
 \nabla^b\,Y^{l}_b\left(S^{7-p}\right)\:=\:0.
\end{equation}
Considering also that the vector harmonic $Y^{l}_a\left(S^{7-p}\right)$ is an eigenfunction of the Hodge-de Rham operator
$\Delta$ defined in \eqref{ImodNabl}
\begin{equation}\eqlabel{ImodHodgeVect}
 \Delta\,Y^{l}_a\left(S^{7-p}\right)\:=\:-(l+1)(l+5-p)Y^{l}_a\left(S^{7-p}\right),
\end{equation}
the third equation in \eqref{ImodEom} takes the form
\begin{equation}\eqlabel{ImodEom2}
 0\:=\:\left(\Box_{s+1}-M_{\tilde{a}}^2\right)\tilde{a},
\end{equation}
where $\Box_{s+1}$ is the Dalambertian operator in $AdS_{s+1}$ with $s\,=\,2-\hat q=2(4-p)/(5-p)$ and the mass $M^2_{\tilde{a}}$ is
\begin{equation}\eqlabel{ImodMass}
 M^2_{\tilde{a}}\:=\:\frac{4}{(5-p)^2}\left[(l+1)(l+5-p)\right].
\end{equation}
The behaviour of the normalizable and non-normalizable modes 
$\tilde{a}\,\sim\,\tilde{a}_{-}\,\rho^{2\alpha_{-}}+\tilde{a}_{+}\,\rho^{2\alpha_{+}}$
in a neighbourhood of the boundary is given by
\begin{equation}\eqlabel{ImodNnN}
 M^2_{\tilde{a}}\:=\:2\alpha\left(2\alpha-s\right)\quad\Longrightarrow\quad
 \left\{
  \begin{array}{l}
   \alpha_{-}\:=\:-\frac{1}{5-p}\left(l+1\right),\\
    \phantom{\ldots}\\
   \alpha_{+}\:=\:\frac{1}{5-p}\left(l+5-p\right)
  \end{array}
 \right.
\end{equation}
In this case 
\beq
\tilde\alpha_{a}\,=\, \alpha_{+}\,-\,\alpha_{-}\,=\,{2l+1\over 5-p}\,+\,1\,\,.
\eeq
Proceeding in a similar fashion to the analysis in the previous cases, it is easy to find that the two-point function
of the operator dual to these fluctuations has the following form
\begin{equation}\eqlabel{Imod2ptCF}
 \begin{split}
  &\langle\hat{\mathcal{O}}_{a}\hat{\mathcal{O}}_{a}\rangle\:\sim\:
    \frac{\left[g^2_{\mbox{\tiny eff}}(t)\right]^{\frac{1}{5-p}}}{|t|^{2\Delta_{a}}},
    \hspace{2.6cm}\tilde{\alpha}_{a}\mbox{ non-integer}\\
  &\langle\hat{\mathcal{O}}_{a}\hat{\mathcal{O}}_{a}\rangle\:\sim\:
    \mathcal{R}\left(\frac{\left[g^2_{\mbox{\tiny eff}}(t)\right]^{\frac{1}{5-p}}}{|t|^{2\Delta_{a}}}\right)
    %\log{|2\pi t|^{-2\tilde{\alpha}}},
    \hspace{1.7cm}\tilde{\alpha}_{a}\mbox{ integer},
 \end{split}
\end{equation}
with $\Delta_{a}$ being given by
\beq
\Delta_{a}\:=\:\tilde{\alpha}_{a}+1-{1\over 5-p}\,=\,{2\over 5-p}\,l\,+\,1
\eeq 
Notice that $\Delta_{a}=\Delta_{\chi}$  and for $p\,=\,3$ the exponent 
$\Delta_{a}$ coincides with the conformal dimension of the operator $\hat{\mathcal{O}}_{a}$ in the $AdS_5\times S^5$ background, which is $\Delta_{a}\,=\,l+2$, as it should \cite{Camino:2001at,Faraggi:2011ge,Harrison:2011fs}.

\section{Conclusion}\label{sec:Concl}

In this paper we investigated the insertion of impurities in $(p+1)$-dimensional Supersymmetric Yang-Mills theories,
which are (trivially) non-conformal. The ambient theory is holographically described by the near-horizon geometry
generated by a stack of $N$ D$p$-branes, while the impurities are added by introducing probe D$(8-p)$-branes in this
background in such a way that the induced world-volume metric is conformally $AdS_{2}\times S^{7-p}$. The
background RR $(7-p)$-form potential induces an electric gauge-field on the world-volume of the probe branes, giving rise to a bundle
of strings stretching in the radial direction  and forming a flux tube.

We analysed in some detail two possible classes of configurations for such systems. In the first one, the flux tube
is straight -- which corresponds to  a constant embedding function for the probe branes --, while in the second one
two flux tubes get connected in the bulk. From the gauge theory point of view such configurations respectively correspond
to single impurities and to dimers.

For these systems we studied the basic thermodynamic properties such as the free energy, entropy and specific heat. Interestingly, 
the impurity entropy turns out to be in general non-analytic in the filling fraction, except for the case of the D$4$/D$4$ systems for which we had been able to  find a 
closed form. This latter system is actually the only one showing a positive specific heat: for $p\,=\,3$, the specific heat is zero, while
for $p\,<\,3$ it is negative. This is a signature of a thermodynamic instability. We also computed the impurity susceptibility, which is constant for $p\,=\,4$.

We then analysed the case of the hanging flux tubes, for which the position of the probe branes is fixed in the transverse direction, while their embedding
in the conformal-$AdS$ manifold is controlled by a scalar. At finite temperature, there are two possible configurations: in one the two
flux tubes are connected in the bulk and lie outside the black hole, while in the second one the two flux tubes end
into the black hole. The transition from the second configuration to the first one is of  first order and it
corresponds to the dimerisation transition. We analysed the thermodynamics of these connected configurations. All the thermodynamic functions (free energy,
entropy, internal energy, specific heat and latent heat) can be written in terms of hypergeometric functions. We studied the competition between
the two configurations and determined the temperatures at which the dimerisation occurs. This phase transition turns out to be of  first order. Furthermore,
the specific heat for the dimer configuration is always positive for any $p\,<\,3$ and vanishes as $T\,\rightarrow\,0$. As mentioned before, the specific
heat for the straight flux-tube configurations is negative for $p\,<\,3$. One possible interpretation of this result can be that such a configuration is just not allowed
for $p\,<\,3$ and that  the flux-tubes are forced to reconnect and form a dimer. 

We further studied the stability of the systems  at zero temperature by analysing the fluctuations of the probe branes in the straight configuration (impurity fluctuations). 
The fluctuations decouple into two channels: one contains just the fluctuations of the Cartesian coordinates, the other channel instead contains the coupled
fluctuations of the angular embedding function and of the world-volume gauge field. The modes in the second channel can be conveniently decoupled. Interestingly,
all these modes satisfy the equations of motion of free massive scalars in a higher-dimensional $AdS$-space, where the enhancement of dimensions is due to the
presence of a dilaton with a non-trivial profile. These modes are dual to irrelevant operators. In order to make sense of the correlators of such operators,
one needs to holographic renormalise the action for these modes perturbatively up to order $n$, if one needs to compute the $n$-point correlators. For the aim
of characterising the theory, it may be enough just to compute the one- and two-point correlators. Therefore we needed to renormalise the quadratic action which,
as we have mentioned, is just the action for a free massive scalar in a higher-dimensional $AdS$-space. Then,  we computed the correlators and wrote them in
such a way that the underlying generalised conformal symmetry is manifest. From such an expression we can extract the generalised scaling dimension for all
the operators we are interested in. Remarkably, such generalised scaling dimensions are fractional for $p\,<\,3$ since the term which depends
on the KK quantum number $l$ gets multiplied  by the factor $2/(5-p)$, which is just $1$  for $p\,=\,3$. It would be interesting to understand this
scaling from a purely field theoretical point of view. From the discussion about the generalised conformal structure, the most naive expectation is that the impurity
action has a factor with a suitable power of the dimensionful coupling constant.

It would be interesting to repeat this fluctuation analysis for the case of the dimer configuration. One would expect that some of the dimer fluctuations 
satisfy the Heun equation,
as it happens in the ABJM case \cite{Benincasa:2011zu}, which is connected to integrable models. If on one side  the emergence of this integrable structure could be understandable
for the case $p\,=\,3$, it might be a bit counter-intuitive  in the general  $p\,<\,5$ system. However, as we discussed, the modes for $p\,\neq\,3$
can be generally seen as modes propagating in a higher-dimensional $AdS$, so that the features of these systems may be deduced from conformal systems.

As we mentioned in this paper, while it has been showed \cite{Gomis:2006sb} that the introduction of D$5$-branes in a D$3$-brane background such that the world-volume 
induced metric is $AdS_2\times S^4$ is dual to a $1/2$-BPS Wilson loop operator in $\mathcal{N}\,=\,4$ SYM in the anti-symmetric representation of the gauge group 
$U(N)$, it is not yet clear whether a similar interpretation holds in the case of D$(8-p)$-branes in the D$p$-brane background with the same type of embedding. Therefore, a 
natural thing to do would be to clarify this issue. In this spirit, one can also study fermionic excitations in the D$p$/D$(8-p)$-system as done in 
\cite{Faraggi:2011ge} for the conformal case $p\,=\,3$.

The analysis we carried out in this paper can be generalised to the case in which the probe D$(8-p)$-branes wrap different submanifold of $S^{(8-p)}$, as in 
\cite{Karaiskos:2011kf}. More interestingly, one can think to introduce chemical potential to the system by considering Reissner-Nordstr{\"o}m type of backgrounds
and analyse the configuration allowed and the phase structure. Importantly, one could compute the resistivity as a function of the temperature and to study whether the holographic impurities
induce a minimum similar to the one that occurs  in  the condensed matter models reviewed in section  \ref{sec:GenKondo}. 

A final -- and longer term -- direction concerns the effect of the backreaction of the D$(8-p)$-branes. Presumably, this would require dealing with a bubbling geometry, generalizing the one  found in \cite{Harrison:2011fs} for the conformal $p=3$ case.

\section*{Acknowledgements}

We are grateful to  Eduardo Conde, Niko Jokela, Wolfgang M{\"u}ck,  Diego Rodriguez-Gomez and Balt van Rees for useful discussions. P.B. would also like to thank
the developers of SAGE \cite{sage}, Maxima \cite{maxima}, Numpy and Scipy \cite{scipy}.
This work is funded in part by MICINN under grants FPA2008-01838 and 
FPA2011-22594, by the Spanish 
Consolider-Ingenio 2010 Programme CPAN (CSD2007-00042) and by Xunta de Galicia (Conseller{\'i}a de Educac{\'i}on, grant 
INCITE09 206 121 PR and grant PGIDIT10PXIB206075PR) and by FEDER. P.B. is supported as well by the MInisterio de Ciencia e 
INNovaci{\'on} through the Juan de la Cierva program.

\appendix

\section{Polar angles for stable embeddings}\label{app:angles}

In this appendix we analyze the main properties of the polar angles of the flux tube configurations. We begin by listing the different functions $\Lambda_{p,n}(\theta)$  for $0\le p\le 5$, namely
\begin{equation}\eqlabel{Lambda2}
 \begin{split}
  &\Lambda_{0,n}(\theta)\:=\:-\frac{2}{5}\,\left[\,\cos\theta\,\left(\,3\sin^4\theta\,+\,4\sin^2\theta\,+\,8\,\right)\,
    +\,8\,\left(\,2\,\frac{n}{N}\,-\,1\right)\,\right],\\
  &\Lambda_{1,n}(\theta)\:=\:-\frac{5}{4}\,\left[\,\cos\theta\,\left(\,\sin^3\theta\,+\,\frac{3}{2}\,\sin\theta\,\right)\,
    +\,\frac{3}{2}\,\left(\,\frac{n}{N}\,\pi\,-\,\theta\,\right)\,\right], \\
  &\Lambda_{2,n}(\theta)\:=\:-\frac{4}{3}\,\left[\,\cos\theta\,\left(\,\sin^2\theta\,+\,2\,\right)\,+\,2\,\left(\,
    2\,\frac{n}{N}\,-\,1\right)\,\right],\\
  &\Lambda_{3,n}(\theta)\:=\:-\frac{3}{2}\,\left[\,\cos\theta\,\sin\theta\,+\,\frac{n}{N}\,\pi\,-\,\theta\,\right],\\
  &\Lambda_{4,n}(\theta)\:=\:- 2\,\left[\,\cos\theta\,+\,2\,\frac{n}{N}\,-\,1\right],\\
  &\Lambda_{5,n}(\theta)\:=\:\theta\,-\,\frac{n}{N}\,\pi\,\,.
 \end{split}
\end{equation}
As mentioned in the main text the functions $\Lambda_{p,n}(\theta)$ depend on the quantization number $n$ and on the number of colors $N$ through their ratio $\nu=n/N$ (the filling fraction). Moreover, one can check explicitly from (\ref{Lambda2}) that  these functions satisfy
\beq
\Lambda_{p,n}(\theta)\,=\,-\Lambda_{p,N-n}(\pi-\theta)\,\,,
\eeq
from which (\ref{particle-hole}) follows immediately.  As shown in \cite{Camino:2001at}, the $\Lambda_{p,n}(\theta)$ are monotonically increasing functions of $\theta$ in the interval $0<\theta<\pi$. Moreover, one can check that:
\beq
\Lambda_{p,n}(0)\,=\,
-2\sqrt{\pi}\,\,
{\Gamma\Bigl(\,{8-p\over 2}\Bigr)\over
\Gamma\Bigl(\,{7-p\over 2}\Bigr)}\,\,
{n\over N}\,\,,
\qquad\qquad
\Lambda_{p,n}(\pi)\,=\,2\sqrt{\pi}\,\,
{\Gamma\Bigl(\,{8-p\over 2}\Bigr)\over
\Gamma\Bigl(\,{7-p\over 2}\Bigr)}\,\,
(\,1\,-\,{n\over N}\,)\,\,.
\label{Lambda-zero-pi}
\eeq
and, therefore, $\Lambda_{p,n}(0)<0$ if $n>0$ and $\Lambda_{p,n}(\pi)>0$ if $n<N$. Thus, it follows that there exists only one solution $\bar\theta_{(p,n)}\,\in \,(0,\pi)$ of the equation $\Lambda_{p,n}(\bar\theta_{(p,n)})\,=\,0$ for each $n$ in the interval $0<n<N$, \ie\ there are exactly $N-1$ angles which correspond to non-singular wrappings of the D(8-p)-brane probe on the $S^{7-p}$ sphere. Let us work out the expressions of some of them for different values of $p$. In the conformal case $p=3$, the angles $\bar\theta_{(3,n)}$ are the solutions of the equation:
\beq
\bar\theta_{(3,n)}\,-\,\cos\bar\theta_{(3,n)}\,\sin\,\bar\theta_{(3,n)}\,=\,
{n\over N}\,\pi\,\,.
\label{theta-3}
\eeq
For $p=4$ the angles $\bar\theta_{(4,n)}$ can be immediately obtained from (\ref{Lambda2}), with the result
\beq
{\rm cos}\,\bar\theta_{(4,n)}\,=\,1\,-\,2\,{n\over N}\,\,.
\label{theta-4}
\eeq
The analytic expression of the angles $\bar \theta_{(p,n)}$ for $p\le 2$ is more difficult to obtain.  Let us analyse in detail the equation which determines the angles for $p\,=\,2$, namely the solutions of the equation $\Lambda_{2,n}(\theta)=0$. In terms of the
filling fraction $\nu$, after  using the third expression in
\eqref{Lambda2}, this equation becomes
\begin{equation}\eqlabel{Anglep2}
 \cos^3\bar\theta_{(2,n)}\,-\,3\cos\bar\theta_{(2,n)}\,=\,4\nu-2\,\, ,
\end{equation}
which is a cubic equation in $\cos{\bar\theta_{(2,n)}}$.  In general, an equation of the type
\begin{equation}\eqlabel{cubic}
 x^3\,+\,r\,x\,=\,s\,\,,
\end{equation}
can be solved for $x$ by means of the so-called Vieta's substitution, namely:
\begin{equation}\eqlabel{Vieta}
x\,=\,w\,-\,\frac{r}{3 w}\,\,.
\end{equation}
Indeed, by substituting \eqref{Vieta}) into the cubic equation \eqref{cubic}, one obtains the following quadratic equation 
for $w^3$:
\begin{equation}\eqlabel{cubic2}
(w^3)^2\,-\,s\,(w^3)\,-\,{r^3\over 27}\,=\,0\,\,.
\end{equation}
In our case $x=\cos\bar\theta_{(2,n)}$, with $r\,=\,-3$ and $s\,=\,4\nu-2$. The two solutions for $w^3$ are just
\begin{equation}\eqlabel{w3}
w^3\:=\:2\nu-1\,\pm 2\,i\,\sqrt{\nu(1-\nu)}\,\,.
\end{equation}
Remarkably, the right-hand side of (\ref{w3}) is a complex number of modulus one and, therefore, can be represented as
\begin{equation}\eqlabel{w3b}
w^3\,=\,e^{i\alpha}\,\,\qquad\qquad\Longrightarrow\qquad\qquad w\,=\,e^{i{\alpha\over 3}}\,\,,
\end{equation}
where $\alpha$ is an angle such that
\begin{equation}\eqlabel{alpha}
\cos\alpha\,=\,2\nu-1\,\,,\qquad\qquad \sin\alpha\,=\,\pm 2\,\sqrt{\nu(1-\nu)}\,\,.
\end{equation}
Notice that the relation between $x=\cos\theta$ and $w$ is, in our case, given by (\ref{Vieta}) with $r=-3$, namely
\begin{equation}\eqlabel{x}
x\,=\,w+w^{-1}\,=\,\cos\bar\theta_{(2,n)}\,.
\end{equation}
It follows that $\bar\theta_{(2,n)}$ is given by
\begin{equation}\eqlabel{Anglep2sol}
\cos\bar\theta_{(2,n)}\,=\,2\cos \left(\frac{\alpha}{3}\right)\,\,,\qquad\qquad\cos\alpha\,=\,2\nu\,-\,1\,=\,\frac{2n-N}{N}.
\end{equation}
As a further example, let us point out that the angles for $p=1$ can be obtained by solving the following transcendental equation
\beq
\bar\theta_{(1,n)}\,-\,{2\over 3}\,\,\sin (2\bar\theta_{(1,n)})\,\Big[\,1\,-\,{1\over 4}\,\,
\sin (2\bar\theta_{(2,n)})\,\Big]\,=\,{n\over N}\,\,\pi\,\,.
\eeq

\section{Dimer integrals}\label{app:dimerintegrals}

The purpose of this appendix is to derive the integrals needed in the calculations of section \ref{sec:Hanging} of the thermodynamic properties of the dimer configurations. First of all, we define the following two integrals $I_1(\alpha, \gamma)$ and $I_2(\alpha, \gamma)$ as
\bear
&&I_1(\alpha, \gamma)\,\overset{\mbox{\tiny def}}{=}\,\int_1^{\infty}\,{dz\over \sqrt{(z^{\alpha}-1)(z^{\alpha}-\gamma)}}\,\,,
\rc\rc
&&I_2(\alpha, \gamma)\,\overset{\mbox{\tiny def}}{=}\,\int_1^{\infty}\,{z^{\alpha}\,\,dz\over \sqrt{(z^{\alpha}-1)(z^{\alpha}-\gamma)^3}}\,\,,
\eear
where $\alpha$ and $\gamma$ are real numbers such that $\alpha>2$ and 
$|\,\gamma\,|<1$. These integrals can be performed in terms of hypergeometric functions, namely:
\bear
&&I_1(\alpha, \gamma)\,=\,{1\over \alpha}\,B\Big(1-{1\over \alpha}, {1\over 2}\Big)\,
F\Big({1\over 2}\,,\,1-{1\over \alpha}\,;\,{3\over 2}-{1\over \alpha};\gamma\,\Big)\,\,,
\rc\rc
&&I_2(\alpha, \gamma)\,=\,{1\over \alpha}\,B\Big(1-{1\over \alpha}, {1\over 2}\Big)\,
F\Big({3\over 2}\,,\,1-{1\over \alpha}\,;\,{3\over 2}-{1\over \alpha};\gamma\,\Big)\,\,,
\eear
where $B\big(x, y \big)$ is the Euler Gamma function : $B\big(x, y \big)=\Gamma(x)\,\Gamma(y)/\Gamma(x+y)$. Let us next consider the integral:
\beq
J(\alpha, \gamma)\,\overset{\mbox{\tiny def}}{=}\,\lim_{R\to\infty}\,
\Bigg[\,\int_1^{R}\,
\sqrt{{z^{\alpha}-\gamma\over z^{\alpha}-1}}\,dz\,-\,R\,\Bigg]\,\,.
\label{J-integral}
\eeq
To evaluate this integral we shall proceed as follows \footnote{We are grateful to Wolfgang M{\"u}ck for correspondence about this method to compute the integral \ref{J-integral}.}. First of all, let us rewrite the integrand in (\ref{J-integral}) as
\bear
&&\sqrt{{z^{\alpha}-\gamma\over z^{\alpha}-1}}\,=\,{d\over dz}\,
\Bigg[\,z\,\sqrt{{z^{\alpha}-1\over z^{\alpha}-\gamma}}\,\Bigg]
\,-\,{\alpha(1-\gamma)\over 2}\,\,{z^{\alpha}\over 
\sqrt{(z^{\alpha}-1)(z^{\alpha}-\gamma)^3}}\,-\,\rc\rc
&&
\qquad\qquad\qquad\qquad\qquad\qquad
-\,{1-\gamma\over \sqrt{(z^{\alpha}-1)(z^{\alpha}-\gamma)}}
\eear
Using this result it is straightforward to relate $J(\alpha, \gamma)$ to the following combination of the integrals $I_1(\alpha, \gamma)$ and $I_2(\alpha, \gamma)$ defined above
\beq
J(\alpha, \gamma)\,=\,(1-\gamma)\,
\Big[\,I_1(\alpha, \gamma)\,-\,{\alpha\over 2}\,\,I_2(\alpha, \gamma)\,\Big]\,\,.
\eeq
Moreover, by using identities satisfied by the hypergeometric functions, one can show that (for $\alpha>2$)
\beq
J(\alpha, \gamma)\,=\,-{\alpha-2\over 2\alpha}\,
B\Big(1-{1\over \alpha}, {1\over 2}\Big)\,
F\Big(-{1\over 2}\,,\,-{1\over \alpha}\,;\,{1\over 2}-{1\over \alpha};\gamma\,\Big)\,\,.
\label{J-int-value}
\eeq

\bibliographystyle{utphys}
\bibliography{gaugegravityrefs}	

\providecommand{\href}[2]{#2}\begingroup\raggedright\begin{thebibliography}{10}

\bibitem{Maldacena:1997re}
J.~M. Maldacena, ``{The large N limit of superconformal field theories and
  supergravity},'' {\em Adv. Theor. Math. Phys.} {\bf 2} (1998)  231--252,
\href{http://arxiv.org/abs/hep-th/9711200}{{\tt arXiv:hep-th/9711200}}.
%%CITATION = HEP-TH/9711200;%%.

\bibitem{Gubser:1998bc}
S.~S. Gubser, I.~R. Klebanov, and A.~M. Polyakov, ``{Gauge theory correlators
  from non-critical string theory},''
  \href{http://dx.doi.org/10.1016/S0370-2693(98)00377-3}{{\em Phys. Lett.} {\bf
  B428} (1998)  105--114},
\href{http://arxiv.org/abs/hep-th/9802109}{{\tt arXiv:hep-th/9802109}}.
%%CITATION = HEP-TH/9802109;%%.

\bibitem{Witten:1998qj}
E.~Witten, ``{Anti-de Sitter space and holography},'' {\em Adv. Theor. Math.
  Phys.} {\bf 2} (1998)  253--291,
\href{http://arxiv.org/abs/hep-th/9802150}{{\tt arXiv:hep-th/9802150}}.
%%CITATION = HEP-TH/9802150;%%.

\bibitem{Aharony:1999ti}
O.~Aharony, S.~S. Gubser, J.~M. Maldacena, H.~Ooguri, and Y.~Oz, ``{Large N
  field theories, string theory and gravity},''
  \href{http://dx.doi.org/10.1016/S0370-1573(99)00083-6}{{\em Phys. Rept.} {\bf
  323} (2000)  183--386},
\href{http://arxiv.org/abs/hep-th/9905111}{{\tt arXiv:hep-th/9905111}}.
%%CITATION = HEP-TH/9905111;%%.

\bibitem{Itzhaki:1998dd}
N.~Itzhaki, J.~M. Maldacena, J.~Sonnenschein, and S.~Yankielowicz,
  ``{Supergravity and the large N limit of theories with sixteen
  supercharges},'' \href{http://dx.doi.org/10.1103/PhysRevD.58.046004}{{\em
  Phys. Rev.} {\bf D58} (1998)  046004},
\href{http://arxiv.org/abs/hep-th/9802042}{{\tt arXiv:hep-th/9802042}}.
%%CITATION = HEP-TH/9802042;%%.

\bibitem{Duff:1994fg}
M.~J. Duff, G.~W. Gibbons, and P.~K. Townsend, ``{Macroscopic superstrings as
  interpolating solitons},''
  \href{http://dx.doi.org/10.1016/0370-2693(94)91260-2}{{\em Phys. Lett.} {\bf
  B332} (1994)  321--328},
\href{http://arxiv.org/abs/hep-th/9405124}{{\tt arXiv:hep-th/9405124}}.
%%CITATION = HEP-TH/9405124;%%.

\bibitem{Boonstra:1997dy}
H.~J. Boonstra, B.~Peeters, and K.~Skenderis, ``{Duality and asymptotic
  geometries},'' \href{http://dx.doi.org/10.1016/S0370-2693(97)01008-3}{{\em
  Phys. Lett.} {\bf B411} (1997)  59--67},
\href{http://arxiv.org/abs/hep-th/9706192}{{\tt arXiv:hep-th/9706192}}.
%%CITATION = HEP-TH/9706192;%%.

\bibitem{Boonstra:1998yu}
H.~J. Boonstra, B.~Peeters, and K.~Skenderis, ``{Brane intersections, anti-de
  Sitter spacetimes and dual superconformal theories},''
  \href{http://dx.doi.org/10.1016/S0550-3213(98)00512-4}{{\em Nucl. Phys.} {\bf
  B533} (1998)  127--162},
\href{http://arxiv.org/abs/hep-th/9803231}{{\tt arXiv:hep-th/9803231}}.
%%CITATION = HEP-TH/9803231;%%.

\bibitem{Boonstra:1998mp}
H.~J. Boonstra, K.~Skenderis, and P.~K. Townsend, ``{The domain wall/QFT
  correspondence},'' {\em JHEP} {\bf 01} (1999)  003,
\href{http://arxiv.org/abs/hep-th/9807137}{{\tt arXiv:hep-th/9807137}}.
%%CITATION = HEP-TH/9807137;%%.

\bibitem{Jevicki:1998yr}
A.~Jevicki and T.~Yoneya, ``{Space-time uncertainty principle and conformal
  symmetry in D particle dynamics},''
  \href{http://dx.doi.org/10.1016/S0550-3213(98)00578-1}{{\em Nucl.Phys.} {\bf
  B535} (1998)  335--348}, \href{http://arxiv.org/abs/hep-th/9805069}{{\tt
  arXiv:hep-th/9805069 [hep-th]}}.

\bibitem{Jevicki:1998qs}
A.~Jevicki, Y.~Kazama, and T.~Yoneya, ``{Quantum metamorphosis of conformal
  transformation in D3-brane Yang-Mills theory},''
  \href{http://dx.doi.org/10.1103/PhysRevLett.81.5072}{{\em Phys.Rev.Lett.}
  {\bf 81} (1998)  5072--5075}, \href{http://arxiv.org/abs/hep-th/9808039}{{\tt
  arXiv:hep-th/9808039 [hep-th]}}.

\bibitem{Jevicki:1998ub}
A.~Jevicki, Y.~Kazama, and T.~Yoneya, ``{Generalized conformal symmetry in
  D-brane matrix models},''
  \href{http://dx.doi.org/10.1103/PhysRevD.59.066001}{{\em Phys. Rev.} {\bf
  D59} (1999)  066001},
\href{http://arxiv.org/abs/hep-th/9810146}{{\tt arXiv:hep-th/9810146}}.
%%CITATION = HEP-TH/9810146;%%.

\bibitem{Skenderis:1999dq}
K.~Skenderis, ``{Field theory limit of branes and gauged supergravities},''
  {\em Fortsch. Phys.} {\bf 48} (2000)  205--208,
\href{http://arxiv.org/abs/hep-th/9903003}{{\tt arXiv:hep-th/9903003}}.
%%CITATION = HEP-TH/9903003;%%.

\bibitem{Skenderis:2002wp}
K.~Skenderis, ``{Lecture notes on holographic renormalization},''
  \href{http://dx.doi.org/10.1088/0264-9381/19/22/306}{{\em Class. Quant.
  Grav.} {\bf 19} (2002)  5849--5876},
\href{http://arxiv.org/abs/hep-th/0209067}{{\tt arXiv:hep-th/0209067}}.
%%CITATION = HEP-TH/0209067;%%.

\bibitem{Wiseman:2008qa}
T.~Wiseman and B.~Withers, ``{Holographic renormalization for coincident
  Dp-branes},'' \href{http://dx.doi.org/10.1088/1126-6708/2008/10/037}{{\em
  JHEP} {\bf 10} (2008)  037},
\href{http://arxiv.org/abs/0807.0755}{{\tt arXiv:0807.0755 [hep-th]}}.
%%CITATION = 0807.0755;%%.

\bibitem{Kanitscheider:2008kd}
I.~Kanitscheider, K.~Skenderis, and M.~Taylor, ``{Precision holography for
  non-conformal branes},''
  \href{http://dx.doi.org/10.1088/1126-6708/2008/09/094}{{\em JHEP} {\bf 09}
  (2008)  094},
\href{http://arxiv.org/abs/0807.3324}{{\tt arXiv:0807.3324 [hep-th]}}.
%%CITATION = 0807.3324;%%.

\bibitem{Kanitscheider:2009as}
I.~Kanitscheider and K.~Skenderis, ``{Universal hydrodynamics of non-conformal
  branes},'' \href{http://dx.doi.org/10.1088/1126-6708/2009/04/062}{{\em JHEP}
  {\bf 0904} (2009)  062}, \href{http://arxiv.org/abs/0901.1487}{{\tt
  arXiv:0901.1487 [hep-th]}}.

\bibitem{Karch:2002sh}
A.~Karch and E.~Katz, ``{Adding flavor to AdS/CFT},'' {\em JHEP} {\bf 06}
  (2002)  043,
\href{http://arxiv.org/abs/hep-th/0205236}{{\tt arXiv:hep-th/0205236}}.
%%CITATION = HEP-TH/0205236;%%.

\bibitem{Pawelczyk:2000hy}
J.~Pawelczyk and S.-J. Rey, ``{Ramond-Ramond flux stabilization of D-branes},''
  \href{http://dx.doi.org/10.1016/S0370-2693(00)01159-X}{{\em Phys. Lett.} {\bf
  B493} (2000)  395--401},
\href{http://arxiv.org/abs/hep-th/0007154}{{\tt arXiv:hep-th/0007154}}.
%%CITATION = HEP-TH/0007154;%%.

\bibitem{Camino:2001at}
J.~Camino, A.~Paredes, and A.~Ramallo, ``{Stable wrapped branes},'' {\em JHEP}
  {\bf 0105} (2001)  011, \href{http://arxiv.org/abs/hep-th/0104082}{{\tt
  arXiv:hep-th/0104082 [hep-th]}}.

\bibitem{Kachru:2009xf}
S.~Kachru, A.~Karch, and S.~Yaida, ``{Holographic Lattices, Dimers, and
  Glasses},'' \href{http://dx.doi.org/10.1103/PhysRevD.81.026007}{{\em
  Phys.Rev.} {\bf D81} (2010)  026007},
\href{http://arxiv.org/abs/0909.2639}{{\tt arXiv:0909.2639 [hep-th]}}.
%%CITATION = ARXIV:0909.2639;%%.

\bibitem{Kachru:2010dk}
S.~Kachru, A.~Karch, and S.~Yaida, ``{Adventures in Holographic Dimer
  Models},'' \href{http://dx.doi.org/10.1088/1367-2630/13/3/035004}{{\em New
  J.Phys.} {\bf 13} (2011)  035004},
\href{http://arxiv.org/abs/1009.3268}{{\tt arXiv:1009.3268 [hep-th]}}.
%%CITATION = ARXIV:1009.3268;%%.

\bibitem{Mueck:2010ja}
W.~M{\"u}ck, ``{The Polyakov Loop of Anti-symmetric Representations as a
  Quantum Impurity Model},''
  \href{http://dx.doi.org/10.1103/PhysRevD.83.066006}{{\em Phys. Rev.} {\bf
  D83} (2011)  066006},
\href{http://arxiv.org/abs/1012.1973}{{\tt arXiv:1012.1973 [hep-th]}}.
%%CITATION = 1012.1973;%%.

\bibitem{Harrison:2011fs}
S.~Harrison, S.~Kachru, and G.~Torroba, ``{A maximally supersymmetric Kondo
  model},''
\href{http://arxiv.org/abs/1110.5325}{{\tt arXiv:1110.5325 [hep-th]}}.
%%CITATION = ARXIV:1110.5325;%%.

\bibitem{Faraggi:2011ge}
A.~Faraggi, W.~M{\"u}ck, and L.~A. Pando~Zayas, ``{One-loop Effective Action of
  the Holographic Antisymmetric Wilson Loop},''
\href{http://arxiv.org/abs/1112.5028}{{\tt arXiv:1112.5028 [hep-th]}}.
%%CITATION = 1112.5028;%%.

\bibitem{Benincasa:2011zu}
P.~Benincasa and A.~V. Ramallo, ``{Fermionic Impurities in Chern-Simons-Matter
  Theories},'' \href{http://dx.doi.org/10.1007/JHEP02(2012)076}{{\em JHEP} {\bf
  1202} (2012)  076},
\href{http://arxiv.org/abs/1112.4669}{{\tt arXiv:1112.4669 [hep-th]}}.
%%CITATION = ARXIV:1112.4669;%%.

\bibitem{Benincasa:2009ze}
P.~Benincasa, ``{A Note on Holographic Renormalization of Probe D-Branes},''
\href{http://arxiv.org/abs/0903.4356}{{\tt arXiv:0903.4356 [hep-th]}}.
%%CITATION = 0903.4356;%%.

\bibitem{Karch:2005ms}
A.~Karch, A.~O'Bannon, and K.~Skenderis, ``{Holographic renormalization of
  probe D-branes in AdS/CFT},'' {\em JHEP} {\bf 04} (2006)  015,
\href{http://arxiv.org/abs/hep-th/0512125}{{\tt arXiv:hep-th/0512125}}.
%%CITATION = HEP-TH/0512125;%%.

\bibitem{vanRees:2011fr}
B.~C. van Rees, ``{Holographic renormalization for irrelevant operators and
  multi-trace counterterms},''
  \href{http://dx.doi.org/10.1007/JHEP08(2011)093}{{\em JHEP} {\bf 1108} (2011)
   093},
\href{http://arxiv.org/abs/1102.2239}{{\tt arXiv:1102.2239 [hep-th]}}.
%%CITATION = ARXIV:1102.2239;%%.

\bibitem{Kondo:1964aa}
J.~Kondo, ``Resistance minimum in dilute magnetic alloys,'' {\em Prog. Theor.
  Phys.} {\bf 32} (1964)  37.

\bibitem{Affleck:1995aa}
I.~Affleck, ``Conformal field theory approach to the kondo effect,'' {\em Acta
  Phys. Polon.} {\bf B26} (1995)  1869,
  \href{http://arxiv.org/abs/cond-mat/9512099}{{\tt arXiv:cond-mat/9512099}}.

\bibitem{Hewson:1997aa}
A.~C. Hewson, ``{The Kondo Model to Heavy Fermions},'' {\em Cambridge
  University Press} (1993)  .

\bibitem{Affleck:2008aa}
I.~Affleck, ``{Quantum Impurity Problems in Condensed Matter Physics},''
  \href{http://arxiv.org/abs/0809.3474}{{\tt arXiv:0809.3474 [cond-mat]}}.

\bibitem{Anderson:1970aa}
P.~W. Anderson, ``{ A poor man's derivation of scaling laws for the Kondo
  problem},'' {\em J. Phys.} {\bf C3} (1970)  2346.

\bibitem{Wilson:1975aa}
K.~G. Wilson, ``{ The renormalization group: Critical phenomena and the Kondo
  problem},'' {\em Rev. Mod. Phys.} {\bf 47} (1975)  773.

\bibitem{Nozieres:1975aa}
P.~Nozi{\'e}res {\em Proc. of 14th Int. Conf. on Low Temp. Phys.} {\bf 5}
  (1975)  339.

\bibitem{Nozieres:1980aa}
P.~Nozi{\`e}res and A.~Blandin, ``{Kondo Effect in Real Materials},'' {\em J.
  Phys. (Paris)} {\bf 41} (1980)  193.

\bibitem{Cox:1997aa}
D.~Cox and A.~Zawadowski, ``{Exotic Kondo Effects in Metals: Magnetic Ions in a
  Crystalline Electric Field and Tunneling Centers},'' {\em Adv. Phys.} {\bf
  47} (1998)  599, \href{http://arxiv.org/abs/cond-mat/9704103}{{\tt
  arXiv:cond-mat/9704103}}.

\bibitem{Parcollet:1997aa}
O.~Parcollet, A.~Georges, G.~Kotliar, and A.~Sengupta, ``{Overscreened
  Multichannel SU(N) Kondo Model: Large-N Solution and Conformal Field
  Theory},'' {\em Phys. Rev. B} {\bf 58} (1998)  3794--3813,
  \href{http://arxiv.org/abs/cond-mat/9711192}{{\tt arXiv:cond-mat/9711192}}.

\bibitem{Tomonaga:1950aa}
S.-i. Tomonaga, ``Remarks on bloch's method of sound waves applied to
  many-fermion problems,'' \href{http://dx.doi.org/10.1143/PTP.5.544}{{\em
  Progress of Theoretical Physics} {\bf 5} (1950) no.~4, 544--569}.
  \url{http://ptp.ipap.jp/link?PTP/5/544/}.

\bibitem{Luttinger:1963aa}
J.~Luttinger, ``An exactly soluble model of a many-fermion system,'' {\em J.
  Math. Phys.} {\bf 4} (1963)  1154.

\bibitem{Mattis:1965aa}
D.~C. Mattis and E.~H. Lieb, ``{Exact Solution of a Many-Fermion System and Its
  Associated Boson Field},'' {\em J. Math. Phys.} {\bf 6} (1965)  304--312.

\bibitem{Haldane:1981aa}
F.~D.~M. Haldane, ``{`Luttinger Liquid Theory' of One-Dimensional Quantum
  Liquid: I. Properties of the Luttinger Model and Their Extension to the
  General 1D Interacting Spinless Fermi Gas},'' {\em J. Phys. C} {\bf 14}
  (1981)  2585--2609.

\bibitem{Schulz:1998aa}
H.~J. Schulz, G.~Cuniberti, and P.~Pieri, ``{Fermi Liquids and Luttinger
  Liquids},'' {\em G. Morandi {\it et al} eds, Springer} (2000)  ,
  \href{http://arxiv.org/abs/cond-mat/9807366}{{\tt arXiv:cond-mat/9807366}}.

\bibitem{Giamarchi:2009aa}
T.~Giamarchi, ``Quantum physics in one-dimensions,'' {\em Oxford University
  Press} (2003)  .

\bibitem{Frojdh:1995aa}
P.~Frojdh and H.~Johanneson, ``Kondo effect in a luttinger liquid: Exact
  results from conformal field theory,''
  \href{http://dx.doi.org/10.1103/PhysRevLett.75.300}{{\em Phys. Rev. Lett.}
  {\bf 75} (1995)  300}, \href{http://arxiv.org/abs/cond-mat/9505100}{{\tt
  arXiv:cond-mat/9505100}}.

\bibitem{Gomis:2006sb}
J.~Gomis and F.~Passerini, ``{Holographic Wilson loops},''
  \href{http://dx.doi.org/10.1088/1126-6708/2006/08/074}{{\em JHEP} {\bf 08}
  (2006)  074},
\href{http://arxiv.org/abs/hep-th/0604007}{{\tt arXiv:hep-th/0604007}}.
%%CITATION = HEP-TH/0604007;%%.

\bibitem{Yamaguchi:2006tq}
S.~Yamaguchi, ``{Wilson loops of anti-symmetric representation and D5-
  branes},'' \href{http://dx.doi.org/10.1088/1126-6708/2006/05/037}{{\em JHEP}
  {\bf 05} (2006)  037},
\href{http://arxiv.org/abs/hep-th/0603208}{{\tt arXiv:hep-th/0603208}}.
%%CITATION = HEP-TH/0603208;%%.

\bibitem{vanRees:2011ir}
B.~C. van Rees, ``{Irrelevant deformations and the holographic Callan-Symanzik
  equation},'' \href{http://dx.doi.org/10.1007/JHEP10(2011)067}{{\em JHEP} {\bf
  1110} (2011)  067},
\href{http://arxiv.org/abs/1105.5396}{{\tt arXiv:1105.5396 [hep-th]}}.
%%CITATION = ARXIV:1105.5396;%%.

\bibitem{Petkou:1999fv}
A.~Petkou and K.~Skenderis, ``{A Nonrenormalization theorem for conformal
  anomalies},'' \href{http://dx.doi.org/10.1016/S0550-3213(99)00514-3}{{\em
  Nucl.Phys.} {\bf B561} (1999)  100--116},
\href{http://arxiv.org/abs/hep-th/9906030}{{\tt arXiv:hep-th/9906030
  [hep-th]}}.
%%CITATION = HEP-TH/9906030;%%.

\bibitem{Kruczenski:2003be}
M.~Kruczenski, D.~Mateos, R.~C. Myers, and D.~J. Winters, ``{Meson spectroscopy
  in AdS / CFT with flavor},'' {\em JHEP} {\bf 0307} (2003)  049,
  \href{http://arxiv.org/abs/hep-th/0304032}{{\tt arXiv:hep-th/0304032
  [hep-th]}}.

\bibitem{Karaiskos:2011kf}
N.~Karaiskos, K.~Sfetsos, and E.~Tsatis, ``{Brane embeddings in sphere
  submanifolds},'' \href{http://dx.doi.org/10.1088/0264-9381/29/2/025011}{{\em
  Class.Quant.Grav.} {\bf 29} (2012)  025011},
\href{http://arxiv.org/abs/1106.1200}{{\tt arXiv:1106.1200 [hep-th]}}.
%%CITATION = ARXIV:1106.1200;%%.

\bibitem{sage}
W.~Stein {\em et al.}, {\em {S}age {M}athematics {S}oftware ({V}ersion 4.6.1)}.
\newblock The Sage Development Team, 2011.
\newblock {\tt http://www.sagemath.org}.

\bibitem{maxima}
Maxima, ``Maxima, a computer algebra system. version 5.25.1,'' 2011.
\newblock \url{http://maxima.sourceforge.net/}. {\tt
  http://maxima.sourceforge.net/}.

\bibitem{scipy}
E.~Jones, T.~Oliphant, P.~Peterson, {\em et al.}, ``{SciPy}: Open source
  scientific tools for {Python},'' 2001--.
\newblock \url{http://www.scipy.org/}. http://www.scipy.org/.

\end{thebibliography}\endgroup

\end{document}